\documentclass[11pt,a4paper]{article}
\usepackage{jheppub}   % loads amsmath, amssymb, graphicx, natbib, hyperref
\allowdisplaybreaks

\renewcommand{\thesection}{\arabic{section}}

\newcommand{\nn}{\nonumber}

\usepackage{color}
\hypersetup{colorlinks=true,linkcolor=blue,citecolor=blue,urlcolor=cyan}
\newcommand{\arxiv}[1]{\mbox{[\href{https://arxiv.org/abs/#1}{\textcolor{magenta}{\ttfamily arXiv:#1}}]}}

\makeatletter
\renewcommand{\@email}[1]{\href{mailto:#1}{\small\tt #1}}
\makeatother

\title{\boldmath Carrollian minimal massive gravity and the fate of third-way
consistency}

\author[a]{B\"u\c{s}ra Dedeo\u{g}lu,}
\author[a]{\"Ozg\"ur Sar{\i}o\u{g}lu,}
\author[b]{and Utku Zorba}

\affiliation[a]{Department of Physics, Faculty of Arts and Sciences,
Middle East Technical University, 06800 Ankara, T\"urkiye}
\affiliation[b]{Department of Engineering Sciences,
\.{I}stanbul University-Cerrahpa\c{s}a, Avc{\i}lar, 34320 \.{I}stanbul, T\"urkiye}

\emailAdd{busrad@metu.edu.tr}
\emailAdd{sarioglu@metu.edu.tr}
\emailAdd{utku.zorba@iuc.edu.tr}

\abstract{We construct the Carrollian limit of minimal massive gravity. The limit is singular for
the algebra and regular for the Chern--Simons-like data. The one divergent bracket points
along the null direction of the contracted bilinear form. The Lagrangian is therefore
finite while the brackets are not. The contracted curvature still obeys a cubic closure
condition, forced on shell rather than identically. The third-way coupling itself,
however, is removable. For generic parameters an invertible redefinition of the auxiliary
field and the boost connection takes it to Carrollian topologically massive gravity up to
a boundary term. The same redefinition fails before the limit. We solve for an
anti-de~Sitter vacuum and a Carrollian black hole, whose rotation parameter leaves the
frame and survives only in the boost connection. The degenerate vector flavor metric
obstructs the standard charge formula, which we replace by a canonical one. The boundary
charges, obtained in two independent ways, carry no trace of the deformation. A Dirac
analysis gives nine first-class constraints and no local degrees of freedom on any
background. The massive graviton is removed. The asymptotic symmetry algebra is
$\mathfrak{bms}_3$, with central charges carried by the Einstein and
Lorentz--Chern--Simons terms. The black hole has two sets of thermodynamic data, each with
its own first law. One obeys the Bekenstein--Hawking area law, the other keeps the
Chern--Simons entropy.}


\begin{document}
\maketitle
\flushbottom
%--- display spacing, set here because font-size changes reset it ------
\setlength{\parskip}{0pt plus 1.5pt}
\setlength{\abovedisplayskip}{7pt plus 3pt minus 4pt}
\setlength{\belowdisplayskip}{7pt plus 3pt minus 4pt}
\setlength{\abovedisplayshortskip}{3pt plus 2pt}
\setlength{\belowdisplayshortskip}{4pt plus 2pt minus 2pt}

%=======================================================================
%  SECTION 1 :  INTRODUCTION
%=======================================================================
\section{Introduction}\label{intro}

Gravity in three spacetime dimensions carries no local degrees of freedom. The
cosmological Einstein theory is nevertheless far from empty. It has black holes
\cite{Banados:1992wn,Banados:1992gq} and a boundary algebra of charges with a non-zero
central extension \cite{Brown:1986nw}, and it can be recast as a Chern--Simons gauge
theory \cite{Achucarro:1986uwr,Witten:1988hc}. Giving the graviton a mass without losing
any of that is what the three-dimensional massive gravities were built to do. Adding
curvature-squared terms generically brings in ghosts. Topologically massive gravity (TMG)
\cite{Deser:1981wh,Deser:1982vy} avoids them. It adds the Lorentz--Chern--Simons term
instead, and it propagates a single massive graviton. Theories of this kind are no longer
Chern--Simons theories. They retain enough of the structure to be written in a
Chern--Simons-like form, with a fixed number of one-form flavors and a flavor metric in
place of an invariant bilinear form
\cite{Hohm:2012vh,Bergshoeff:2014bia,Afshar:2014ffa,Merbis:2014vja}.

%--- third-way consistency and MMG -------------------------------------
TMG pays for its graviton at the boundary. The sign of the Einstein term that gives the
bulk mode positive energy is the sign that makes one of the boundary central charges
negative \cite{Li:2008dq}. Minimal massive gravity (MMG) \cite{Bergshoeff:2014pca} removes
that clash, and it does so by giving up something else. The divergence of its metric field
equation does not vanish off shell. It vanishes only after the field equation itself is
substituted back into it, and no metric Lagrangian returns that equation by ordinary
variation \cite{Bergshoeff:2015zga,Ozkan:2018cxj}. Models with this
property are called \emph{third-way consistent}
\cite{Bergshoeff:2014pca,Bergshoeff:2015zga}. MMG is the smallest of a family of them,
whose Chern--Simons-like formulations use more one-form flavors
\cite{Ozkan:2018cxj,Dedeoglu:2024ies,Dedeoglu:2026ceb}.

%--- the Carroll limit -------------------------------------------------
The Carrollian limit of a Lorentz-invariant theory is what survives when the speed of
light is sent to zero ($c\to0$ or $1/c\to\infty$). Limits of this kind have attracted
much interest recently in view of their potential applications; see
e.g.\ \cite{Bergshoeff:2022eog,Bagchi:2025vri} for reviews. They are not unique. A Lorentz-invariant
theory admits an electric and a magnetic Carroll contraction \cite{Henneaux:2021yzg}, and
for gravity the magnetic one is the theory obtained by gauging the Carroll algebra
\cite{Hartong:2015xda,Campoleoni:2022ebj,Concha:2025vhd}. It is the magnetic limit that we
take here.

%--- the question, and why MMG -----------------------------------------
This paper asks what becomes of third-way consistency in that limit.\footnote{The
restriction to three dimensions is a restriction on gravity. A third-way consistent
Yang--Mills theory exists in three dimensions \cite{Arvanitakis:2015oga}, and interacting
$(d-2)$-form theories with the same property have been constructed for any $d\geq3$
\cite{Broccoli:2021pvv}; see \cite{Deger:2021ojb} for a review. What singles out three
dimensions is that a $(d-2)$-form is a one-form there, hence a connection, which is what
lets the deformation act on the spin connection of a first-order theory. No third-way
consistent gravity model is known for $d>3$.} MMG is the place to ask it. It is the
original third-way consistent gravity and the smallest one. Its Chern--Simons-like
formulation needs only three one-form flavors, so every structure can be exhibited in
closed form rather than estimated. It also contains TMG as the limit in which the
deformation responsible for third-way consistency is switched off, and that lets the
effect of the deformation be isolated at every step of the calculation.

%--- the tension between the two ingredients ---------------------------
The question has some bite, because the two ingredients pull against each other. A Carroll
contraction is an operation on a Lie algebra. One assigns weights in a contraction
parameter to the generators and keeps the leading term, and it is the Jacobi identity that
guarantees the result is again a Lie algebra. Third-way consistency, in the first-order
formulation, is the statement that the Jacobi identity \emph{fails}. The one-form fields of
a Chern--Simons-like model are not the gauge fields of any Lie algebra. The curvature built
from them obeys $DF=\frac12[A,[A,A]]$, which is the Jacobiator evaluated on three copies of
the connection. The closure condition $DF=0$ is therefore not an identity. It is an
algebraic constraint forced on every solution of the field equations. There is thus no
reason to expect the contraction to be defined at all, and none to expect the result to
retain the property.

%--- what the answer turns out to be -----------------------------------
What happens is more interesting than either. The contraction is singular at the level
of the brackets and regular at that of the Chern--Simons-like data. The only
divergent bracket points along the null direction of the contracted bilinear form. The
structure constants with an upper flavor index have no limit, while their
lowered counterparts $f_{RST}=g_{RU}f^{U}{}_{ST}$ do. The Lagrangian survives the limit
although the algebra does not, and the closure condition with it. The contracted
curvature still satisfies a non-trivial cubic condition, holding on every solution and not
identically. What does not survive is the coupling that carries the deformation,
and that is the second half of the answer.

%--- summary of results ------------------------------------------------
Our results are as follows. The Carroll contraction of the MMG Lagrangian is finite and
defines a theory we call Carrollian MMG. We solve its field equations for an
anti-de~Sitter (AdS) vacuum and for a Carrollian counterpart of the BTZ black hole. The
rotation parameter disappears from the Carrollian frame and survives only in the boost
connection. The black hole admits a one-parameter deformation of its auxiliary field. It
is not a diffeomorphism, and carries neither energy nor angular momentum. It is an improper gauge transformation, charged only under a generator that
moves the radial coordinate. Third-way consistency holds in the first-order sense above. The deformation itself, however, does not survive the contraction as a
coupling. For generic parameters an invertible redefinition of the auxiliary field and the
boost connection removes it, so that Carrollian MMG and Carrollian TMG are one theory. The
redefinition has no relativistic counterpart, and it is what lies behind the repeated
disappearance of the deformation below.

The Chern--Simons-like structure of the contracted theory is governed by two flavor
metrics, one on the rotation-scalar sector and one on the spatial-vector sector, and the
second of these is degenerate. That single fact organizes everything that follows. It
produces a four-dimensional kernel in the matrix of primary constraint brackets, and it
forces the temporal component of the auxiliary field to vanish. It is the reason the
relativistic formula for asymptotic charges, which requires the inverse flavor metric, is
unavailable and must be replaced by a canonical construction. Together with the
redefinition just described, it is also what keeps the MMG deformation out of every
boundary charge the theory possesses.

The boundary charges are computed twice, by routes that share no intermediate step: from
the canonical charge of the Chern--Simons-like form, and from the boundary term of the
Castellani generator built out of the first-class constraints. They agree, and neither the energy nor the angular momentum
depends on the couplings that make the model third-way consistent. A Dirac analysis
returns nine first-class constraints and hence no local degrees of freedom, on every
background rather than at a single vacuum. The same holds for Carrollian TMG, whose massive graviton the contraction therefore removes, and by the
identification above it could not have come out otherwise.

The absence of propagating modes should not be read as a sign of triviality.
Off shell Carrollian MMG is not a Chern--Simons theory. Its vector flavor metric is
degenerate, its field equations are third-way consistent, and its charges are non-trivial.
The failure of the Jacobi identity is not by itself what distinguishes it, since the
relativistic parents share that. On shell, on the maximally symmetric branch, the content
does reduce to an AdS--Carroll Chern--Simons theory, and Section~\ref{bms} uses that
reduction to obtain the central charges. The asymptotic symmetry algebra is
$\mathfrak{bms}_3$ with two central charges, one supported by the Einstein term and one by
the Lorentz--Chern--Simons term. The Carrollian black hole obeys a first law with a
vanishing angular velocity, its temperature fixed by an intrinsic Carroll surface gravity
and its entropy obeying the Bekenstein--Hawking area law. A second set of thermodynamic
data, inherited from the inner horizon of the relativistic solution, satisfies a first law
of its own and keeps the Chern--Simons contribution to the entropy. We obtain that second
set twice, once as a limit and once from the holonomy of the Carrollian Chern--Simons
connection. With no local degrees of freedom, these boundary data are not one part of the
physical content of the theory but the whole of it.

%--- outline -----------------------------------------------------------
The paper is organized as follows. Section~\ref{mmgsec} sets up MMG and performs the
Carroll contraction of its Lagrangian and of its underlying algebra, recording why the
first survives while the second does not. It also derives the Carrollian field equations
\eqref{carroll-eom} that every later section uses, writes them covariantly, shows that
genericity forces the temporal auxiliary field to vanish, $\bar{\tau}=0$, sets up
third-way consistency in first-order form, and identifies the
limit as magnetic. Section~\ref{adsbtz} constructs the AdS vacuum, the Carrollian BTZ
solution, and a one-parameter deformation of its auxiliary field.
Section~\ref{third-way} carries third-way consistency through the contraction.
Section~\ref{CCSlike} gives the general Carrollian Chern--Simons-like framework, identifies
the flavor data, and supplies the constraints and the boundary-charge formula
used in Sections~\ref{bdrycharges}, \ref{Dirac} and \ref{asympt}.
Section~\ref{bdrycharges} computes the boundary charges of the Carrollian black hole and
compares them with their relativistic counterparts. Section~\ref{Dirac} carries out the
Hamiltonian analysis and counts the degrees of freedom, twice, first by an argument that
selects no background and then by evaluating the rank at the flat vacuum. It then runs
the same count for Carrollian TMG. Section~\ref{asympt} obtains the
asymptotic charges canonically, which is the second route, and compares the two.
Section~\ref{bms} derives the asymptotic symmetry algebra and its central charges, and
Section~\ref{thermo} the thermodynamics of the black hole. We conclude in
Section~\ref{conclusions}.

%--- conventions -------------------------------------------------------
\emph{Conventions.} The Lorentz metric is $\eta_{ab}=\mathrm{diag}(-,+,+)$, with indices
$a,b,c=0,1,2$, so that $V_0=-V^{0}$ and $V_i=V^{i}$ for any Lorentz vector $V^{a}$, and
$\epsilon_{012}=+1$. After the Carroll split the internal spatial
indices $i,j=1,2$ are raised and lowered with $\delta_{ij}$ and written down throughout. Both the internal symbol $\epsilon_{ij}$ and the spatial coordinate symbol
$\epsilon_{KL}$, with $K,L=1,2$, are normalized as $\epsilon_{12}=+1$. The relativistic
dreibein, spin connection and auxiliary field are in typewriter font,
$(\mathtt{e}^{a},\mathtt{w}^{a},\mathtt{h}^{a})$, their Carrollian counterparts in italic.
Wedge products are frequently suppressed, so that
$\tau\mathcal{R}(\omega)$ stands for $\tau\wedge\mathcal{R}(\omega)$. Throughout we take
$\sigma=\pm1$, and the overall normalization of the action is $1/(8\pi G)$. The flavor
labels $I,J$ and $r,s$ appear in Section~\ref{CCSlike}, the canonical indices in
Section~\ref{Dirac}.
Relativistic couplings and solution parameters carry a tilde and their Carrollian
counterparts none, as in $(\tilde{\sigma},\tilde{\gamma},\tilde{\alpha},\tilde{\Lambda}_0,
\tilde{\mu},\tilde{q},\tilde{c})$ against $(\sigma,\gamma,\alpha,\Lambda_0,\mu,q,c)$. What
the two theories share carries none either: the mass $M$, the AdS radius $\ell$, Newton's
constant $G$, the coordinates $r$ and $\phi$, and the ratio
$\alpha/\gamma=\tilde{\alpha}/\tilde{\gamma}$, which \eqref{rescaling-fields} leaves fixed.
A prime on a Carrollian field marks a field redefinition, and a hat on a flavor label a
basis vector of flavor space.

%=======================================================================
%  SECTION 2 :  MMG AND ITS CARROLL CONTRACTION
%=======================================================================
\section{MMG and its Carroll contraction}\label{mmgsec}
Let us start by writing a slightly modified version of the
original MMG Lagrangian \cite{Bergshoeff:2014pca}
%--- 2.1  the MMG Lagrangian and the dimensionful couplings ------------
\begin{equation}\label{mmg-lagrangian}
L_{\text{MMG}}
=-\tilde{\sigma} \mathtt{e}^a \mathtt{R}_a
+\frac{\tilde{\Lambda}_0}{6}\epsilon^{abc}
\mathtt{e}_a\mathtt{e}_b\mathtt{e}_c
+\tilde{\gamma}\mathtt{h}^a D\mathtt{e}_a
+\frac{1}{2\tilde{\mu}}\mathtt{w}^a
\left(d\mathtt{w}_a+\frac{1}{3}\epsilon_{abc}
\mathtt{w}^b\mathtt{w}^c\right)
+\frac{\tilde{\alpha}}{2}\epsilon^{abc}
\mathtt{e}_a\mathtt{h}_b\mathtt{h}_c\,.
\end{equation}
In the original article \cite{Bergshoeff:2014pca}, the parameter $\tilde{\gamma}$ was taken to be $1$ since both the
dreibein and the Schouten tensor were assumed to be dimensionless. In SI units, with
$[\mathtt{e}^a]=L$ and $[\mathtt{h}^a]=L^{-1}$, it follows that $[\tilde{\Lambda}_0]  = L^{-2}$ and
$[\tilde{\mu}] = L^{-1}$. Therefore one needs a dimensionful parameter for the coefficient of the
$\mathtt{h}\cdot D\mathtt{e}$ term, with $[\tilde{\gamma}]=L$ and $[\tilde{\alpha}]=L^2$.
Nothing beyond this bookkeeping depends on $\tilde{\gamma}$. The rescaling
$\mathtt{h}^a\rightarrow \mathtt{h}^a/\tilde{\gamma}$ returns the coefficient of
$\mathtt{h}\cdot D\mathtt{e}$ to unity and sends $\tilde{\alpha}\rightarrow\tilde{\alpha}/\tilde{\gamma}^{2}$, so
the relativistic theory sees $\tilde{\alpha}$ and $\tilde{\gamma}$ only through $\tilde{\alpha}/\tilde{\gamma}^{2}$. We
keep $\tilde{\gamma}$ explicit because the contraction below treats $\mathtt{h}^0$ and
$\mathtt{h}^i$ differently, and the dimensions are easier to follow with it in place.

%--- 2.2  Carroll rescaling of the fields and of the couplings ---------
To obtain Carrollian MMG, the fields are rescaled such that
\begin{align} \label{rescaling-fields}
  \mathtt{e}^a=(\mathtt{e}^0,\mathtt{e}^i) \rightarrow   \left( \frac{1}{\rho} \tau,e_i \right) \,, \quad
  \mathtt{w}^a=(\mathtt{w}^0,\mathtt{w}^i) \rightarrow   \left(  \omega,\frac{1}{\rho}\varpi_i \right) \,, \quad
  \mathtt{h}^a=(\mathtt{h}^0,\mathtt{h}^i) \rightarrow   \left(  \bar{\tau},\frac{1}{\rho}h_i \right) \,.
\end{align}
Here $\rho$ is a dimensionless contraction parameter, and on the right-hand sides
$(\sigma,\gamma,\alpha,\Lambda_0,\mu)$ denote the \emph{Carrollian} couplings, held
fixed as $\rho\to\infty$; the couplings of \eqref{mmg-lagrangian} are
$\tilde{\sigma}=\rho\sigma$, $\tilde{\gamma}=\rho\gamma$,
$\tilde{\alpha}=\rho\alpha$, $\tilde{\Lambda}_0=\rho\Lambda_0$ and
$\tilde{\mu}=\mu$. This assignment is fixed by the structure of the limit rather
than chosen. Under the field redefinition \eqref{rescaling-fields} alone, every term
of \eqref{mmg-lagrangian} except the Lorentz--Chern--Simons term is of order
$\rho^{-1}$, while the Chern--Simons term is already of order $\rho^{0}$. Hence
finiteness of the limit requires precisely one power of $\rho$ on each of
$\sigma,\gamma,\alpha,\Lambda_0$ and none on $\mu$.\footnote{Note that
these also imply that $\tilde{q}=\rho q$ is automatic rather than an independent choice, where
$q$ is the combination \eqref{genericity} below.
Equivalently, up to an overall factor of $\rho$ in the action, the same limit may be
presented as \eqref{rescaling-fields} together with $\tilde{\mu}=\rho\mu$ at fixed
$(\sigma,\gamma,\alpha,\Lambda_0)$. That presentation makes the conventional normalization
$\sigma=\pm1$ manifest. We use the form above instead, so that the overall normalization
$1/(8\pi G)$ in the action is left untouched.}

%--- 2.3  standing conventions and genericity assumptions --------------
We take $\gamma\neq0$ and $\mu\neq0$, as \eqref{mmg-lagrangian} already
requires. In what follows, two generic
assumptions are used repeatedly and are collected here for convenience. The first is
\begin{equation} \label{genericity}
q:=\frac{\mu}{\gamma}\left(\gamma^{2}+\sigma\alpha\right)\neq0
\qquad\Longleftrightarrow\qquad \gamma^{2}+\sigma\alpha\neq0\,,
\end{equation}
which is what makes the model Chern--Simons-\emph{like} rather than Chern--Simons: at
$q=0$ the algebra of Section~\ref{mmgalg} closes, the Jacobi identity is restored, and
several of the arguments of Sections~\ref{Dirac} and \ref{rankcheck} cease to apply. The
second is that the spatial zweibein $e_{iK}$ is invertible, with $K=1,2$ a spatial
coordinate index, and that $\tau$ is nowhere vanishing, so that $(\tau,e_1,e_2)$ is a
basis of one-forms. This is what allows a two-form annihilated by wedging with $\tau$ and
with both $e_i$ to be set to zero, a step used in Sections~\ref{adsbtz},
\ref{third-way} and \ref{Dirac}. The two halves of the assumption come apart on the
black hole of Section~\ref{btzsub}, where Section~\ref{thermo} finds that $\tau$
vanishes on one locus while $e_{iK}$ stays invertible.

%--- 2.4  the rescaled Lagrangian and the rho -> infinity limit --------
With these, the rescaled Lagrangian becomes, with no total derivative discarded,
\begin{align} \label{rescaled-lagrangian}
    L_{\text{rescaled}}=&\sigma \tau \mathcal{R}(\omega)-\sigma e_i \mathcal{R}_i(\varpi)-\gamma \bar{\tau} \mathcal{R}(\tau)+\gamma h_i \mathcal{R}_i(e)-\frac{1}{2\mu}\omega \mathcal{R}(\omega)+\frac{1}{2\mu \rho^2}\varpi_i \mathcal{R}_i(\varpi) \nn \\
    &-\frac{\sigma}{2\rho^2}\epsilon_{ij} \tau \varpi_i \varpi_j +\frac{\Lambda_0}{2}\epsilon_{ij}\tau e_i e_j +\frac{\gamma}{\rho^2} \epsilon_{ij}h_i \varpi_j \tau +\alpha \epsilon_{ij}e_ih_j \bar{\tau}+\frac{\alpha}{2\rho^2}\epsilon_{ij}\tau h_ih_j\,,
\end{align}
where
\begin{align} \label{curvature-defs}
\mathcal{R}(\tau):=&d\tau-\epsilon_{ij}e_i\varpi_j\,, &
\mathcal{R}(\bar{\tau}):=&d\bar{\tau}-\frac{1}{\rho^2}\epsilon_{ij}h_i\varpi_j\,, \nn \\
\mathcal{R}_i(e):=&de_i- \epsilon_{ij} \omega e_j\,, &
\mathcal{R}_i(h):=&dh_i- \epsilon_{ij} \omega h_j\,, \nn \\
\mathcal{R}(\omega):=&d\omega\,, &
\mathcal{R}_i(\varpi):=&d\varpi_i- \epsilon_{ij} \omega \varpi_j \,.
\end{align}
The subscript on $\mathcal{R}_i$ carries the free spatial index and the argument names
the field whose curvature is taken: $\mathcal{R}(\omega)$ belongs to the Carrollian
rotation connection $\omega$, and $\mathcal{R}_i(\varpi)$ to the boost connection
$\varpi_i$. Throughout, $\omega$ and $\varpi_i$ are independent fields; the two
symbols are never used for components of a single object.
As $\rho \rightarrow \infty$, the Lagrangian \eqref{rescaled-lagrangian} becomes
\begin{equation} \label{carrollian-mmg-lagrangian}
\boxed{\;
\begin{aligned}
    L_{\text{CMMG}}=\;&\sigma \tau \mathcal{R}(\omega)-\sigma e_i \mathcal{R}_i(\varpi)
     -\gamma \bar{\tau} \mathcal{R}(\tau)+\gamma h_i \mathcal{R}_i(e)
     -\frac{1}{2\mu}\omega \mathcal{R}(\omega) \\
    &+\frac{\Lambda_0}{2}\epsilon_{ij}\tau e_i e_j+\alpha \epsilon_{ij}e_ih_j \bar{\tau}\,.
\end{aligned}
\;}
\end{equation}
This is the Lagrangian of the Carrollian MMG model we examine in what follows.
%--- 2.7  the Carrollian field equations -------------------------------
The corresponding field equations that follow from \eqref{carrollian-mmg-lagrangian} are
\begin{subequations} \label{carroll-eom}
\begin{align}
    \delta \tau \, : & \quad 0=\sigma \mathcal{R}(\omega) +\frac{\Lambda_0}{2}\epsilon_{ij}e_ie_j-\gamma d\bar{\tau}\,, \label{eom-tau} \\
    \delta\bar{\tau} \, : & \quad 0=\mathcal{R}(\tau)-\frac{\alpha}{\gamma}\epsilon_{ij}e_ih_j\,, \label{eom-bartau} \\
    \delta \omega \, : & \quad 0=\sigma\mathcal{R}(\tau)-\frac{1}{\mu}\mathcal{R}(\omega)+\gamma\epsilon_{ij}e_ih_j\,, \label{eom-omega} \\
    \delta e_i \, : & \quad 0=\sigma\mathcal{R}_i(\varpi)-\gamma\mathcal{R}_i(h)+\Lambda_0\epsilon_{ij}\tau e_j+\gamma \epsilon_{ij}\bar{\tau}\varpi_j+\alpha\epsilon_{ij}\bar{\tau}h_j\,, \label{eom-e} \\
    \delta h_i \, : & \quad 0=\gamma\mathcal{R}_i(e)+\alpha\epsilon_{ij}e_j \bar{\tau}\,, \label{eom-h} \\
    \delta \varpi_i \, : & \quad 0=\sigma\mathcal{R}_i(e)+\gamma\epsilon_{ij}\bar{\tau}e_j\,. \label{eom-varpi}
\end{align}
\end{subequations}

%--- 2.8  the deformation is removable after the contraction -----------
One property of \eqref{carrollian-mmg-lagrangian} is worth recording at once, because it
organizes much of what follows. For $q\neq0$ set
\begin{equation} \label{alpha-removal}
\varpi'_i:=\varpi_i+\frac{\alpha}{\gamma}h_i\,, \qquad
h'_i:=\frac{q}{\mu\gamma}h_i\,,
\end{equation}
a change of variables that is invertible precisely because $q\neq0$. Writing
$L_{\text{CMMG}}[\alpha]$ for \eqref{carrollian-mmg-lagrangian} at that value of the
coupling, one finds identically and off shell
\begin{equation} \label{cmmg-is-ctmg}
L_{\text{CMMG}}[\alpha]\left(\tau,\bar{\tau},\omega,e_i,h_i,\varpi_i\right)
=L_{\text{CMMG}}[0]\left(\tau,\bar{\tau},\omega,e_i,h'_i,\varpi'_i\right)
-\frac{\sigma\alpha}{\gamma}\,d\left(e_ih_i\right)\,.
\end{equation}
At $\alpha=0$ the auxiliary field is a Lagrange multiplier, the relativistic theory is TMG
and the right-hand side of \eqref{cmmg-is-ctmg} is its Carroll contraction. This is taken
up again in Section~\ref{Fcount}. A further rescaling
$\left(\bar{\tau},h'_i\right)\rightarrow\left(\bar{\tau},h'_i\right)/\gamma$
removes $\gamma$ as well. So in the bulk, and for $q\neq0$, Carrollian MMG is Carrollian
TMG, and the couplings it carries are $\sigma$, $\Lambda_0$ and $\mu$ alone.
The combination $\varpi'_i$ is not a new object. It is the spatial part of the
torsion-free connection $\Omega:=\mathtt{w}+\frac{\tilde{\alpha}}{\tilde{\gamma}}\mathtt{h}$, which
Section~\ref{adsbtz} uses to construct the solutions. What \eqref{alpha-removal} says is
that after the contraction one may pass to that connection in the spatial directions and
leave the Lagrangian alone up to a boundary term.

The same redefinition fails before the contraction, and the reason is worth a line. Its
relativistic counterpart shifts the whole connection,
$\mathtt{w}^a\rightarrow \mathtt{w}^a+\frac{\tilde{\alpha}}{\tilde{\gamma}}\mathtt{h}^a$. The
Lorentz--Chern--Simons term of \eqref{mmg-lagrangian} is cubic in $\mathtt{w}^a$, so the
shift hands $\mathtt{h}^a$ a Chern--Simons kinetic term
$\frac{\tilde{\alpha}^{2}}{2\tilde{\mu}\tilde{\gamma}^{2}}\mathtt{h}^a d\mathtt{h}_a$ of its own, and
\eqref{mmg-lagrangian} at $\tilde{\alpha}=0$ holds nothing that could cancel it. After the
contraction the Chern--Simons term is $-\frac{1}{2\mu}\omega \mathcal{R}(\omega)$ and
involves the rotation connection alone. The two terms that would have obstructed the
shift, the boost kinetic term $\frac{1}{2\mu\rho^{2}}\varpi_i \mathcal{R}_i(\varpi)$ and
the quadratic $\frac{\alpha}{2\rho^{2}}\epsilon_{ij}\tau h_ih_j$ of
\eqref{rescaled-lagrangian}, both carry $\rho^{-2}$ and are gone. What is left is linear
in $\varpi_i$ and linear in $h_i$, and a field that enters linearly can be shifted for
free. This is also why only the spatial half of the shift to $\Omega$ is available. The
temporal half would move $\omega$ by $\frac{\alpha}{\gamma}\bar{\tau}$, and $\omega$ is
the one field the Chern--Simons term still carries. Accompanying \eqref{alpha-removal} by
that shift destroys the identity \eqref{cmmg-is-ctmg}.

Two consequences are used repeatedly below. First, everything the model does in the bulk
is independent of $\alpha$ and $\gamma$. The field equations \eqref{carroll-eom}, the
constraint structure of Section~\ref{Dirac} and the count of local degrees of freedom are
all insensitive to a total derivative, so each of them is a Carrollian TMG statement
rewritten. Boundary quantities are not insensitive to it in general, and we follow the
boundary term where it matters. On the solutions of Section~\ref{adsbtz} it is absent as
well, since $h_i=c\,e_i$ there gives $e_ih_i=0$. Second, the genericity assumption
\eqref{genericity} gains a sharper reading. The condition $q\neq0$ is exactly the
condition under which the deformation can be removed, so the critical theory $q=0$ is not
a degenerate limit of the generic one but a different and more symmetric model.

%--- 2.10  the field equations in covariant form -----------------------
The system \eqref{carroll-eom} becomes more transparent once the Carrollian
structure group is made explicit. The Lagrangian
\eqref{carrollian-mmg-lagrangian} is invariant, up to a total derivative, under
local spatial rotations with parameter $\lambda$, under local Carroll boosts with
parameters $\kappa_i$, and under a Carrollian time translation with parameter
$\varsigma$,
\begin{align} \label{rot-boost}
\delta_{\lambda}\omega&=d\lambda\,, &
\delta_{\lambda}V_i&=\epsilon_{ij}\lambda V_j \ \ (V=e,h,\varpi)\,, &
\delta_{\lambda}\tau&=\delta_{\lambda}\bar{\tau}=0\,, \nn \\
\delta_{\kappa}\varpi_i&=d\kappa_i-\epsilon_{ij}\omega\kappa_j\,, &
\delta_{\kappa}\tau&=\epsilon_{ij}\kappa_ie_j\,, &
\delta_{\kappa}h_i&=-\epsilon_{ij}\kappa_j\bar{\tau}\,, \nn \\
\delta_{\varsigma}\tau&=d\varsigma\,, &
\delta_{\varsigma}\varpi_i&=-\frac{\Lambda_0}{\sigma}\epsilon_{ij}e_j\varsigma\,, &&
\end{align}
every field not listed being inert. In particular $e_i$, $\bar{\tau}$ and $\omega$ are
boost inert, while $\tau$ shifts by $e_i$ and $h_i$ by $\bar{\tau}$. Thus $\omega$ and
$\varpi_i$ are the connections of the homogeneous Carroll group. The third transformation
is the time translation generated by $H$. It exists for every $\Lambda_0$, and at constant
$\varsigma$ it reduces to a shift of $\varpi_i$ alone, which is what
Section~\ref{betadef} uses. The associated covariant derivatives are
\begin{align} \label{Dcov}
\mathcal{D}\tau&:=d\tau-\epsilon_{ij}\varpi_ie_j=\mathcal{R}(\tau)\,, &
\mathcal{D}\bar{\tau}&:=d\bar{\tau}=\mathcal{R}(\bar{\tau})\,, \nn \\
\mathcal{D}e_i&:=de_i-\epsilon_{ij}\omega e_j=\mathcal{R}_i(e)\,, &
\mathcal{D}h_i&:=dh_i-\epsilon_{ij}\omega h_j+\epsilon_{ij}\varpi_j\bar{\tau}\,,
\end{align}
the first three of which are already among the curvatures
\eqref{curvature-defs} at $\rho\to\infty$, while the last differs from
$\mathcal{R}_i(h)$ by the boost connection term. With \eqref{Dcov} the field
equations \eqref{carroll-eom} read, in the same order,
\begin{subequations} \label{carroll-eom-cov}
\begin{align}
0&=\sigma\mathcal{R}(\omega)-\gamma\,\mathcal{D}\bar{\tau}
   +\frac{\Lambda_0}{2}\epsilon_{ij}e_ie_j\,, \label{eom-tau-cov} \\
0&=\mathcal{D}\tau-\frac{\alpha}{\gamma}\epsilon_{ij}e_ih_j\,, \label{eom-bartau-cov} \\
0&=\sigma\mathcal{D}\tau-\frac{1}{\mu}\mathcal{R}(\omega)
   +\gamma\epsilon_{ij}e_ih_j\,, \label{eom-omega-cov} \\
0&=\sigma\mathcal{R}_i(\varpi)-\gamma\,\mathcal{D}h_i
   +\Lambda_0\epsilon_{ij}\tau e_j+\alpha\epsilon_{ij}\bar{\tau}h_j\,, \label{eom-e-cov} \\
0&=\gamma\,\mathcal{D}e_i-\alpha\epsilon_{ij}\bar{\tau}e_j\,, \label{eom-h-cov} \\
0&=\sigma\,\mathcal{D}e_i+\gamma\epsilon_{ij}\bar{\tau}e_j\,. \label{eom-varpi-cov}
\end{align}
\end{subequations}
The rearrangement is exact. The term $\gamma\epsilon_{ij}\bar{\tau}\varpi_j$ of
\eqref{eom-e} is precisely the boost connection term of $\mathcal{D}h_i$, and
once it is absorbed no bare $\varpi_i$ appears anywhere outside a curvature or a
covariant derivative. In this form the last two equations are two statements
about the single object $\mathcal{D}e_i$. Eliminating it returns
$\left(\gamma^{2}+\sigma\alpha\right)\epsilon_{ij}\bar{\tau}e_j=0$, which for
$q\neq0$ gives $\epsilon_{ij}\bar{\tau}e_j=0$. Expanding
$\bar{\tau}=a\,\tau+b_i\,e_i$ in the basis $(\tau,e_1,e_2)$, the two components of this
read $a\,\tau e_2+b_1e_1e_2=0$ and $-a\,\tau e_1+b_2e_1e_2=0$, so that $a=b_1=b_2=0$ and
hence $\bar{\tau}=0$. The Carrollian torsion then vanishes as well,
$\mathcal{D}e_i=0$. This is the covariant counterpart of the
canonical constraint $\bar{\tau}_K\approx0$ of \eqref{bartauK}, obtained in
Section~\ref{Dirac}.

%--- 2.13  a fourth local symmetry: the shift in the ratio gamma/sigma -
There is one more local symmetry, and it is invisible in \eqref{carroll-eom-cov} because
it moves two fields at once. Shift the two vector flavors in the fixed ratio
$\gamma/\sigma$,
\begin{equation} \label{null-shift}
\delta_S h_i=S_{ij}e_j\,, \qquad
\delta_S \varpi_i=\frac{\gamma}{\sigma}S_{ij}e_j\,,
\end{equation}
with $S_{ij}(x)$ an arbitrary matrix of functions and every other field inert. A direct
computation gives
\begin{equation} \label{null-shift-boundary}
\delta_S L_{\text{CMMG}}=\gamma\,d\left(e_i\wedge S_{ij}e_j\right)
 +\frac{\gamma q}{\sigma\mu}\,S_{kk}\;\bar{\tau}e_1e_2\,,
\end{equation}
off shell and on any background. The first term is a total derivative. The second is not,
and for $q\neq0$ it forces
\begin{equation} \label{traceless}
S_{kk}=0\,.
\end{equation}
A traceless $S_{ij}$ has three components, and the parameter enters \eqref{null-shift}
undifferentiated. So \eqref{null-shift} is a local symmetry with three parameters that
generates no secondary constraint. It should not be confused with \eqref{alpha-removal},
which moves $\varpi_i$ by $h_i$ rather than moving both by $e_i$, and which is a change of
variables rather than a gauge transformation. The ratio $\gamma/\sigma$ is not arbitrary. It is the
null direction of the degenerate bilinear form \eqref{vector-bilinear-block} below, and of
the vector flavor metric \eqref{flavor-metrics}, and Section~\ref{Dirac} identifies these
three parameters with three of the nine first-class constraints of the model.

%--- 2.16  the shift at the critical locus -----------------------------
The restriction to $\lambda_i=S_{ij}e_j$ matters only while $q\neq0$. For an arbitrary
one-form $\lambda_i$ the same computation gives
\begin{equation} \label{null-shift-general}
\delta_{\lambda} h_i=\lambda_i\,, \qquad
\delta_{\lambda} \varpi_i=\frac{\gamma}{\sigma}\lambda_i\,, \qquad
\delta_{\lambda} L_{\text{CMMG}}=\gamma\,d\left(e_i\lambda_i\right)
 +\frac{\gamma q}{\sigma\mu}\,\epsilon_{ij}\bar{\tau}e_i\lambda_j\,,
\end{equation}
which returns \eqref{null-shift-boundary} at $\lambda_i=S_{ij}e_j$. On the critical locus
$q=0$ the second term is absent for every $\lambda_i$, and the auxiliary field may then be
shifted away altogether. It enters $L_{\text{CMMG}}$ only through the total derivative
$\gamma\,d(e_ih_i)$, and every equation of \eqref{carroll-eom} sees the pair
$(h_i,\varpi_i)$ through the invariant combination $\varpi_i-\frac{\gamma}{\sigma}h_i$
alone. That is also why \eqref{alpha-removal} fails there. What degenerates is $h'_i$,
not the change of variables.

%--- which Carroll limit this is ---------------------------------------
Before deriving the Carrollian Lagrangian \eqref{carrollian-mmg-lagrangian} algebraically,
we comment briefly on the nature of the resulting model.
A Lorentz-invariant theory admits two inequivalent Carroll limits, an electric and a
magnetic one, distinguished by the rescaling of the canonical variables
\cite{Henneaux:2021yzg}. The contraction \eqref{rescaling-fields} is carried out on the
first-order form of \eqref{mmg-lagrangian}, that is by rescaling the gauge fields of the
relativistic algebra, and the magnetic limit of Einstein gravity is precisely the theory
obtained by gauging the Carroll algebra in this way \cite{Campoleoni:2022ebj}. The
resulting equations keep the spatial curvature: \eqref{eom-tau} fixes
$\mathcal{R}(\omega)=d\omega$ in terms of $\epsilon_{ij}e_ie_j$ and $\Lambda_0$, so the
spatial curvature is carried by a field equation rather than lost. The electric theory
loses that term from its Hamiltonian constraint. The same statement in canonical form,
that the temporal constraints retain $\epsilon_{KL}\partial_K\omega_L$, is
\eqref{temporal-constraints} of Section~\ref{Dirac}. Carrollian MMG is therefore of magnetic type.
%=======================================================================
\subsection{MMG algebra}\label{mmgalg}
%=======================================================================
%--- 2.17  the undeformed algebra --------------------------------------
Here we will examine how the Lagrangian \eqref{carrollian-mmg-lagrangian} can be constructed
algebraically. Let us start by considering the algebra in \cite{Dedeoglu:2024ies,Bergshoeff:2021tbz} with
$J^{(0)}_a=\mathtt{J}_a$ as the generators of dual Lorentz transformations,
$P^{(0)}_a=\mathtt{P}_a$ as the generators of translations and
$P^{(1)}_a=\mathtt{Q}_a$:
\begin{align} \label{1stalg}
    [\mathtt{J}_a,\mathtt{J}_b] &=\epsilon_{abc}\mathtt{J}^c\,, & [\mathtt{J}_a,\mathtt{P}_b] &= \epsilon_{abc}\mathtt{P}^c\,, & [\mathtt{J}_a,\mathtt{Q}_b] &=\epsilon_{abc}\mathtt{Q}^c\,, \nn \\
    [\mathtt{P}_a,\mathtt{P}_b] &=0\,, & [\mathtt{P}_a,\mathtt{Q}_b]&=\epsilon_{abc}\mathtt{J}^c\,, & [\mathtt{Q}_a,\mathtt{Q}_b]&=0\,.
\end{align}
This is the standard algebra whose curvature \eqref{defF} gives the torsion-free equations.
%--- 2.18  gauge field and group-theoretical curvature -----------------
The generators and the gauge fields are associated as follows
\begin{equation} \label{A(ewh)}
A=\mathtt{e}^a\mathtt{P}_a+\mathtt{w}^a\mathtt{J}_a+\mathtt{h}^a\mathtt{Q}_a\,,
\end{equation}
which yields the group-theoretical curvature that gives the field equations
\begin{align} \label{defF}
    F=&dA+\frac{1}{2}[A,A] \nn \\
 =&\mathtt{P}_aD\mathtt{e}^a+\mathtt{J}_a\left(\mathtt{R}+\mathtt{e}\times \mathtt{h}\right)^a
   +\mathtt{Q}_aD\mathtt{h}^a=0\,.
\end{align}

%--- 2.20  MMG field equations and the deformed algebra ----------------
The original MMG field equations \cite{Bergshoeff:2014pca}, written with our
$\tilde{\gamma}=1$, are
\begin{align} \label{mmg-eqn}
    0&=D\mathtt{e}+\tilde{\alpha}_1 \mathtt{e} \times \mathtt{h}\,, \nn \\
    0&=\mathtt{R}+\tilde{q}_1 \mathtt{e} \times \mathtt{h}\,, \nn \\
    0&=D\mathtt{h}+\tilde{\sigma} \tilde{q}_1 \mathtt{e} \times \mathtt{h}+\frac{\tilde{\Lambda}_0}{2}\mathtt{e} \times \mathtt{e}+\frac{\tilde{\alpha}_1}{2} \mathtt{h} \times \mathtt{h}\,,
\end{align}
where $\tilde{\alpha}_1:=\tilde{\alpha}/\tilde{\gamma}^{2}$ and
$\tilde{q}_1:=\tilde{\mu}(1+\tilde{\sigma}\tilde{\alpha}_1)$.\footnote{For general
$\tilde{\gamma}$, equations \eqref{mmg-eqn}--\eqref{Fdeformed} are the ones carried by the
unit-coupling normalization: the $\mathtt{h}^a$ appearing in them is the auxiliary field
after the rescaling $\mathtt{h}^a\rightarrow\mathtt{h}^a/\tilde{\gamma}$ described below
\eqref{mmg-lagrangian}, which sends $\tilde{\alpha}$ to $\tilde{\alpha}_1$, and
correspondingly the $\mathtt{Q}_a$ appearing in them is $\mathtt{Q}_a/\tilde{\gamma}$ in
terms of the generator of \eqref{3rdalg}.}
There is a deformation of \eqref{1stalg} whose curvature reproduces \eqref{mmg-eqn},
given specifically as
\begin{align} \label{2ndalg}
     [\mathtt{J}_a,\mathtt{J}_b] &=\epsilon_{abc}\mathtt{J}^c\,, & [\mathtt{J}_a,\mathtt{P}_b] &= \epsilon_{abc}\mathtt{P}^c\,, & [\mathtt{J}_a,\mathtt{Q}_b] &=\epsilon_{abc}\mathtt{Q}^c\,, \nn \\
    [\mathtt{P}_a,\mathtt{P}_b] &=\tilde{\Lambda}_0\epsilon_{abc}\mathtt{Q}^c\,, & [\mathtt{P}_a,\mathtt{Q}_b]&=\epsilon_{abc}(\tilde{q}_1\mathtt{J}^c+\tilde{\alpha}_1 \mathtt{P}^c+\tilde{\sigma} \tilde{q}_1 \mathtt{Q}^c)\,, & [\mathtt{Q}_a,\mathtt{Q}_b]&=\tilde{\alpha}_1 \epsilon_{abc} \mathtt{Q}^c\,.
\end{align}
The group-theoretical curvature now reads
\begin{align} \label{Fdeformed}
    F=&\mathtt{P}_a\left(D\mathtt{e}+\tilde{\alpha}_1 \mathtt{e} \times \mathtt{h}\right)^a
      +\mathtt{J}_a\left(\mathtt{R}+\tilde{q}_1 \mathtt{e}\times \mathtt{h}\right)^a \nn \\
     &+\mathtt{Q}_a\left(D\mathtt{h}+\tilde{\sigma} \tilde{q}_1 \mathtt{e} \times \mathtt{h}
      +\frac{\tilde{\Lambda}_0}{2} \mathtt{e} \times \mathtt{e}
      + \frac{\tilde{\alpha}_1}{2}\mathtt{h} \times \mathtt{h} \right)^a=0\,,
\end{align}
which yields field equations identical to \eqref{mmg-eqn}, as advertised.

%--- 2.23  the gamma-dressed algebra and its bilinear form -------------
As already
noted, we need the parameter $\tilde{\gamma}$ as the coefficient of the torsion constraint in the Lagrangian
\eqref{mmg-lagrangian}, so one needs to dress up \eqref{2ndalg} by $\tilde{\gamma}$ factors, which yields
the final form of the algebra we will work with:
\begin{align} \label{3rdalg}
     [\mathtt{J}_a,\mathtt{J}_b] &=\epsilon_{abc}\mathtt{J}^c\,, & [\mathtt{J}_a,\mathtt{P}_b] &= \epsilon_{abc}\mathtt{P}^c\,, & [\mathtt{J}_a,\mathtt{Q}_b] &=\epsilon_{abc}\mathtt{Q}^c\,, \nn \\
    [\mathtt{P}_a,\mathtt{P}_b] &=\frac{\tilde{\Lambda}_0}{\tilde{\gamma}}\epsilon_{abc}\mathtt{Q}^c\,, & [\mathtt{P}_a,\mathtt{Q}_b]&=\epsilon_{abc}\left(\tilde{q}\mathtt{J}^c+\frac{\tilde{\alpha}}{\tilde{\gamma}} \mathtt{P}^c+\frac{\tilde{\sigma} \tilde{q}}{\tilde{\gamma}} \mathtt{Q}^c\right)\,, & [\mathtt{Q}_a,\mathtt{Q}_b]&=\frac{\tilde{\alpha}}{\tilde{\gamma}} \epsilon_{abc}\mathtt{Q}^c\,,
\end{align}
with $\tilde{q}:=\frac{\tilde{\mu}}{\tilde{\gamma}}\left(\tilde{\gamma}^{2}+\tilde{\sigma}\tilde{\alpha}\right)=\tilde{\gamma}\tilde{q}_1$ the combination \eqref{genericity} in relativistic couplings. The non-degenerate bilinear forms are
\begin{equation} \label{bilfor}
\langle \mathtt{J}_a,\mathtt{J}_b \rangle=\frac{1}{2\tilde{\mu}}\eta_{ab}\,, \quad \langle\mathtt{J}_a,\mathtt{P}_b\rangle=-\frac{\tilde{\sigma}}{2}\eta_{ab}\,, \quad \langle\mathtt{P}_a,\mathtt{Q}_b\rangle=\frac{\tilde{\gamma}}{2} \eta_{ab} \,.
\end{equation}

%--- 2.25  DF is the Jacobiator ----------------------------------------
With the definition \eqref{defF}, $DF := dF + [A, F]$, and substituting \eqref{defF} back in,
and recognizing that the $dA \wedge A$ terms identically cancel out, one is left with
\begin{equation} \label{DFleft}
DF=\frac12\big[A,[A,A]\big] \,.
\end{equation}
Since $A$ is a one-form, $A^{A} \wedge A^{B} \wedge A^{C}$ is totally antisymmetric in its flavor
labels, so this contracts $f_{AE}{}^{D}f_{BC}{}^{E}$ against a totally antisymmetric
object. Hence
$DF$ is nothing but the Jacobiator evaluated on three copies of $A$, and in particular contains
no derivatives of $A$ at all. Since $d^{2}=0$, any solution of $F=0$ must satisfy $DF=0$:
writing $F=0$ as $dA=-\tfrac12[A,A]$ and acting with $d$ returns exactly $\tfrac12[A,[A,A]]=0$.
For a Lie algebra the Jacobi identity makes this automatic and nothing is learned. Here
the Jacobi identity fails, so $DF$ is a non-vanishing cubic expression in the fields and
 $DF=0$ is a genuine algebraic condition; i.e.\ it is not imposed by hand, but forced on
every solution of the field equations. This is what we mean by \emph{third-way consistency in
first-order form}: it closes only on shell.

%--- 2.26  the relativistic Jacobiators and ad-invariance --------------
Up to permutations, the Jacobiator vanishes except in the
$(\mathtt{P}, \mathtt{P}, \mathtt{Q})$ and $(\mathtt{P}, \mathtt{Q}, \mathtt{Q})$ sectors,
where (using $\tilde{q} \tilde{\gamma}/\tilde{\mu} =(\tilde{\gamma}^2+\tilde{\sigma} \tilde{\alpha})$)
\begin{align} \label{rel-jacobiators}
\mathrm{Jac}(\mathtt{P}_0,\mathtt{P}_1,\mathtt{Q}_0)
 &=-\frac{\tilde{q}^{2}\tilde{\sigma}}{\tilde{\gamma}}\mathtt{J}_1-\frac{\tilde{q}^{2}}{\tilde{\gamma}\tilde{\mu}}\mathtt{P}_1
   -\frac{\tilde{q}^{2}\tilde{\sigma}^{2}}{\tilde{\gamma}^{2}}\mathtt{Q}_1\,, \nn \\
\mathrm{Jac}(\mathtt{P}_0,\mathtt{Q}_0,\mathtt{Q}_1)&=-\frac{\tilde{q}^{2}}{\tilde{\gamma}\tilde{\mu}}\mathtt{Q}_1\,,
\end{align}
where all coefficients are proportional to $\tilde{q}^2$, which we assume is non-zero in the generic case.
Conversely, as $\tilde{q} \to 0$, the algebra \eqref{3rdalg} does close and the model does become an
ordinary Chern--Simons theory. Note also that the bilinear form \eqref{bilfor} \emph{is}
ad-invariant; indeed invariance is what fixes $\tilde{q}=\frac{\tilde{\mu}}{\tilde{\gamma}}(\tilde{\gamma}^{2}+\tilde{\sigma}\tilde{\alpha})$,
the obstruction $\frac{\tilde{\sigma}\tilde{\alpha}}{2\tilde{\gamma}}+\frac{\tilde{\gamma}}{2}-\frac{\tilde{q}}{2\tilde{\mu}}$ vanishing
precisely for that value. The invariance therefore holds while the Jacobi identity fails
whenever $\tilde{q}\neq0$, and what we have is precisely a Chern--Simons-like theory that is
not a Chern--Simons theory. This is not peculiar to MMG. At $\tilde{\alpha}=0$, which is TMG, one
has $\tilde{q}=\tilde{\mu}\tilde{\gamma}\neq0$, so \eqref{rel-jacobiators} does not vanish there either. The
failure of the Jacobi identity is the signature of being Chern--Simons-like rather than of
the third-way deformation.

%=======================================================================
\subsection{Carrollian MMG algebra}
%=======================================================================

%--- 2.27  generator rescaling and the contracted algebra --------------
To construct the Carrollian counterpart of MMG, the generators
$\left\{\mathtt{J}_a,\mathtt{P}_a,\mathtt{Q}_a\right\}$ are decomposed by splitting the
Lorentz index as $a=(0,i)$ with $i=1,2$. The Carrollian limit is obtained through the rescaling of the generators
\begin{equation} \label{rescaling-generators}
\begin{aligned}
\mathtt{J}_0 &\rightarrow J\,, & \mathtt{J}_i &\rightarrow \rho K_i\,, &
\mathtt{P}_0 &\rightarrow \rho H\,, \\
\mathtt{P}_i &\rightarrow P_i \,, & \mathtt{Q}_0 &\rightarrow  \bar{H}\,, &
\mathtt{Q}_i &\rightarrow \rho \mathcal{P}_i\,,
\end{aligned}
\end{equation}
which, together with $\tilde{q}=\rho q$, takes \eqref{3rdalg} explicitly to the form:
\begin{align} \label{algdec}
\boxed{
\begin{gathered}
\begin{aligned}
 \relax[J,K_i]&=\epsilon_{ij}K_j\,, &
 [J,P_i]&=\epsilon_{ij}P_j\,, &
 [J,\mathcal{P}_i]&=\epsilon_{ij}\mathcal{P}_j\,, \\
 [K_i,P_j]&=-\epsilon_{ij}H \,, &
 [K_i,\mathcal{P}_j]&=-\tfrac{1}{\rho^2}\epsilon_{ij}\bar{H}\,, &
 [K_i,K_j]&=-\tfrac{1}{\rho^2}\epsilon_{ij}J\,, \\
 [P_i,P_j]&=-\tfrac{\Lambda_0}{\gamma}\epsilon_{ij}\bar{H}\,, &
 [\mathcal{P}_i,\mathcal{P}_j]&=-\tfrac{\alpha}{\gamma \rho^2}\epsilon_{ij}\bar{H}\,, &
 [H,K_i]&=\tfrac{1}{\rho^2}\epsilon_{ij}P_j\,, \\
 [H,P_i]&=\tfrac{\Lambda_0}{\gamma}\epsilon_{ij}\mathcal{P}_j \,, &
 [\bar{H},K_i]&=\epsilon_{ij}\mathcal{P}_j\,, &
 [\bar{H},\mathcal{P}_i]&=\tfrac{\alpha}{\gamma}\epsilon_{ij}\mathcal{P}_j \,,
\end{aligned} \\[4pt]
\begin{aligned}
 \relax[P_i,\mathcal{P}_j]&=-\epsilon_{ij}\left(qJ+\tfrac{\alpha}{\gamma} H
   +\tfrac{\sigma q}{\gamma}\bar{H}\right)\,, \\
 [H,\mathcal{P}_i]&=\epsilon_{ij}\left(qK_j+\tfrac{\alpha}{\rho^2 \gamma}P_j
   +\tfrac{q\sigma}{\gamma}\mathcal{P}_j\right)\,, \\
 [\bar{H},P_i]&=\epsilon_{ij}\left(q\rho^2 K_j+\tfrac{\alpha}{\gamma}P_j
   +\tfrac{q\sigma \rho^2}{\gamma}\mathcal{P}_j\right)\,.
\end{aligned}
\end{gathered}
}
\end{align}
%--- 2.29  Jacobi still fails after the contraction --------------------
The Jacobi identity continues to fail after the contraction: for instance
\begin{align} \label{carroll-jacobiators}
\mathrm{Jac}(H,\bar H,\mathcal{P}_i)&=-\frac{q^{2}}{\gamma\mu}\mathcal{P}_i\,, &
\mathrm{Jac}(P_1,\bar H,\mathcal{P}_1)&=-\frac{q^{2}}{\gamma\mu}\bar H\,, \nn \\
\mathrm{Jac}(H,P_1,\mathcal{P}_1)&=-\frac{q^{2}\sigma}{\gamma}J
  -\frac{q^{2}}{\gamma\mu}H-\frac{q^{2}\sigma^{2}}{\gamma^{2}}\bar H\,, &&
\end{align}
so the non-Lie character is genuinely Carrollian and not an artifact the limit removes.

%--- 2.30  the divergent structure constants lie in the kernel ---------
Since the generator weights are already locked by field weights and
$\tilde{q}=\rho q$ is already forced by our choice of parameter rescalings (see the discussion
after \eqref{rescaling-fields}), naively taking the dimensionless contraction parameter
$\rho \to \infty$ heralds trouble since e.g.\ $[\bar{H},P_i] \xrightarrow{\rho \to \infty} \infty$
in \eqref{algdec}. However, the divergent structure constants lie in the kernel of the
contracted bilinear form, so they never reach the Lagrangian, e.g.
\begin{align} \label{kernel-pairings}
\langle[\bar{H},P_i],P_k\rangle&=\epsilon_{ij}\left(q\rho^{2}\langle K_j,P_k\rangle
+\frac{q\sigma\rho^{2}}{\gamma}\langle\mathcal{P}_j,P_k\rangle\right)
=\epsilon_{ik}\left(-\frac{q\sigma\rho^{2}}{2}+\frac{q\sigma\rho^{2}}{2}\right)=0\,, \nn \\
\langle[\bar{H},P_i],K_k\rangle&=\epsilon_{ij}\left(q\rho^{2}\langle K_j,K_k\rangle
+\frac{\alpha}{\gamma}\langle P_j,K_k\rangle\right)
=\epsilon_{ik}\left(\frac{q}{2\mu}-\frac{\alpha\sigma}{2\gamma}\right)
=\frac{\gamma}{2}\epsilon_{ik}\,.
\end{align}
So the contraction is singular at the level of the bracket but regular at the level of the
Chern--Simons-like data, that is the flavor metrics $g_{rs}$ and the cubic flavor tensor
$f_{rsI}$ of Section~\ref{CCSlike}. That is exactly
why \eqref{carrollian-mmg-lagrangian} is finite while \eqref{algdec} is not.\footnote{The
divergence is not merely orthogonal to the kernel of the contracted bilinear form, it lies
\emph{along} it. Inspecting \eqref{algdec}, the only bracket that diverges is $[\bar{H},P_i]$,
and its divergent part is
$\frac{q\rho^{2}}{\gamma}\epsilon_{ij}\left(\gamma K_j+\sigma\mathcal{P}_j\right)$, i.e.\ it
points exactly along the null direction $\gamma K_i+\sigma\mathcal{P}_i$ identified below.
Consequently the structure constants carrying an upper flavor index, $f^{R}{}_{ST}$, have no
$\rho\to\infty$ limit, whereas the lowered combination $f_{RST}=g_{RU}f^{U}{}_{ST}$ does,
because contracting with $g$ annihilates precisely the divergent direction. Every statement
below is therefore made with lowered flavor indices, and the charge formula
\eqref{rel-asymptotic}, which needs $f^{R}{}_{ST}$, is unavailable for the same reason.}

%--- 2.31  the contracted bilinear forms; scalar vs vector sector ------
The bilinear forms \eqref{bilfor} become
\begin{align} \label{contracted-bilinears}
    \langle J,J\rangle=&-\frac{1}{2\mu}\,, & \langle K_i,K_j\rangle=&\frac{1}{2\mu \rho^2}\delta_{ij}\,, & \langle J,H\rangle=&\frac{\sigma}{2}\,, \nn \\
    \langle K_i,P_j\rangle=&-\frac{\sigma}{2}\delta_{ij}\,, & \langle H,\bar{H}\rangle=&-\frac{\gamma}{2 }\,, & \langle P_i,\mathcal{P}_j\rangle=&\frac{\gamma}{2}\delta_{ij} \,.
\end{align}
After contraction, it is easy to see that the scalar sector $(J, H, \bar{H})$ of the bilinear forms
remains non-degenerate since
\begin{align} \label{scalar-bilinear-block}
\langle\,\cdot\,,\,\cdot\,\rangle\big|_{(J,\,H,\,\bar{H})}&=
\begin{pmatrix}
-\dfrac{1}{2\mu} & \dfrac{\sigma}{2} & 0 \\[6pt]
\dfrac{\sigma}{2} & 0 & -\dfrac{\gamma}{2} \\[6pt]
0 & -\dfrac{\gamma}{2} & 0
\end{pmatrix}\,,
&\det&=\frac{\gamma^{2}}{8\mu}\neq 0\,,
\end{align}
whereas the vector sector $(K_i, P_i, \mathcal{P}_i)$ is degenerate:
\begin{align} \label{vector-bilinear-block}
\langle\,\cdot\,,\,\cdot\,\rangle\big|_{(K_i,\,P_i,\,\mathcal{P}_i)}&=
\begin{pmatrix}
\dfrac{1}{2\mu\rho^{2}} & -\dfrac{\sigma}{2} & 0 \\[6pt]
-\dfrac{\sigma}{2} & 0 & \dfrac{\gamma}{2} \\[6pt]
0 & \dfrac{\gamma}{2} & 0
\end{pmatrix}\,,
&\det&=-\frac{\gamma^{2}}{8\mu\rho^{2}}
\ \xrightarrow{\;\rho\to\infty\;}\ 0\,,
\end{align}
where the null direction is along $\gamma K_i + \sigma \mathcal{P}_i$. This is the ratio
that \eqref{null-shift} moves along. The same degeneracy returns in the vector flavor
metric \eqref{flavor-metrics} and in the null vectors \eqref{4solns} of the Dirac
analysis of Section~\ref{Dirac}.

%--- 2.34  the gauge field in Carrollian variables ---------------------
To construct the CS-Lagrangian on the Carrollian MMG algebra, one needs to split the
space and time indices of the gauge one-form \eqref{A(ewh)}
\begin{align} \label{A-carroll-split}
    A=&\mathtt{e}^0 \mathtt{P}_0 + \mathtt{e}^k\mathtt{P}_k+ \mathtt{h}^0 \mathtt{Q}_0 +\mathtt{h}^k\mathtt{Q}_k+\mathtt{w}^0 \mathtt{J}_0+\mathtt{w}^k\mathtt{J}_k \nn \\
    =&\tau H + \bar{\tau} \bar{H} + e^kP_k+h^k\mathcal{P}_k+\omega J+\varpi^kK_k\,,
\end{align}
which is invariant under the simultaneous scale transformations \eqref{rescaling-fields} and \eqref{rescaling-generators}.
The Lagrangian is then \eqref{carrollian-mmg-lagrangian}, obtained above by splitting the
indices and rescaling the fields.

%=======================================================================
%  SECTION 3 :  CARROLLIAN ADS AND BTZ SOLUTIONS
%=======================================================================
\section{Carrollian AdS and BTZ solutions}\label{adsbtz}

This section solves the Carrollian field equations \eqref{carroll-eom}. Two configurations
come out of them. The first is a maximally symmetric vacuum, and it fixes the sign of
$\Lambda_0$, the curvature of the spatial slices and the constant that relates the
auxiliary field to the spatial zweibein. The second is a Carrollian counterpart of the BTZ
black hole. Neither is imported. The relativistic vacuum enters only at the end of
Section~\ref{adsvac}, as a check, and the relativistic black hole enters only in
Section~\ref{btzsub}, as a starting ansatz that the Carrollian equations then confirm and
complete.

%=======================================================================
\subsection{The maximally symmetric vacuum}\label{adsvac}
%=======================================================================

%--- 3.1  the two scalar curvatures ------------------------------------
It is convenient to name two combinations,
\begin{equation}  \label{carroll-defs}
K:=\epsilon_{ij}e_ih_j\,, \qquad \bar{S}_i:=\epsilon_{ij}\bar{\tau}e_j\,.
\end{equation}
The scalar equations \eqref{eom-bartau} and \eqref{eom-omega} then read
\begin{equation} \label{Rtau-Romega}
\mathcal{R}(\tau)=\frac{\alpha}{\gamma}K\,, \qquad
\mathcal{R}(\omega)=qK\,,
\end{equation}
the second following from the first because \eqref{eom-omega} gives
$\mathcal{R}(\omega)=\mu\sigma\mathcal{R}(\tau)+\gamma\mu K$ and
\mbox{$\mu\sigma\alpha/\gamma+\gamma\mu=q$}. Both curvatures are thus carried by the single
two-form $K$.

%--- 3.3  the vector equations force taubar = 0 ------------------------
The vector equations come next. Combining \eqref{eom-varpi} and \eqref{eom-h} gives
\begin{equation} \label{Si=0}
\mathcal{R}_i(e)=\frac{\alpha}{\gamma}\bar{S}_i\,, \qquad
(\gamma^2+\alpha \sigma)\bar{S}_i=0\,.
\end{equation}
These are \eqref{eom-h-cov} and \eqref{eom-varpi-cov} in the present notation. By
\eqref{genericity} the second relation forces $\bar{S}_i=0$, hence $\bar{\tau}=0$ by the
basis argument given below \eqref{carroll-eom-cov}. The first relation then leaves
$\mathcal{R}_i(e)=0$, so the spatial torsion vanishes and $\omega$ is the spin connection
of the zweibein $e_i$.

%--- 3.4  Lambda_0 fixes the spatial geometry --------------------------
With $\bar{\tau}=0$, equation \eqref{eom-tau} and the second of \eqref{Rtau-Romega} give
\begin{equation} \label{K-from-Lambda0}
0=\sigma \mathcal{R}(\omega)+\frac{\Lambda_0}{2}\epsilon_{ij}e_ie_j
\qquad\Longrightarrow\qquad
K=-\frac{\Lambda_0\sigma}{2q}\epsilon_{ij}e_ie_j \,.
\end{equation}
Take $\sigma\Lambda_0<0$ and define the Carrollian radius $\ell$ by
\begin{equation} \label{Lambda0-ell}
\Lambda_0=-\frac{\sigma}{\ell^{2}}\,.
\end{equation}
The first relation in \eqref{K-from-Lambda0} then reads
$\mathcal{R}(\omega)=\frac{1}{2\ell^{2}}\epsilon_{ij}e_ie_j$. The spatial slices therefore
have constant negative curvature $-\ell^{-2}$ and are hyperbolic of radius $\ell$. The
spatial geometry is fixed by $\Lambda_0$ and is not free. Section~\ref{bms} recovers the
same statement in \eqref{flat-spatial}.

%--- 3.6  the auxiliary field and the constant c -----------------------
Take the spatial auxiliary field proportional to the zweibein with a constant $c$,
\begin{equation} \label{h-is-ce}
h_i=c\,e_i \qquad\Longrightarrow\qquad K=c\,\epsilon_{ij}e_ie_j\,,
\end{equation}
which is the maximally symmetric choice. Comparing with the second relation in
\eqref{K-from-Lambda0} fixes
\begin{equation} \label{c-value}
c=\frac{1}{2q\ell^{2}}\,.
\end{equation}
One equation is left. At $\bar{\tau}=0$ and $h_i=c\,e_i$ one has
$\mathcal{D}h_i=c\,\mathcal{D}e_i=0$, so \eqref{eom-e-cov} reduces to
$\sigma\mathcal{R}_i(\varpi)=-\Lambda_0\epsilon_{ij}\tau e_j$, that is
\begin{equation} \label{Romega-i}
\mathcal{R}_i(\varpi)=\frac{1}{\ell^2} \tau e_j \epsilon_{ij}\,.
\end{equation}
The vacuum is thereby complete. Only \eqref{carroll-eom} was used.

%--- 3.9  the relativistic check ---------------------------------------
The relativistic theory gives the same answer, which is worth recording as a check. The
MMG field equations are
\begin{equation} \label{mmg-eom-rel}
0=D(\mathtt{w})\mathtt{e}+\frac{\tilde{\alpha}}{\tilde{\gamma}}\mathtt{e}\times \mathtt{h}\,, \quad
0=\mathtt{R}(\mathtt{w})+\tilde{q}\, \mathtt{e}\times \mathtt{h}\,, \quad
0=D(\mathtt{w})\mathtt{h}+\frac{\tilde{\sigma} \tilde{q}}{\tilde{\gamma}}\mathtt{e} \times \mathtt{h}
 +\frac{\tilde{\Lambda}_0}{2\tilde{\gamma}}\mathtt{e} \times \mathtt{e}
 +\frac{\tilde{\alpha}}{2\tilde{\gamma}}\mathtt{h} \times \mathtt{h}\,.
\end{equation}
The first of them says that $\Omega:=\mathtt{w}+\frac{\tilde{\alpha}}{\tilde{\gamma}}\mathtt{h}$ is torsion
free, and in terms of $\Omega$ the system becomes
\begin{equation} \label{torsionless-eom}
0=D(\Omega)\mathtt{e} \,, \quad
0=\mathtt{R}(\Omega)+\frac{\tilde{\alpha} \tilde{\Lambda}_0}{2\tilde{\gamma}^2}\mathtt{e}\times \mathtt{e}
 +\frac{\tilde{q}^2}{\tilde{\gamma} \tilde{\mu}}\mathtt{e}\times \mathtt{h}\,, \quad
0=D(\Omega)\mathtt{h}-\frac{\tilde{\alpha}}{2\tilde{\gamma}}\mathtt{h} \times \mathtt{h}
 +\frac{\tilde{\sigma} \tilde{q}}{\tilde{\gamma}}\mathtt{e} \times \mathtt{h}
 +\frac{\tilde{\Lambda}_0}{2\tilde{\gamma}}\mathtt{e}\times \mathtt{e} \,.
\end{equation}
A maximally symmetric AdS background satisfies
\begin{equation} \label{maximally-sym1}
0=\mathtt{R}(\Omega)+\frac{1}{2\ell^2}\mathtt{e}\times \mathtt{e} \,,
\end{equation}
on which the auxiliary field is proportional to the dreibein, $\mathtt{h}=\tilde{c}\,\mathtt{e}$,
with
\begin{equation} \label{relc}
\tilde{c}=\frac{\tilde{\mu}\left(\tilde{\gamma}^{2}-\tilde{\alpha}\ell^{2}\tilde{\Lambda}_0\right)}{2\tilde{q}^{2}\tilde{\gamma}\ell^{2}}\,.
\end{equation}
Imposing \eqref{Lambda0-ell} in \eqref{relc} with $\tilde{q}=\rho q$ returns
$\tilde{c}=c/\rho$, the Carrollian value \eqref{c-value}.

%--- 3.13  the one feature the limit does not keep ---------------------
One relativistic feature does not survive. On the background above the torsion of
$\mathtt{w}$ is constant and proportional to the area two-form,
\begin{equation}  \label{torsion-rel}
D(\mathtt{w}) \mathtt{e} = D(\Omega) \mathtt{e}
 - \frac{\tilde{\alpha}}{\tilde{\gamma}}\mathtt{e} \times \mathtt{h}
 =-\frac{\tilde{c}\tilde{\alpha}}{\tilde{\gamma}}\mathtt{e} \times \mathtt{e}\,,
\end{equation}
whereas the Carrollian torsion vanishes outright for $q\neq0$ by
\eqref{carroll-eom-cov}. The mechanism is $\bar{\tau}=0$: it makes the two rotation
connections coincide, as \eqref{omega-from-Omega} shows.

%=======================================================================
\subsection{The Carrollian BTZ solution}\label{btzsub}
%=======================================================================

The black hole is built the same way. The $\rho\to\infty$ limit of the relativistic BTZ
solution supplies an ansatz for the frame and a starting point for the connection. The
Carrollian equations then fix the lapse, fix the part of the connection the ansatz leaves
open, and confirm the rest. The rotation parameter ends up in the connection and
nowhere else.

%--- 3.14  the relativistic solution and the Carroll scaling -----------
The relativistic BTZ metric \cite{Banados:1992wn,Banados:1992gq} is
\begin{align} \label{BTZ-metric}
ds^2&=-\tilde{N}^2d\tilde{t}^2+\tilde{N}^{-2}dr^2
 +r^2(\tilde{N}^{\phi}d\tilde{t}+d\phi)^2\,, \nn \\
\tilde{N}^2&:=-M+\frac{r^2}{\ell^2}+\frac{\tilde{\mathcal{J}}^2}{4r^2}\,,
 \qquad
\tilde{N}^{\phi}:=-\frac{\tilde{\mathcal{J}}}{2r^2}\,.
\end{align}
For compatibility with the Carroll scaling we take $\tilde{t}=t/\rho$ and
\begin{equation} \label{BTZ-carroll-scaling}
\mathtt{e}^a=(\mathtt{e}^{0},\mathtt{e}^{i}) \rightarrow
  \left(\frac{1}{\rho}\tau,e_i\right)\,, \qquad
\Omega^a=(\Omega^{0},\Omega^{i}) \rightarrow
  \left(\Omega,\frac{1}{\rho}\Omega_i\right)\,,
\end{equation}
and the rotation parameter must scale as
$\tilde{\mathcal{J}}=\mathcal{J}/\rho$.\footnote{We quote the finite Carrollian fields
directly throughout, $\tau:=\rho\,\mathtt{e}^{0}$ and
$\Omega_i:=\rho\,\Omega^{i}$. The exponent is fixed by the connection, not by the
metric. With $\tilde{\mathcal{J}}=\rho^{-p}\mathcal{J}$ both $\tilde{N}^{2}$ and $e_2$
stay finite for every $p>0$, whereas
$\Omega_{2}\supset-\rho^{1-p}(\mathcal{J}/(2r))\,d\phi$ diverges for $p<1$ and loses
$\mathcal{J}$ for $p>1$. The charges select the same value: at $p=1$ the relativistic
angular momentum \eqref{rel-charges} tends to $J_{\rm C}$ of \eqref{final-charges} term by
term.}

%--- 3.16  the Carrollian frame ----------------------------------------
The last term of $\tilde{N}^{2}$ in \eqref{BTZ-metric} is then
$\mathcal{O}(\rho^{-2})$ and does not survive. The Carrollian lapse is free of the
rotation parameter,
\begin{equation} \label{lapse}
N^{2}=-M+\frac{r^{2}}{\ell^{2}}\,,
\end{equation}
and the finite fields of the frame are
\begin{equation} \label{BTZ-soln-of-vielbein}
\tau=Ndt\,, \qquad e_1=N^{-1}dr\,, \qquad e_2=rd\phi\,.
\end{equation}
The temporal component of the auxiliary field vanishes, $\bar{\tau}=0$, and its spatial
components are taken proportional to the zweibein as in \eqref{h-is-ce},
\begin{equation} \label{BTZ-soln-of-h}
h_1=cN^{-1}dr\,, \qquad h_2=crd\phi\,.
\end{equation}

%--- 3.19  what fixes the torsion-free connection ----------------------
The torsion-free connection $\Omega$ follows from the Cartan structure equation
\begin{equation} \label{cartan-tf}
0=d\mathtt{e}^a+\epsilon^{abc}\Omega_b\mathtt{e}_c \,.
\end{equation}
Its spatial components are six equations for the three components of $\Omega$, and they
determine $\Omega$ uniquely when they are consistent, as they are for
\eqref{BTZ-soln-of-vielbein}. Its temporal component is the single two-form condition
$d\tau=\epsilon_{ij}\Omega_ie_j$, which imposes only three relations on the six components
of $\Omega_i$. Torsion freedom alone therefore does not determine $\Omega_i$. What
determines it is \eqref{eom-e}, which on this background reduces to \eqref{Romega-i}. The
$\rho\to\infty$ limit of the relativistic BTZ connection supplies a convenient starting
point for that equation, and in the contracted frame $\mathtt{e}^{0}=\tau/\rho$ it reads
\begin{equation} \label{BTZ-Omega}
\Omega=Nd\phi\,, \qquad \Omega_1=\frac{\mathcal{J}}{2Nr^2}dr\,, \qquad \Omega_2=\frac{r}{\ell^2}dt-\frac{\mathcal{J}}{2r}d\phi\,.
\end{equation}

%--- 3.21  the lapse is an output --------------------------------------
The condition $d\tau=\epsilon_{ij}\Omega_ie_j$ is now a statement about $N$. The
$\mathcal{J}$-dependent terms cancel between $\Omega_1e_2$ and $\Omega_2e_1$, and what
survives is
\begin{equation} \label{lapse-ode}
\ell^{2}NN'=r\,,
\end{equation}
the prime denoting $d/dr$, which integrates to \eqref{lapse} with $M$ as the integration constant. The lapse is thus
fixed by the Carrollian equations and not imported with the rest of the configuration. So
is its independence of $t$ and $\phi$. Allow $N$ to depend on all three coordinates, and
$\mathcal{J}$ on $t$ and $\phi$, in
\eqref{BTZ-soln-of-vielbein}--\eqref{BTZ-solution-of-omega}. Then, besides
\eqref{lapse-ode}, the system \eqref{carroll-eom} returns
$\partial_tN=\partial_{\phi}N=0$ and $\partial_t\mathcal{J}=\partial_{\phi}\mathcal{J}=0$.
The first pair comes from \eqref{eom-tau}, \eqref{eom-bartau} and \eqref{eom-omega}, the
second from \eqref{eom-e}. Both $N=N(r)$ and the constancy of $\mathcal{J}$ are outputs
rather than assumptions.

%--- 3.22  the torsional connection ------------------------------------
The torsional connection follows from the relation between the two,
\begin{equation} \label{omega-from-Omega}
\mathtt{w}^a=\Omega^a-\frac{\tilde{\alpha}}{\tilde{\gamma}}\mathtt{h}^a \implies
\omega=\Omega-\frac{\alpha}{\gamma}\bar{\tau}=\Omega\,, \quad
\varpi_i=\Omega_i-\frac{\alpha}{\gamma}h_i=\Omega_i-\frac{c\alpha}{\gamma}e_i \,.
\end{equation}
The two components must be taken separately, because $\mathtt{h}=\tilde{c}\,\mathtt{e}$ does not
survive the limit as a Lorentz-covariant statement. The temporal component obeys
$\bar{\tau}=0$ while $\tau\neq0$, so only the spatial relation $h_i=c\,e_i$ holds. That
spatial relation is \eqref{alpha-removal} evaluated on this solution,
$\Omega_i=\varpi'_i$, which is why \eqref{BTZ-Omega} is free of $\alpha$ while
\eqref{BTZ-solution-of-omega} is not. The BTZ solution of Carrollian MMG is then
\begin{equation} \label{BTZ-solution-of-omega}
\boxed{\;
\omega=Nd\phi\,, \quad \varpi_1=\left(\frac{\mathcal{J}}{2Nr^2}-\frac{c\alpha}{N\gamma}\right)dr\,, \quad \varpi_2=\frac{r}{\ell^2}dt-\left(\frac{\mathcal{J}}{2r}+\frac{c\alpha}{\gamma}r\right)d\phi \,.
\;}
\end{equation}
One verifies directly that \eqref{BTZ-soln-of-vielbein}, \eqref{BTZ-soln-of-h} and
\eqref{BTZ-solution-of-omega}, together with $\bar{\tau}=0$, satisfy all six equations
\eqref{carroll-eom} exactly, provided $\Lambda_0=-\sigma/\ell^{2}$ and
$c=1/(2q\ell^{2})$. The relativistic limit is a starting point and nothing more.

The mass $M$ enters only through $N$. The rotation parameter $\mathcal{J}$ is absent from
the Carrollian frame $(\tau,e_i)$ and survives only in the spatial connection $\varpi_i$.
We refer to \eqref{BTZ-soln-of-vielbein}, \eqref{BTZ-soln-of-h} and
\eqref{BTZ-solution-of-omega} as the Carrollian BTZ solution, and to the geometry it
describes as the Carrollian black hole. The two names denote the same configuration, and
the relativistic solution \eqref{BTZ-metric} is named as such wherever it is meant. In
what sense \eqref{BTZ-soln-of-vielbein} is a black hole is taken up in
Section~\ref{thermo}.

%--- where this solution sits in the literature ------------------------
Rotation carried by the connection rather than by the coframe is not peculiar to this
model. Rotating solutions of magnetic Carroll gravity are built in \cite{Ecker:2026noz}
from the freedom in the Carroll compatible connection, and two features of that
construction recur here. The metric data are those of the static solution, and the locus
where the Carroll vector diverges does not move when the rotational charge is switched on.
The counterpart of the second is that $\mathcal{J}$ is absent from $N^{2}$ of
\eqref{lapse}, as Section~\ref{thermo} uses. Those configurations have no Lorentzian
counterpart, whereas $\mathcal{J}$ here is inherited from the relativistic BTZ solution.
Rotational data can also survive beyond leading Carrollian order, as the small-$c$
expansion of \cite{Bal:2026xup} shows.

Three further points of contact should be recorded. A Carroll black hole is defined in
\cite{Ecker:2023uwm} as a solution of Carroll gravity with Carroll thermal properties and
a Carroll extremal surface, a criterion Section~\ref{thermo} returns to for the solution
above. The asymptotic structure of three-dimensional Carroll gravity with a negative
cosmological constant is analyzed in \cite{Aviles:2025ygw}, whose boundary conditions give
a generalized $\mathfrak{bms}_3$ with four functions on the circle. The algebra
\eqref{bms3} is its subalgebra, and the difference lies in \eqref{bc} rather than in the
bulk theory. Finally \cite{Kolar:2025ebv} shows that a stationary axisymmetric solution of
Carrollian general relativity is static in every dimension above three, with $d=3$ as the
exception, and obtains a rotating Carroll BTZ solution from a static one by a boost and a
re-identification of $\phi$. The solution above carries angular momentum in three
dimensions, so it falls in the case that theorem leaves open. In any case the staticity
there constrains the metric data and is silent about the connection \cite{Ecker:2026noz}.
Section~\ref{thermo} settles the relation between the two rotations.

%=======================================================================
\subsection{A one-parameter deformation}\label{betadef}
%=======================================================================

%--- 3.24  the shifted auxiliary field ---------------------------------
The solution just constructed is not isolated. Let $\beta$ be a constant and replace
\begin{equation} \label{beta-shift}
h_i\;\longrightarrow\;h_i+\beta\,\epsilon_{ij}e_j\,, \qquad
\varpi_i\;\longrightarrow\;\varpi_i-\frac{\alpha\beta}{\gamma}\,\epsilon_{ij}e_j\,,
\end{equation}
leaving $\tau$, $\bar{\tau}$, $e_i$ and $\omega$ untouched. The second shift is the first
one carried through $\varpi_i=\Omega_i-\frac{\alpha}{\gamma}h_i$, so that the torsion-free
connection $\Omega_i$ of \eqref{BTZ-Omega} does not move. Equivalently, in the variables
of \eqref{alpha-removal} the deformation is $\delta_{\beta}\varpi'_i=0$ and
$\delta_{\beta}h'_i=\frac{q\beta}{\mu\gamma}\epsilon_{ij}e_j$. It lives entirely in
the auxiliary field of the Carrollian TMG description, and \eqref{beta-shift} is what that
single shift looks like once $\varpi'_i$ is traded back for $\varpi_i$. Written
out, the deformed Carrollian BTZ solution is
\begin{align} \label{BTZ-beta}
h_1&=\frac{c}{N}dr+\beta r\,d\phi\,, &
h_2&=c\,r\,d\phi-\frac{\beta}{N}dr\,, \nn \\
\varpi_1&=\left(\frac{\mathcal{J}}{2Nr^2}-\frac{c\alpha}{N\gamma}\right)dr
   -\frac{\alpha\beta r}{\gamma}\,d\phi\,, &
\varpi_2&=\frac{r}{\ell^2}dt
   -\left(\frac{\mathcal{J}}{2r}+\frac{c\alpha}{\gamma}r\right)d\phi
   +\frac{\alpha\beta}{\gamma N}\,dr\,,
\end{align}
with $\tau$, $e_i$, $\omega$ and $\bar{\tau}=0$ as in \eqref{BTZ-soln-of-vielbein} and
\eqref{BTZ-solution-of-omega}, and with $\Lambda_0=-\sigma/\ell^{2}$ and
$c=1/(2q\ell^{2})$ unchanged. One verifies directly that \eqref{BTZ-beta} satisfies all
six equations \eqref{carroll-eom} exactly, for every constant $\beta$.

%--- 3.26  why it works ------------------------------------------------
The reason is visible in \eqref{carroll-eom-cov}. The combination that the scalar
equations constrain is $K=\epsilon_{ij}e_ih_j$, and it does not feel the shift,
\begin{equation} \label{K-invariant}
\epsilon_{ij}e_i\left(h_j+\beta\epsilon_{jn}e_n\right)
 =\epsilon_{ij}e_ih_j-\beta\,e_ie_i=\epsilon_{ij}e_ih_j\,,
\end{equation}
so \eqref{eom-tau-cov}, \eqref{eom-bartau-cov} and \eqref{eom-omega-cov} are untouched,
and with them the value of $c$. Equations \eqref{eom-h-cov} and \eqref{eom-varpi-cov}
contain no $h_i$ at all. The only one that could notice is \eqref{eom-e-cov}, where the
shift enters through
$\mathcal{D}\left(\beta\epsilon_{ij}e_j\right)=\beta\epsilon_{ij}\mathcal{D}e_j$, and that
vanishes with the Carrollian torsion.

%--- 3.27  what it changes ---------------------------------------------
Two things do change. The auxiliary field is no longer proportional to the spatial
zweibein, so \eqref{BTZ-beta} is not of the maximally symmetric form \eqref{h-is-ce}. And
the two-form $\Psi:=e_ih_i-\tau\bar{\tau}$ of Section~\ref{third-way} no longer vanishes,
\begin{equation} \label{Psi-beta}
\Psi=e_ih_i=\beta\,\epsilon_{ij}e_ie_j=2\beta\,e_1e_2\,,
\end{equation}
a point taken up after \eqref{ehzero} in Section~\ref{third-way}.

The deformation is not a diffeomorphism. For constant $c$ and $\beta$ the ansatz gives
$\mathcal{L}_{\xi}h_i=c\,\mathcal{L}_{\xi}e_i+\beta\epsilon_{ij}\mathcal{L}_{\xi}e_j$, so
a transformation that leaves $e_i$ inert, as a change of $\beta$ alone does, would have to
leave $h_i$ inert as well, and cannot produce
$\delta_{\beta}h_i=\delta\beta\,\epsilon_{ij}e_j$. Matching Lie derivatives component by
component on \eqref{BTZ-beta} gives the same answer, returning $\delta\beta=0$ whenever
$\alpha\neq0$. What \eqref{beta-shift} does to the boundary charges is recorded at the end
of Section~\ref{bdrycharges}.

%--- 3.28  the deformation is an improper gauge transformation ---------
It is, however, a gauge transformation. Two of the local symmetries listed in
Section~\ref{mmgsec} combine to produce it at constant parameters. The first is the shift
\eqref{null-shift}, with $S_{ij}$ traceless. The second is the time translation
$\varsigma$ of \eqref{rot-boost}, whose $\delta_{\varsigma}\tau=d\varsigma$ vanishes at
constant $\varsigma$ while $\delta_{\varsigma}\varpi_i$ does not. One checks on each field
that
\begin{equation} \label{beta-split}
\delta_{\beta}=\delta_S\Big|_{S_{ij}=\beta\epsilon_{ij}}
 -\delta_{\varsigma}\Big|_{\varsigma=-q\beta/(\mu\Lambda_0)}\,.
\end{equation}
The shift supplies $\delta h_i=\beta\epsilon_{ij}e_j$ and
$\frac{\gamma\beta}{\sigma}\epsilon_{ij}e_j$ in $\varpi_i$, while the time translation
removes $\frac{\sigma q\beta}{\mu}\epsilon_{ij}e_j$ from the latter. Since
$\frac{\sigma q}{\mu}=\frac{\alpha}{\gamma}+\frac{\gamma}{\sigma}$, what is left in
$\varpi_i$ is $-\frac{\alpha\beta}{\gamma}\epsilon_{ij}e_j$, which is
\eqref{beta-shift}. So \eqref{BTZ-beta} is a gauge copy of
\eqref{BTZ-solution-of-omega} and not a new solution.

Two things keep this from making \eqref{beta-shift} empty. The decomposition needs
$\Lambda_0\neq0$, since $\delta_{\varsigma}\varpi_i$ carries a factor of $\Lambda_0$. On
the flat branch \eqref{beta-shift} is a rigid symmetry and not a gauge transformation. And
both parameters are constant, so neither vanishes at the boundary. A gauge transformation
of that kind is improper and may carry charge, which is what \eqref{beta-charge} computes.
The factor $q/\mu$ of \eqref{beta-split} is the one that reappears there. The deformation
is thus invisible in the bulk and visible at the boundary, and then only to a generator
that moves the radial coordinate.

%=======================================================================
%  SECTION 4 :  CARROLLIAN THIRD-WAY CONSISTENCY
%=======================================================================
\section{Carrollian third-way consistency}\label{third-way}

%--- 4.1  covariant derivative of a Lorentz vector, split --------------
As discussed at the end of Section~\ref{mmgalg}, third-way consistency is the
statement that the field equations cannot be obtained by varying an action for the
metric alone. Its signature in the Chern--Simons-like form is the failure of the Jacobi
identity.

Let us first separate the space and time indices of a Lorentz vector-valued one-form $V^a$,
\begin{align} \label{Dvec}
DV^a&=dV^a+\epsilon^{abc}\mathtt{w}_bV_c \nn \\
a=0\,:\quad DV^0&=dV^0-\epsilon_{ij} \mathtt{w}_iV_j\,, \nn \\
a=i\,:\quad DV^i&=dV^i-\epsilon_{ij} \mathtt{w}_jV_0+\epsilon_{ij}\mathtt{w}_0V_j\,.
\end{align}
Recall the rescaling of the fields introduced earlier in \eqref{rescaling-fields} as
\begin{equation} \label{rescale}
\mathtt{e}^a=(\mathtt{e}^0,\mathtt{e}^i)=\left(\frac{\tau}{\rho},e_i\right)\,, \quad
\mathtt{w}^a=(\mathtt{w}^0,\mathtt{w}^i)=\left(\omega,\frac{1}{\rho}\varpi_i\right)\,, \quad
\mathtt{h}^a=(\mathtt{h}^0,\mathtt{h}^i)=\left(\bar{\tau},\frac{1}{\rho}h_i\right)\,.
\end{equation}

%--- 4.3  torsion, curvature, Schouten field strength; decomposition ---
The torsion $T^a$, curvature $\mathtt{R}^a$, and the Schouten field strength $\bar{T}^a$ are
defined in terms of relativistic fields $(\mathtt{e},\mathtt{w},\mathtt{h})$ such that
\begin{align} \label{TRTbar}
T^a&=D\mathtt{e}^a=d\mathtt{e}^a+\epsilon^{abc}\mathtt{w}_b\mathtt{e}_c\,, &
\mathtt{R}^a&=d\mathtt{w}^a+\frac{1}{2}\epsilon^{abc}\mathtt{w}_b\mathtt{w}_c\,, &
\bar{T}^a&=D\mathtt{h}^a=d\mathtt{h}^a+\epsilon^{abc}\mathtt{w}_b\mathtt{h}_c\,.
\end{align}
Decomposing the Lorentz indices into time and space components, followed by the rescaling \eqref{rescale}, yields
\begin{align} \label{decomposed}
T^0&=\frac{1}{\rho}\left(d\tau-\epsilon_{ij}\varpi_i e_j\right)=\frac{1}{\rho}\mathcal{R}(\tau)\,, \nn \\
T_i&=de_i-\epsilon_{ij}\omega e_j+\frac{1}{\rho^{2}}\epsilon_{ij}\varpi_j\tau=\mathcal{R}_i(e)+\frac{1}{\rho^{2}}\epsilon_{ij}\varpi_j\tau\,, \nn \\
\mathtt{R}^0&=d\omega-\frac{1}{2\rho^{2}}\epsilon_{ij}\varpi_i\varpi_j=\mathcal{R}(\omega)-\frac{1}{2\rho^{2}}\epsilon_{ij}\varpi_i\varpi_j\,, \nn \\
\mathtt{R}_i&=\frac{1}{\rho}\left(d\varpi_i-\epsilon_{ij}\omega\varpi_j\right)=\frac{1}{\rho}\mathcal{R}_i(\varpi)\,, \nn \\
\bar{T}^0&=d\bar{\tau}-\frac{1}{\rho^{2}}\epsilon_{ij}\varpi_ih_j=\mathcal{R}(\bar{\tau})\,, \nn \\
\bar{T}_i&=\frac{1}{\rho}\left(dh_i-\epsilon_{ij}\omega h_j+\epsilon_{ij}\varpi_j\bar{\tau}\right)=\frac{1}{\rho}\mathcal{R}_i(h)+\frac{1}{\rho}\epsilon_{ij}\varpi_j\bar{\tau}\,.
\end{align}
%--- 4.5  the field equations in rescaled form -------------------------
Now the field equations in terms of rescaled fields read
\begin{equation} \label{Fsplit}
0=F=E(\tau)H+E(\bar{\tau})\bar{H}+E(\omega)J
    +E_i(e)P_i+E_i(\varpi)K_i+E_i(h)\mathcal{P}_i\,,
\end{equation}
where
\begin{align} \label{Fcomponents}
E(\tau)      &:=\rho T^0-\frac{\alpha}{\gamma}\epsilon_{ij}e_ih_j\,, \nn \\
E(\bar{\tau})&:=\bar{T}^0-\frac{\alpha}{2\rho^{2}\gamma}\epsilon_{ij}h_ih_j
               -\frac{\Lambda_0}{2\gamma}\epsilon_{ij}e_ie_j
               -\frac{\sigma q}{\gamma}\epsilon_{ij}e_ih_j\,, \nn \\
E(\omega)    &:=\mathtt{R}^0-q\,\epsilon_{ij}e_ih_j\,, \nn \\
E_i(e)     &:=T_i-\frac{\alpha}{\rho^{2}\gamma}\epsilon_{ij}\tau h_j
               -\frac{\alpha}{\gamma}\epsilon_{ij}\bar{\tau}e_j\,, \nn \\
E_i(\varpi)&:=\rho\,\mathtt{R}_i-q\,\epsilon_{ij}\tau h_j
               -q\rho^{2}\epsilon_{ij}\bar{\tau}e_j\,, \nn \\
E_i(h)     &:=\rho\bar{T}_i-\frac{\Lambda_0}{\gamma}\epsilon_{ij}\tau e_j
               -\frac{\sigma q}{\gamma}\epsilon_{ij}\tau h_j
               -\frac{\sigma q\rho^{2}}{\gamma}\epsilon_{ij}\bar{\tau}e_j
               -\frac{\alpha}{\gamma}\epsilon_{ij}\bar{\tau}h_j\,.
\end{align}

%--- 4.7  epsilon identities and the dictionary ------------------------
We shall need the identities
\begin{align} \label{epsids}
\epsilon^{abc}(u\times v)_b w_c&=u^a(v\cdot w)+v^a(u\cdot w)\,, &
\epsilon^{abc}u_b(v\times w)_c&=v^a(u\cdot w)+w^a(u\cdot v)\,,
\end{align}
valid for Lorentz vector-valued one-forms, where $(u\cdot v):=u^a\wedge v_a$. Note
that $(\mathtt{e}\cdot \mathtt{e})$ and $(\mathtt{h}\cdot \mathtt{h})$ vanish identically. Writing $\mathcal{E}^a,\mathcal{F}^a,\mathcal{G}^a$ for the
$\mathtt{P}_a$, $\mathtt{J}_a$ and $\mathtt{Q}_a$ components of $F$, and comparing with \eqref{Fsplit}, one finds that the two sets of components are
related by
\begin{equation} \label{dictionary}
E(\tau)=\rho\,\mathcal{E}^0\,, \quad E_i(e)=\mathcal{E}^i\,, \quad
E(\omega)=\mathcal{F}^0\,, \quad E_i(\varpi)=\rho\,\mathcal{F}^i\,, \quad
E(\bar{\tau})=\mathcal{G}^0\,, \quad E_i(h)=\rho\,\mathcal{G}^i\,,
\end{equation}
the factors of $\rho$ following from the generator rescalings \eqref{rescaling-generators}.

%--- 4.9  D(e x h) and the covariant form of DF ------------------------
Now using $D\mathtt{R}^a=0$ together with
$D^2 \mathtt{e}^a=\epsilon^{abc} \mathtt{R}_b \mathtt{e}_c$ and
$D^2 \mathtt{h}^a=\epsilon^{abc} \mathtt{R}_b \mathtt{h}_c$,
one finds first
\begin{equation} \label{Dcross}
D(\mathtt{e}\times \mathtt{h})^a=\epsilon^{abc}\left(T_b\mathtt{h}_c-\mathtt{e}_b\bar{T}_c\right)
 =\frac{\tilde{\sigma} \tilde{q}}{\tilde{\gamma}}\,\mathtt{e}^a(\mathtt{e}\cdot \mathtt{h})\,,
\end{equation}
the $\mathtt{h}^a(\mathtt{e}\cdot \mathtt{h})$ contributions canceling between the two terms
and the $\tilde{\Lambda}_0$ contribution dropping by $(\mathtt{e}\cdot \mathtt{e})=0$.
Substituting \eqref{Dcross} back into the covariant derivatives of the three equations, and using
$1+\tilde{\sigma}\tilde{\alpha}/\tilde{\gamma}^{2}=\tilde{q}/(\tilde{\mu}\tilde{\gamma})$, gives
\begin{align} \label{DFcov}
D\mathcal{E}^a&=\frac{\tilde{q}^{2}}{\tilde{\mu}\tilde{\gamma}}\,\mathtt{e}^a(\mathtt{e}\cdot \mathtt{h})\,, \nn \\
D\mathcal{F}^a&=\frac{\tilde{q}^{2}\tilde{\sigma}}{\tilde{\gamma}}\,\mathtt{e}^a(\mathtt{e}\cdot \mathtt{h})\,, \nn \\
D\mathcal{G}^a&=-\frac{\tilde{q}^{2}}{\tilde{\mu}\tilde{\gamma}}\,\mathtt{h}^a(\mathtt{e}\cdot \mathtt{h})+\frac{\tilde{q}^{2}\tilde{\sigma}^{2}}{\tilde{\gamma}^{2}}\,\mathtt{e}^a(\mathtt{e}\cdot \mathtt{h})\,.
\end{align}
Every component is proportional to $\mathtt{e}\cdot \mathtt{h}$, and every coefficient to $\tilde{q}^{2}$:
the failure of $DF$ to vanish identically is controlled by the same $\tilde{q}\neq0$ that
obstructs the Jacobi identity.

%--- 4.11  decomposition, and the closure condition --------------------
Decomposing \eqref{DFcov} with $\mathtt{e}^0=\tau/\rho$, $\mathtt{e}^i=e_i$,
$\mathtt{h}^0=\bar{\tau}$, $\mathtt{h}^i=h_i/\rho$ and
$\mathtt{e}\cdot \mathtt{h}=\Psi/\rho$, where
$\Psi:=e_ih_i-\tau\bar{\tau}=\rho\,\mathtt{e}^{a}\mathtt{h}_{a}$, the six components of
$DF$ read
\begin{align} \label{DFcomponents}
DE(\tau)&=\frac{q^{2}}{\mu\gamma}\,\tau\wedge\Psi\,, \nn \\
DE(\bar{\tau})&=-\frac{q^{2}}{\mu\gamma}\,\bar{\tau}\wedge\Psi+\frac{q^{2}\sigma^{2}}{\gamma^{2}}\,\tau\wedge\Psi\,, \nn \\
DE(\omega)&=\frac{q^{2}\sigma}{\gamma}\,\tau\wedge\Psi\,, \nn \\
DE_i(e)&=\frac{q^{2}}{\mu\gamma}\,e_i\wedge\Psi\,, \nn \\
DE_i(\varpi)&=\frac{\rho^{2}q^{2}\sigma}{\gamma}\,e_i\wedge\Psi\,, \nn \\
DE_i(h)&=-\frac{q^{2}}{\mu\gamma}\,h_i\wedge\Psi+\frac{\rho^{2}q^{2}\sigma^{2}}{\gamma^{2}}\,e_i\wedge\Psi\,.
\end{align}
Since $q\neq0$, demanding $DE(\tau) = 0$ and $DE_i(e) = 0$
requires $\tau\wedge\Psi=0$ and $e_i\wedge\Psi=0$, which automatically
ensures that the other four vanish as well so that $DF=0$.
As $(\tau,e_1,e_2)$ is a basis of one-forms these three
conditions are equivalent to the single two-form relation
\begin{equation} \label{ehzero}
\Psi=e_ih_i-\tau\bar{\tau}=0
\qquad\Longleftrightarrow\qquad \mathtt{e}^{a}\mathtt{h}_{a}=0\,,
\end{equation}
that is, $h_{\mu\nu}:=\mathtt{e}_{\mu}{}^{a}\mathtt{h}_{\nu a}$ is symmetric.
The remaining two structures in
\eqref{DFcomponents}, $\bar{\tau}\wedge\Psi$ and $h_i\wedge\Psi$, then vanish
automatically. Note that \eqref{Fsplit}--\eqref{DFcomponents} are exact at finite
$\rho$, the limit $\rho\to\infty$ being taken only at the end; in particular the
$\rho^{2}$ factors in $DE_i(\varpi)$ and $DE_i(h)$ multiply $\Psi$ and are therefore
harmless once \eqref{ehzero} holds.

%--- 4.13  the scope of the closure condition --------------------------
The scope of \eqref{ehzero} deserves care. The calculation above is carried out at
finite $\rho$, where $F=0$ is relativistic MMG written in rescaled variables, so
\eqref{ehzero} is a property of the solutions of that system. The Carrollian field
equations \eqref{carroll-eom} are the $\rho\to\infty$ limit of the Euler--Lagrange
equations of \eqref{rescaled-lagrangian}, and they are weaker. The components
$E_i(\varpi)$ and $E_i(h)$ of \eqref{Fcomponents} carry explicit factors of $\rho^{2}$,
and only the finite combination $\sigma E_i(\varpi)-\gamma E_i(h)$ survives the limit.
Carrollian MMG therefore has more solutions than the limits of relativistic ones, and
not every one of them satisfies \eqref{ehzero}.

What \eqref{carroll-eom} does force is the weaker statement $e_i\wedge\Psi=0$, and that
is already what the closure needs. With the covariant derivative \eqref{Dcov} one finds
\begin{equation} \label{thirdway-cov}
\boxed{\;\mathcal{D}\left(\gamma\,\mathcal{D}e_i-\alpha\epsilon_{ij}\bar{\tau}e_j\right)
 =\frac{q^{2}}{\mu}\;e_i\wedge\Psi\;}
\end{equation}
up to terms proportional to the field equations, with no reference to $\rho$ anywhere.
The left-hand side is the covariant derivative of \eqref{eom-h-cov} and therefore
vanishes on every solution, which gives $e_i\wedge\Psi=0$ there. The right-hand side
does not vanish off shell, and its coefficient is once more the $q^{2}$ of
\eqref{carroll-jacobiators}. Equation \eqref{thirdway-cov} is third-way consistency
stated intrinsically: the closure requirement is a genuine condition rather than an
identity, it is cubic in the fields, and the field equations supply exactly what it
needs.

The same conclusion follows without \eqref{thirdway-cov}, by an argument worth recording
because it says which part of $h_i$ the field equations leave free. At $\bar{\tau}=0$,
equations \eqref{eom-bartau-cov} and \eqref{eom-omega-cov} give
$\mathcal{R}(\omega)=q\,\epsilon_{ij}e_ih_j$, while \eqref{eom-tau-cov} gives
$\mathcal{R}(\omega)=-\frac{\Lambda_0}{2\sigma}\epsilon_{ij}e_ie_j$, so that
\begin{equation} \label{KE}
\epsilon_{ij}e_ih_j=-\frac{\Lambda_0}{2\sigma q}\,\epsilon_{ij}e_ie_j\,.
\end{equation}
Since $(\tau,e_1,e_2)$ is a basis of one-forms, write $h_i=a_i\tau+B_{ij}e_j$. Then $\epsilon_{ij}e_ih_j$ equals
$\epsilon_{ij}a_j\,e_i\tau+B_{kk}\,e_1e_2$, and the right-hand side of
\eqref{KE} has no $e_i\tau$ part, so $a_i=0$ and $h_i=B_{ij}e_j$. Hence
\begin{equation} \label{Psi-B}
\Psi=e_ih_i=\left(B_{12}-B_{21}\right)e_1e_2\,,
\end{equation}
and $e_k\wedge\Psi=0$ for both $k$, because $e_k\wedge e_1e_2=0$. Only the antisymmetric
part of $B_{ij}$ reaches $\Psi$ at all. Its trace is fixed by \eqref{KE} to
$B_{kk}=-\Lambda_0/\sigma q$, which is $2c$ at $\Lambda_0=-\sigma/\ell^{2}$, and its
symmetric traceless part never enters. This is why \eqref{ehzero} is a stronger statement
than the field equations make.

The stronger condition \eqref{ehzero} is not a consequence of \eqref{carroll-eom}. The
deformed black hole \eqref{BTZ-beta} solves all six equations and has
$\Psi=2\beta\,e_1e_2$ by \eqref{Psi-beta}, while satisfying $e_i\wedge\Psi=0$ because
$e_k\wedge e_1e_2=0$. On the undeformed solutions of Section~\ref{adsbtz} one does have
$\Psi=0$: there $\bar{\tau}=0$, so that $\tau\bar{\tau}=0$, while $h_i=c\,e_i$ gives
$e_ih_i=c\,\delta_{ij}e_ie_j=0$, the wedge product being antisymmetric and
$\delta_{ij}$ symmetric. That is a property of the maximally symmetric branch
\eqref{h-is-ce} rather than of the theory. One caution goes with this. By \eqref{Psi-B} the
antisymmetric part of $B_{ij}$ is the whole of $\Psi$, and the local symmetry
\eqref{null-shift} shifts it, $\delta_S\Psi=(S_{12}-S_{21})e_1e_2$.
A non-vanishing $\Psi$ can therefore be removed by a gauge choice, so $\Psi$ is not an
invariant of a solution, and \eqref{BTZ-beta} is itself a gauge copy of
\eqref{BTZ-solution-of-omega} by \eqref{beta-split}. What is being said here is only that \eqref{carroll-eom} does not
force $\Psi=0$.

%--- 4.16  a second condition, at order q ------------------------------
A second condition of the same kind is hidden in \eqref{carroll-eom}, and it enters at
order $q$ rather than $q^{2}$. Taking the exterior derivative of
\eqref{eom-bartau-cov} gives
\begin{equation} \label{dEbtau}
d\left(\mathcal{D}\tau-\frac{\alpha}{\gamma}\epsilon_{ij}e_ih_j\right)
 =\frac{\sigma q}{\mu}\,\epsilon_{ij}\,e_i\wedge\mathcal{D}h_j\,,
\end{equation}
again up to terms proportional to the field equations. Every solution of
\eqref{carroll-eom} therefore satisfies
\begin{equation} \label{eDh}
\epsilon_{ij}\,e_i\wedge\mathcal{D}h_j=0\,.
\end{equation}
At $\bar{\tau}=0$ one may use \eqref{eom-e-cov} to trade $\mathcal{D}h_i$ for
$\mathcal{R}_i(\varpi)$. The term carrying $\Lambda_0$ drops out of the contraction
because $e_i\wedge\tau\wedge e_i=0$, and \eqref{eDh} becomes
$\epsilon_{ij}e_i\wedge\mathcal{R}_j(\varpi)=0$. This is independent of
$e_i\wedge\Psi=0$, neither condition implying the other.

The complete list of conditions that \eqref{carroll-eom} forces is therefore
$\bar{\tau}=0$, $e_i\wedge\Psi=0$ and \eqref{eDh}. The deformed black hole
\eqref{BTZ-beta} satisfies the last of these as well. There
$h_i=c\,e_i+\beta\epsilon_{ij}e_j$, so that
$\mathcal{D}h_i=c\,\mathcal{D}e_i+\beta\epsilon_{ij}\mathcal{D}e_j$, and this vanishes
with the Carrollian torsion. All three conditions collapse at $q=0$. There the Jacobi
identity is restored, the obstructions in \eqref{thirdway-cov} and \eqref{dEbtau}
disappear, and the model becomes an ordinary Chern--Simons theory.

%--- what has and has not been established -----------------------------
What has been established here is the on-shell consistency of the
first-order system \eqref{carroll-eom}, in the intrinsic form \eqref{thirdway-cov}: the
closure requirement is a genuine condition rather than an identity, precisely because
the Jacobi identity fails, and it is satisfied on every solution. Third-way consistency is
more commonly stated one level down: in the relativistic theory one eliminates
$\mathtt{h}^{a}$ in favor of a single equation for the metric, whose divergence vanishes
only when that equation itself holds. The Carrollian analogue would require eliminating
$h_{i}$ and $\bar{\tau}$ and exhibiting the corresponding identity for the Carrollian
geometry; we do not attempt that here, and the results of this section should be read as
the first-order statement of third-way consistency rather than its metric-level
counterpart.

%=======================================================================
%  SECTION 5 :  CARROLLIAN CHERN-SIMONS-LIKE THEORIES
%=======================================================================
\section{Carrollian Chern--Simons-like theories}\label{CCSlike}

%--- 5.1  field content and index conventions --------------------------
Here we show how the Carrollian MMG Lagrangian \eqref{carrollian-mmg-lagrangian}
follows from the generic form of a Chern--Simons-like Lagrangian for three one-form
flavors. The section has three outputs that the rest of the paper uses. It fixes the two
flavor metrics and the cubic flavor tensor of the model, it produces the primary and
temporal constraints, and it gives the canonical boundary-charge formula
\eqref{NLCSL-charge}. These feed Sections~\ref{bdrycharges}, \ref{Dirac} and
\ref{asympt} in turn.

We first introduce two different sets of one-form fields\footnote{The indices $I,J,\ldots$ label the
one-form flavors that transform as scalars under spatial rotations, and $r,s,\ldots$ those
transforming as spatial vectors, the latter carrying an internal index $a,b=1,2$, written
$i,j$ in Sections~\ref{mmgsec} to \ref{third-way}. The spatial coordinate indices $K,L$
are introduced below.}
\begin{equation} \label{oneform-fields}
A^I=A_\mu^I dx^\mu\,, \qquad A^{ra}=A_\mu^{ra}dx^\mu\,.
\end{equation}
For the class of theories considered
here, the Chern--Simons-like Lagrangian three-form is
\begin{equation} \label{NLCSL-wedge}
\boxed{\;
L_{\text{CSL}} = \frac{1}{2}g_{IJ}\,A^I\wedge dA^J
 + \frac{1}{2}g_{rs}\delta_{ab}\,A^{ra}\wedge dA^{sb}
 + \frac{1}{2}f_{rsI}\epsilon_{ab}\,A^I\wedge A^{ra}\wedge A^{sb}\,.
\;}
\end{equation}
The coupling tensors satisfy
\begin{equation} \label{props}
g_{IJ}=g_{JI}\,, \qquad g_{rs}=g_{sr}\,, \qquad f_{rsI}=f_{srI}\,.
\end{equation}
The symmetry of $f_{rsI}$ under $r\leftrightarrow s$ is consistent with the antisymmetry of
both the wedge product and $\epsilon_{ab}$. No kinetic cross term $g_{Ir}$ and no cubic
coupling among three scalar flavors appear in \eqref{NLCSL-wedge}: under the spacetime split of the
Lorentz index the invariant metric contributes $\eta_{00}$ to the scalar sector and
$\delta_{ab}$ to the vector sector, with nothing that mixes the two, while the Levi-Civita
symbol is non-vanishing only when exactly one of its three indices is temporal. Thus
\eqref{NLCSL-wedge} is the general form inherited from a relativistic Chern--Simons-like
theory, not a restriction of it.

%--- 5.4  canonical form, constraints, Hamiltonian ---------------------
Up to a spatial boundary term, the Lagrangian density takes the canonical form\footnote{Here $K,L=1,2$ label the spatial
coordinates, as in Section~\ref{mmgsec}, and a dot denotes a derivative with respect to
time. They are not flavor labels.}
\begin{equation} \label{CSL-canonical}
\mathcal L_{\text{CSL}} = -\frac{1}{2}\epsilon^{KL}
 \left( g_{IJ}A_K^I\dot A_L^J + g_{rs}\delta_{ab}A_K^{ra}\dot A_L^{sb} \right)
 + A_0^I\phi_I + A_0^{ra}\phi_{ra}\,.
\end{equation}
The temporal components $A_0^I$ and $A_0^{ra}$ appear without time
derivatives and therefore act as Lagrange multipliers. They impose the
primary constraints
\begin{align} \label{scalar-constraint}
\phi_I &= \epsilon^{KL}\left( g_{IJ}\partial_K A_L^J
 + \frac{1}{2} f_{rsI}\epsilon_{ab} A_K^{ra}A_L^{sb} \right)\,, &
\phi_{ra} &= \epsilon^{KL}\left( g_{rs}\delta_{ab}\partial_K A_L^{sb}
 + f_{rsI}\epsilon_{ab} A_K^{sb}A_L^I \right)\,,
\end{align}
so that the canonical Hamiltonian density is
$\mathcal H = -A_0^I\phi_I - A_0^{ra}\phi_{ra}$.

%--- 5.6  boundary charge ----------------------------------------------
The canonical boundary-charge variation can then be written as
\begin{equation} \label{NLCSL-charge}
\boxed{\;
\delta Q[\Lambda] = -\frac{1}{8\pi G} \oint_{\partial\Sigma}d\phi
 \left( g_{IJ}\Lambda^I\delta A_\phi^J
      + g_{rs}\delta_{ab}\Lambda^{ra}\delta A_\phi^{sb} \right)\,,
\;}
\end{equation}
where $\Lambda^I$ and $\Lambda^{ra}$ are the gauge parameters. This is the Carroll-split
counterpart of the canonical boundary term of Chern--Simons-like gravity
\cite{Bergshoeff:2014bia}. What matters here is that it is written with the lowered flavor
metrics alone and never calls for their inverse. For a theory whose flavor
metric can be inverted it is the standard result. Here it cannot be inverted, and
\eqref{NLCSL-charge} is used in Sections~\ref{bdrycharges} and \ref{Dirac} before it is
justified. Section~\ref{asympt} justifies it, by deriving the same expression from the
boundary term of the first-class generator.

%=======================================================================
\subsection{The flavor data of Carrollian MMG}\label{flavordata}
%=======================================================================

Matching \eqref{NLCSL-wedge} to \eqref{carrollian-mmg-lagrangian} fixes $g_{IJ}$,
$g_{rs}$ and $f_{rsI}$ uniquely. The output that matters for everything after this section
is the degeneracy of $g_{rs}$.

%--- 5.7  flavor assignment --------------------------------------------
The one-form flavors are identified as
\begin{align} \label{flavor-assignment}
A^I &= \left( A^\tau, A^{\bar\tau}, A^\omega \right) = \left( \tau,\bar\tau,\omega \right)\,, &
A^{ra} &= \left( A^{ea}, A^{\varpi a}, A^{ha} \right) = \left( e^a,\varpi^a,h^a \right)\,.
\end{align}
%--- 5.8  the two flavor metrics ---------------------------------------
In the ordered bases $I=(\tau,\bar\tau,\omega)$ and $r=(e,\varpi,h)$ the two flavor metrics
read
\begin{align} \label{flavor-metrics}
g_{IJ} &= \begin{pmatrix}
        0      & -\gamma & \sigma            \\
       -\gamma &  0      &  0                \\
        \sigma &  0      & -\dfrac{1}{\mu}
\end{pmatrix}\,, &
g_{rs} &= \begin{pmatrix}
        0      & -\sigma & \gamma \\
       -\sigma &  0      &  0     \\
        \gamma &  0      &  0
\end{pmatrix}\,.
\end{align}
The vector flavor metric is degenerate, $\det g_{rs}=0$, with null direction
$\gamma\,\varpi+\sigma\,h$; in particular $g_{\varpi\varpi}=g_{\varpi h}=g_{hh}=0$. The
scalar flavor metric is not, $\det g_{IJ}=\gamma^{2}/\mu$, and neither is the flavor metric
of relativistic MMG, whose determinant is
$-\tilde{\gamma}^{2}/\tilde{\mu}=-\rho^{2}\gamma^{2}/\mu$. What differs is the sign, and it
does so because the spacetime split supplies $\eta_{00}=-1$ in the scalar block. The degeneracy is therefore
produced by the Carroll contraction and is confined to the vector sector. This degeneracy controls the
boundary charges of Sections~\ref{bdrycharges} and \ref{asympt}.

%--- 5.9  cubic flavor tensor ------------------------------------------
The non-vanishing components of the cubic flavor tensor are
\begin{align} \label{cubic-flavor-tensor}
f_{ee\tau}      &= \Lambda_0\,, &
f_{e\varpi\omega}    &= -\sigma\,, &
f_{e\varpi\bar\tau}  &= \gamma\,, \nn \\
f_{eh\omega}    &= \gamma\,, &
f_{eh\bar\tau}  &= \alpha\,. &&
\end{align}
Due to \eqref{props}, one has, e.g.
\begin{equation} \label{f-symmetry}
f_{e\varpi\omega} = f_{\varpi e\omega}\,, \qquad f_{eh\bar\tau} = f_{he\bar\tau}\,.
\end{equation}

%--- 5.11  equivalence with the Carrollian MMG Lagrangian --------------
With these data \eqref{NLCSL-wedge} reproduces the Carrollian MMG Lagrangian
\eqref{carrollian-mmg-lagrangian}: the cubic terms agree identically, and the kinetic
terms agree up to a total derivative, the difference being exactly $d\Upsilon$ with no
residue,
so that $g_{IJ}$, $g_{rs}$ and $f_{rsI}$ are fixed uniquely by
\eqref{carrollian-mmg-lagrangian} rather than chosen to fit it,
\begin{equation} \label{CSL-equals-CMMG}
L_{\text{CSL}} = L_{\text{CMMG}} + d\Upsilon\,, \qquad
\Upsilon = \frac{\gamma}{2}\,\tau\wedge\bar\tau
  - \frac{\sigma}{2}\,\omega\wedge\tau
  - \frac{\sigma}{2}\,\delta_{ab}\,e^a\wedge\varpi^b
  - \frac{\gamma}{2}\,\delta_{ab}\,e^a\wedge h^b\,.
\end{equation}

%=======================================================================
%  SECTION 6 :  BOUNDARY CHARGES FROM THE CANONICAL VARIATION
%=======================================================================
\section{Boundary charges from the canonical variation}\label{bdrycharges}
The charges of the Carrollian BTZ solution are computed twice in what follows, by routes
that share no intermediate step. Here we evaluate the canonical boundary charge
\eqref{NLCSL-charge}, obtained in Section~\ref{CCSlike} by varying the Hamiltonian of the
Chern--Simons-like form, on the solution
\eqref{BTZ-soln-of-vielbein}--\eqref{BTZ-solution-of-omega}.
Section~\ref{asympt} reaches the same charges a second time, as the boundary terms of the
Castellani generator built from the first-class constraints of Section~\ref{Dirac}. The
two routes exercise different parts of the construction, and they are compared at the end
of Section~\ref{asympt}.

%--- canonical charges in Carrollian gravity ---------------------------
Canonical surface charges have been built for Carrollian theories before. The magnetic
and electric contractions of general relativity are treated canonically in
\cite{Perez:2021abf,Perez:2022jpr}, in $3+1$ dimensions, so those are a precedent rather
than a direct comparison. A geometric construction of Carrollian conserved charges, and
its agreement with the covariant phase space, is given in \cite{Ciambelli:2018ojf}.
Closest to what follows are \cite{Aviles:2024llx,Aviles:2025ygw}, where canonical surface
charges are obtained in the Chern--Simons formulation of three-dimensional AdS--Carroll
gravity. What distinguishes
the computation below is that \eqref{NLCSL-charge} belongs to a Chern--Simons-like theory
with a degenerate flavor metric rather than to a Chern--Simons theory, which is why the
result is checked a second way.

%--- 6.1  gauge parameters from a spacetime symmetry -------------------
For a spacetime symmetry generated by a vector field $\xi^\mu$, the
corresponding gauge parameters are chosen as
\begin{equation} \label{gauge-parameters-xi}
\Lambda_\xi^I = \xi^\mu A_\mu^I\,, \qquad \Lambda_\xi^{ra} = \xi^\mu A_\mu^{ra}\,.
\end{equation}
%--- 6.2  Carrollian BTZ angular components ----------------------------
The angular components of the solution are
\begin{align} \label{BTZ-angular-components}
\tau_\phi&=0\,, & \bar\tau_\phi&=0\,, & \omega_\phi&=N\,, \nn \\
e_\phi^a&=(0,r)\,, &
\varpi_\phi^a &= \left( 0, -\frac{\mathcal J}{2r}-\frac{c \alpha}{\gamma}r \right)\,, &
h_\phi^a&=(0,c r)\,.
\end{align}

%--- 6.3  solution-space variations ------------------------------------
The solution-space variation changes $M$ and $\mathcal J$, while
$r$, $\ell$, $G$, and all theory parameters are kept fixed. The gauge parameters
\eqref{gauge-parameters-xi} are built from the solution and so carry $M$ and
$\mathcal{J}$ themselves, but they are the parameters of the transformation whose charge is
being computed and are held fixed under $\delta$, which acts on the fields alone. This is
the usual Regge--Teitelboim prescription, and it is worth stating explicitly here because
\eqref{lapse} makes the dependence of $\Lambda^{\tau}_{E}=N$ on $M$ manifest.
By \eqref{lapse} the lapse carries no $\mathcal{J}$, so that
$N\delta N=-\frac{1}{2}\delta M$. The only
non-vanishing angular variations are therefore
\begin{equation} \label{angular-variations}
\delta\omega_\phi = \delta N = -\frac{\delta M}{2N}\,, \qquad
\delta\varpi_\phi^2 = -\frac{\delta\mathcal J}{2r}\,,
\end{equation}
the remaining part $-\frac{c\alpha}{\gamma}r$ of $\varpi_\phi^2$ being independent of the
state parameters $M$ and $\mathcal J$. All other angular variations vanish,
\begin{equation} \label{vanishing-BTZ-angular-variations}
\delta\tau_\phi = \delta\bar\tau_\phi = \delta e_\phi^a = \delta h_\phi^a
 = \delta\varpi_\phi^1 = 0\,.
\end{equation}

%=======================================================================
\subsection{Energy}\label{energysub}
%=======================================================================

We compute the energy as the charge \eqref{NLCSL-charge} of Section~\ref{CCSlike}
carried by a time translation, evaluated on the Carrollian BTZ solution of
Section~\ref{btzsub}.

%--- 6.5  gauge parameters for xi = d_t --------------------------------
For the time-translation vector $\xi_E^\mu=(1,0,0)$ the gauge parameters are
\begin{align} \label{energy-parameters}
\Lambda_E^\tau &= \tau_t = N\,, &
\Lambda_E^{\bar\tau} &= \bar\tau_t = 0\,, &
\Lambda_E^\omega &= \omega_t = 0\,, \nn \\
\Lambda_E^{ea} &= e_t^a = (0,0)\,, &
\Lambda_E^{\varpi a} &= \varpi_t^a = \left( 0,\frac{r}{\ell^2} \right)\,, &
\Lambda_E^{ha} &= h_t^a = (0,0)\,.
\end{align}

%--- 6.6  assembling the energy ----------------------------------------
Using the scalar flavor metric, the scalar contribution to the charge integrand is
\begin{equation} \label{energy-scalar-integrand}
g_{IJ}\Lambda_E^I\delta A_\phi^J = g_{\tau\omega}\Lambda_E^\tau\delta\omega_\phi
 = \sigma N\delta N\,.
\end{equation}
The vector contribution vanishes because $e_t^a=h_t^a=0$ and
$\delta e_\phi^a=\delta h_\phi^a=0$, while the one surviving pairing,
$\delta_{ab}\Lambda_E^{\varpi a}\delta\varpi_\phi^{b}$, is annihilated
by $g_{\varpi\varpi}=0$. Hence
\begin{equation} \label{deltaE}
\delta E_{\rm C}=\delta Q[\partial_t]
 = -\frac{1}{8\pi G}\int_0^{2\pi}\!\!d\phi\, \sigma N\delta N
 = \frac{\sigma}{8G}\delta M\,,
\end{equation}
and choosing the configuration $M=\mathcal J=0$ to have zero charge gives
\begin{equation} \label{CMMG-Carroll-energy}
E_{\rm C} = Q[\partial_t] = \frac{\sigma M}{8G}\,.
\end{equation}

%=======================================================================
\subsection{Angular momentum}\label{angmomsub}
%=======================================================================

The angular momentum is the same charge carried by a rotation, on the same solution.

%--- 6.9  gauge parameters for xi = -d_phi -----------------------------
We define the angular momentum by $\xi_J^\mu=(0,0,-1)$, that is
$J_{\rm C}=Q[-\partial_\phi]$, for which the gauge parameters are
\begin{align} \label{angular-parameters}
\Lambda_J^\tau&=0\,, & \Lambda_J^{\bar\tau}&=0\,, &
\Lambda_J^\omega&=-N\,, \nn \\
\Lambda_J^{ea} &= -e_\phi^a = (0,-r)\,, &
\Lambda_J^{\varpi a} &= -\varpi_\phi^a
 = \left( 0, \frac{\mathcal J}{2r} + \frac{c \alpha}{\gamma}r \right)\,, &
\Lambda_J^{ha} &= -h_\phi^a = (0,-c r)\,.
\end{align}
%--- 6.10  scalar and vector contributions -----------------------------
The scalar contribution comes from $g_{\omega\omega}$ and, in the vector sector, the only
non-vanishing contribution comes from $g_{e\varpi}=-\sigma$:
\begin{align} \label{angular-integrand}
g_{IJ}\Lambda_J^I\delta A_\phi^J
 &= g_{\omega\omega}\Lambda_J^\omega\delta\omega_\phi
  = \frac{N\delta N}{\mu} = -\frac{\delta M}{2\mu}\,, \nn \\
g_{rs}\delta_{ab}\Lambda_J^{ra}\delta A_\phi^{sb}
 &= -\sigma\Lambda_J^{e2}\delta\varpi_\phi^2
  = -\sigma(-r)\left( -\frac{\delta\mathcal J}{2r} \right)
  = -\frac{\sigma}{2}\delta\mathcal J\,.
\end{align}
Every remaining term carries $\delta e_\phi^a$ or $\delta h_\phi^a$ and so vanishes by
\eqref{vanishing-BTZ-angular-variations}. Among them is
$-\sigma\Lambda_J^{\varpi a}\delta e_{\phi a}$, which is the only pairing that could have
detected the $\alpha$ shift carried by $\Lambda_J^{\varpi 2}$.

%--- 6.11  assembling the angular momentum -----------------------------
Combining the two contributions gives
\begin{equation} \label{deltaJ}
\delta J_{\rm C}=\delta Q[-\partial_\phi]
 = -\frac{1}{8\pi G}\int_0^{2\pi}\!\!d\phi
    \left[ -\frac{\sigma}{2}\delta\mathcal J - \frac{\delta M}{2\mu} \right]
 = \frac{1}{8G}\left( \sigma\delta\mathcal J + \frac{\delta M}{\mu} \right)\,,
\end{equation}
and the same zero-charge reference configuration gives
\begin{equation} \label{CMMG-angular-momentum}
J_{\rm C} = Q[-\partial_\phi] = \frac{1}{8G}\left( \sigma\mathcal J + \frac{M}{\mu} \right)\,.
\end{equation}
%--- 6.13  summary -----------------------------------------------------
The final conserved charges are
\begin{equation} \label{final-charges}
\boxed{\;
E_{\rm C} = \frac{\sigma M}{8G}\,, \qquad
J_{\rm C} = \frac{1}{8G}\left( \sigma\mathcal J+\frac{M}{\mu} \right)\,.
\;}
\end{equation}

%--- 6.14  the relativistic benchmark ----------------------------------
The relativistic MMG charges have the same structure, and it is worth having them for
comparison. A background-subtraction computation of the kind set up in
\cite{Bergshoeff:2019rdb}, which contracts the deviation from a reference solution with a
Killing vector instead of varying a field, gives for the relativistic BTZ solution with
\begin{equation} \label{rel-BTZ}
\tilde{N}^{2}=-M+\frac{r^{2}}{\ell^{2}}+\frac{\tilde{\mathcal{J}}^{2}}{4r^{2}}\,,
 \qquad
\mathtt{e}^{2}=r\left( d\phi+\tilde{N}^{\phi}d\tilde{t} \right)\,, \qquad
\tilde{N}^{\phi}=-\frac{\tilde{\mathcal{J}}}{2r^{2}}\,,
\end{equation}
and the torsional connection $\mathtt{w}^a=\Omega^a-\frac{\tilde{\alpha}}{\tilde{\gamma}}\mathtt{h}^a$
with $\mathtt{h}^a=\tilde{c}\,\mathtt{e}^a$, the charges
\begin{equation} \label{rel-charges}
\tilde{E}=\frac{1}{8G}\left[\left(\tilde{\sigma}+\frac{\tilde{\alpha} \tilde{c}}{\tilde{\gamma}\tilde{\mu}}\right)M
 +\frac{\tilde{\mathcal{J}}}{\tilde{\mu}\ell^{2}}\right]\,, \qquad
\tilde{J}=\frac{1}{8G}\left[\left(\tilde{\sigma}+\frac{\tilde{\alpha} \tilde{c}}{\tilde{\gamma}\tilde{\mu}}\right)
 \tilde{\mathcal{J}}+\frac{M}{\tilde{\mu}}\right]\,,
\end{equation}
which reduce to the torsion-free expressions of \cite{Bergshoeff:2019rdb} at $\tilde{\alpha}=0$.
Comparing with \eqref{final-charges}, the Carrollian charges follow once the weights of
Section~\ref{mmgsec} are inserted: the prefactor
$\tilde{\sigma}+\frac{\tilde{\alpha}\tilde{c}}{\tilde{\gamma}\tilde{\mu}}$ is
$\rho\sigma+\mathcal{O}(\rho^{-1})$, the rotation parameter is
$\tilde{\mathcal{J}}=\mathcal{J}/\rho$, and the $\tilde{\mathcal{J}}/(\tilde{\mu}\ell^{2})$
term is lost from the energy; \eqref{outer-limits} carries this out. On the Carrollian side the
absence of $\alpha$ is guaranteed in advance by \eqref{cmmg-is-ctmg}.

%--- discussion of the charges -----------------------------------------
Two features of \eqref{final-charges} deserve comment. First, the energy depends only on
$M$, even though the solution carries $\mathcal{J}$. By
\eqref{lapse}--\eqref{BTZ-soln-of-h} the rotation parameter has left the Carrollian frame
altogether: it appears in none of $\tau$, $e_i$, $\omega$, $h_i$, and survives only in the
boost connection \eqref{BTZ-solution-of-omega}. It could therefore reach the energy only
through a pairing with $\delta\varpi_{i\phi}$, and \eqref{flavor-metrics} admits three,
$g_{e\varpi}\Lambda_E^{e_i}\delta\varpi_{i\phi}$,
$g_{\varpi\varpi}\Lambda_E^{\varpi_i}\delta\varpi_{i\phi}$ and
$g_{h\varpi}\Lambda_E^{h_i}\delta\varpi_{i\phi}$. The first vanishes because
$e_t^{a}=0$, which is the same fact once more: with $\mathcal{J}$ gone from the frame
there is no shift for $\xi_E=\partial_t$ to pick up. The other two vanish because
$g_{\varpi\varpi}=g_{h\varpi}=0$.

The vanishing of $g_{\varpi\varpi}$ deserves emphasis. Decomposing the relativistic energy by the
Lorentz index shows that its entire $\tilde{\mathcal{J}}$ dependence originates in the
spatial part of the Lorentz--Chern--Simons pairing,
$\mathtt{w}_t^{\,a}\delta\mathtt{w}_{\phi a}$; the temporal part contributes only to the
term proportional to $M$ in \eqref{rel-charges}. In the Carrollian theory that spatial
pairing is $\delta_{ab}\Lambda_E^{\varpi a}\delta\varpi_\phi^{b}$, and although
$\Lambda_E^{\varpi 2}$ and $\delta\varpi_\phi^{2}$ are both non-zero it is annihilated by
$g_{\varpi\varpi}=0$: the coupling $1/\mu$ survives the Carroll contraction only in the
scalar sector, its spatial counterpart being suppressed by $\rho^{-2}$ in
\eqref{rescaled-lagrangian}. The degeneracy of the vector flavor metric noted below
\eqref{flavor-metrics} is thus not a technical blemish but the reason the Carrollian energy
is blind to rotation.

Second, neither charge depends on $\alpha$, $\gamma$ or $c$. By \eqref{cmmg-is-ctmg} this
is no accident. The deformation can be removed from the Lagrangian whenever $q\neq0$, and
the boundary term left behind by that removal vanishes on this solution, where
$h_i=c\,e_i$. The energy is therefore exactly the Einstein value, and the angular momentum
is the Einstein value together with the gravitational Chern--Simons contribution $M/\mu$.
It is still worth seeing how the charge formula arranges this. The $\gamma$ couplings
$g_{\tau\bar\tau}$ and $g_{eh}$ drop out because $\bar\tau=0$ and
$\delta e_\phi^a=\delta h_\phi^a=0$. In the energy the only surviving pairing is
$\sigma N\delta N$ of \eqref{energy-scalar-integrand}, and neither factor can carry
$\alpha$. The lapse \eqref{lapse} is built from $M$ and $\ell$ alone. And
$\omega_\phi=N$ is free of the shift in \eqref{omega-from-Omega}, because the Carrollian
field equations force $\bar\tau=0$ for $q\neq0$. The scalar torsional and torsion-free
connections therefore coincide, $\omega=\Omega$. In the angular momentum,
$\Lambda_J^{\varpi 2}$ does retain the shift
$\varpi_i=\Omega_i-\frac{c\alpha}{\gamma}e_i$. The pairing that would detect it is again
annihilated by $g_{\varpi\varpi}=0$. What survives of that shift in $\varpi_\phi^{2}$ is
state-independent and drops out of the variation. The degeneracy of the vector flavor
metric and the removability \eqref{cmmg-is-ctmg} thus act on the charges together, each
blocking the deformation by its own route.

%--- a check against AdS--Carroll Chern--Simons gravity ----------------
One external check is available for the energy. The Chern--Simons formulation of
three-dimensional AdS--Carroll gravity in \cite{Aviles:2025ygw} identifies the global
charges of time translations and spatial rotations as the energy and the angular momentum.
Its Chern--Simons level is $k=1/(4G_{\rm C})$ and its vacuum carries $-k/2=-1/(8G_{\rm C})$.
Our $M=-1$ configuration is the Carrollian counterpart of global AdS space, and
\eqref{final-charges} gives $E_{\rm C}=-\sigma/(8G)$ there. The two agree for $\sigma=+1$
once the two Newton constants are identified. The agreement stops at the energy. At finite
$\mu$ the angular momentum in \eqref{final-charges} carries the parity-odd term
$M/(8G\mu)$, which that theory has no term to produce. In the other direction, the
relativistic contribution $\tilde{\mathcal{J}}/(\tilde{\mu}\ell^{2})$ to the energy is lost in
the contraction. The rotating configuration of \cite{Aviles:2025ygw} is not to be identified
with the $(M,\mathcal{J})$ family of Section~\ref{btzsub} either, since it is reached with
a different set of boundary conditions and a different charge parametrization.

%--- 6.16  the beta direction ------------------------------------------
The deformation \eqref{beta-shift} of Section~\ref{betadef} is invisible for a related
reason. Of the angular components it moves only $h_{1\phi}$ and $\varpi_{1\phi}$, and the flavor metric pairs
both of them with $\Lambda^{e_1}_{\xi}=\xi^{r}/N$, which vanishes for
$\xi_E=\partial_t$ and for $\xi_J=-\partial_{\phi}$. Neither charge in
\eqref{final-charges} depends on $\beta$. For a general $\xi$ the same two terms give
\begin{equation} \label{beta-charge}
\delta_{\beta}Q[\xi]=-\frac{q}{8\pi G\mu}\oint d\phi\;\frac{r}{N}\,\xi^{r}\,\delta\beta\,,
\end{equation}
evaluated as $r\to\infty$, where $r/N\to\ell$. By \eqref{beta-split} the deformation is an improper gauge transformation, so a non-zero
charge is what one should expect of it. It is detected only by a symmetry that moves the
radial coordinate. The residual transformations \eqref{residual} of Section~\ref{bms}
contain no such generator, so the question is one about the boundary conditions, and
Section~\ref{conclusions} returns to it.

%=======================================================================
%  SECTION 7 :  HAMILTONIAN ANALYSIS
%=======================================================================
\section{Hamiltonian analysis} \label{Dirac}

%--- what this section establishes -------------------------------------
This section puts \eqref{carrollian-mmg-lagrangian} in canonical form, using the flavor
data of Section~\ref{CCSlike}, and counts its local degrees of freedom. The outcome is
nine first-class constraints and no propagating mode, on every background rather than at
a single vacuum. The same count is then run for Carrollian TMG.

%--- index conventions used in this section ----------------------------
Several kinds of index occur below, and we collect them here. As in
Section~\ref{CCSlike}, $I,J=\tau,\bar{\tau},\omega$ label the one-form flavors that are
scalars under spatial rotations and $r,s=e,\varpi,h$ those that are spatial vectors, the
latter carrying an internal index $a,b=1,2$, written $i,j$ in Sections~\ref{adsbtz} and
\ref{third-way}; $R,S$ run over all nine flavors at once. Spacetime indices split as
$\mu=(0,K)$ with $K,L,M,N=1,2$ spatial. In the canonical analysis $\alpha,\beta$ denote
the composite label of a spatial field component together with its spatial leg, running
over the eighteen entries listed in \eqref{ordering-Phi};
$\mathtt{I},\mathtt{J}=1,\ldots,27$ labels the complete constraint set
$\Xi_{\mathtt{I}}=\left(\Phi_{\alpha},\mathcal{C}_R\right)$; and $A=1,\ldots,F$ labels
the first-class combinations \eqref{GammaA}.

%--- 7.1  spacetime split of the one-forms -----------------------------
The spacetime split of the components of the one-form fields of
Section~\ref{CCSlike} is
\begin{equation} \label{separate-everything}
A^{I}=A^{I}_0dt+A^{I}_Kdx^K\,, \qquad A^{ra}=A^{ra}_0dt+A^{ra}_Kdx^K\,,
\end{equation}
and likewise for each of $\tau$, $\bar{\tau}$, $\omega$, $e_i$, $h_i$ and $\varpi_i$
separately.
The spatial slice is left unspecified here; the angular coordinate of
Sections~\ref{CCSlike}--\ref{asympt} is reinstated only where the boundary circle
$\partial\Sigma$ is involved.
%--- 7.2  exterior derivatives -----------------------------------------
In the Lagrangian \eqref{carrollian-mmg-lagrangian} there are $d\tau$, $d\omega$, $de_i$
and $d\varpi_i$ terms; no $d\bar{\tau}$ or $dh_i$ occurs. The exterior derivatives of the
one-forms in \eqref{separate-everything} that are needed are
\begin{equation} \label{exterior-derivatives}
dX=\partial_LX_0\,dx^L\wedge dt+\dot{X}_K\,dt\wedge dx^K
  +\partial_LX_K\,dx^L\wedge dx^K\,, \qquad X\in\{\tau,\omega,e_i,\varpi_i\}\,,
\end{equation}
a dot denoting a derivative with respect to $t$ here and below.
%--- 7.3  the split Lagrangian -----------------------------------------
Substituting these into the Lagrangian \eqref{carrollian-mmg-lagrangian}, and integrating
by parts so that the temporal components appear undifferentiated, gives
\begin{align}  \label{separated-lagrangian}
    L=&\epsilon_{KL} \left(-\sigma\tau_K\dot{\omega}_L+\sigma e_{iK}\dot{\varpi}_{iL}+\gamma \bar{\tau}_K\dot{\tau}_L-\gamma h_{iK}\dot{e}_{iL}+\frac{1}{2\mu}\omega_K\dot{\omega}_L\right) \nn \\
    &+ \tau_0 \mathcal{C}_{\tau_0}+\bar{\tau}_0 \mathcal{C}_{\bar{\tau}_0}+\omega_0 \mathcal{C}_{\omega_0}+e_{i0}\mathcal{C}_{e_{i0}}
+h_{i0}\mathcal{C}_{h_{i0}}+\varpi_{i0}\mathcal{C}_{\varpi_{i0}} \nn \\
    &+ \epsilon_{KL}\partial_K \left(\gamma\bar{\tau}_L\tau_0-\sigma \tau_L\omega_0+\frac{1}{2\mu}\omega_L\omega_0-\gamma h_{iL}e_{i0}+\sigma e_{iL}\varpi_{i0}\right)\,,
    \end{align}
the last line being the boundary term that the integration by parts leaves behind.
%--- 7.4  temporal-field constraints -----------------------------------
The nine temporal field constraints are
\begin{align} \label{temporal-constraints}
\mathcal{C}_{\tau_0}:=&\frac{\partial L}{\partial \tau_0}=\epsilon_{KL}\left(-\gamma \partial_K\bar{\tau}_L+\sigma\partial_K\omega_L+ \frac{\Lambda_0}{2}\epsilon_{ij}e_{iK}e_{jL}\right)\approx 0\,, \nn \\
\mathcal{C}_{\bar{\tau}_0}:=&\frac{\partial L}{\partial \bar{\tau}_0}=\epsilon_{KL}\left(-\gamma\partial_K\tau_L+\gamma\epsilon_{ij}e_{iK}\varpi_{jL}+\alpha \epsilon_{ij}e_{iK}h_{jL} \right)\approx 0\,, \nn \\
\mathcal{C}_{\omega_0}:=&\frac{\partial L}{\partial \omega_0}=\epsilon_{KL}\left(\sigma\partial_K\tau_L-\frac{1}{\mu}\partial_K\omega_L-\sigma\epsilon_{ij}e_{iK}\varpi_{jL}+\gamma\epsilon_{ij}h_{iK}e_{jL}\right)\approx 0\,, \nn \\
\mathcal{C}_{e_{i0}}:=&\frac{\partial L}{\partial e_{i0}}=\epsilon_{KL}\Big(-\sigma\partial_K\varpi_{iL}+\gamma\partial_Kh_{iL} \nn \\
&\hphantom{\frac{\partial L}{\partial e_{i0}}=\epsilon_{KL}\Big(}
 -\epsilon_{ij}\left(\sigma\omega_L\varpi_{jK}+\gamma\bar{\tau}_K\varpi_{jL}-\gamma h_{jK}\omega_L+\Lambda_0\tau_Ke_{jL}+\alpha\bar{\tau}_Kh_{jL}\right)\Big)\approx 0\,, \nn \\
\mathcal{C}_{h_{i0}}:=&\frac{\partial L}{\partial h_{i0}}=\epsilon_{KL}\left(\gamma\partial_Ke_{iL}-\gamma\epsilon_{ij}\omega_Ke_{jL}-\alpha\epsilon_{ij}\bar{\tau}_Ke_{jL}\right)\approx 0\,, \nn \\
\mathcal{C}_{\varpi_{i0}}:=&\frac{\partial L}{\partial \varpi_{i0}}=\epsilon_{KL}\left(-\sigma\partial_Ke_{iL}-\sigma\epsilon_{ij}e_{jK}\omega_L-\gamma\epsilon_{ij}\bar{\tau}_Ke_{jL}\right) \approx 0\,.
\end{align}
The ordering of the temporal field constraints is taken to be
\begin{equation} \label{ordering-C}
\left(\mathcal{C}_{\tau_0},\mathcal{C}_{\bar{\tau}_0},\mathcal{C}_{\omega_0},\mathcal{C}_{e_{10}},\mathcal{C}_{e_{20}},\mathcal{C}_{h_{10}},\mathcal{C}_{h_{20}},\mathcal{C}_{\varpi_{10}},\mathcal{C}_{\varpi_{20}} \right)\,.
\end{equation}

%--- 7.6  Poisson brackets and conjugate momenta -----------------------
Since the flavor metric of Carrollian MMG is degenerate, the reduced brackets cannot be obtained by
inverting it as in \cite{Bergshoeff:2014pca}; the full Dirac procedure \cite{Dirac:1950pj} has to be followed
instead. The fundamental Poisson brackets are
\begin{align} \label{fundamental-PB}
    \{\tau_K(\Vec{x}),\Pi^{\tau}_L(\Vec{y}) \}&=\delta_{KL}\delta^2(\Vec{x}-\Vec{y})\,, & \{e_{iK}(\Vec{x}),\Pi^{e_j}_L(\Vec{y}) \}&=\delta_{KL}\delta_{ij}\delta^2(\Vec{x}-\Vec{y})\,, \nn \\
    \{\bar{\tau}_K(\Vec{x}),\Pi^{\bar{\tau}}_L(\Vec{y}) \}&=\delta_{KL}\delta^2(\Vec{x}-\Vec{y})\,, &  \{h_{iK}(\Vec{x}),\Pi^{h_j}_L(\Vec{y}) \}&=\delta_{KL}\delta_{ij}\delta^2(\Vec{x}-\Vec{y})\,, \nn \\
    \{\omega_K(\Vec{x}),\Pi^{\omega}_L(\Vec{y}) \}&=\delta_{KL}\delta^2(\Vec{x}-\Vec{y})\,, & \{\varpi_{iK}(\Vec{x}),\Pi^{\varpi_j}_L(\Vec{y})\}&=\delta_{KL}\delta_{ij}\delta^2(\Vec{x}-\Vec{y})\,,
\end{align}
where the conjugate momenta read off from \eqref{separated-lagrangian} are
\begin{align} \label{momenta}
    \Pi^{\tau}_L=&\frac{\partial L}{\partial \dot{\tau}_L}=\gamma \bar{\tau}_K\epsilon_{KL}\,, &
    \Pi^{e_i}_L=&\frac{\partial L}{\partial \dot{e}_{iL}}=-\gamma h_{iK}\epsilon_{KL}\,, \nn \\
    \Pi^{\omega}_L=&\frac{\partial L}{\partial \dot{\omega}_L}=\left(\frac{1}{2\mu}\omega_K-\sigma \tau_K\right)\epsilon_{KL}\,, &
    \Pi^{\varpi_i}_{L}=&\frac{\partial L}{\partial \dot{\varpi}_{iL}}=\sigma e_{iK}\epsilon_{KL}\,.
\end{align}
%--- 7.8  primary constraints ------------------------------------------
In \eqref{separated-lagrangian} the time derivatives of the fields
$\bar{\tau}$ and $h_i$ do not appear, so that $\Pi^{\bar{\tau}}_L\approx 0$ and
$\Pi^{h_i}_L \approx 0$ are primary constraints. The remaining conjugate momenta are
functions of the fields; therefore there are $18$ primary constraints
\begin{align} \label{primary-constraints}
    \Phi^{\tau}_L&=\Pi^{\tau}_L-\gamma \bar{\tau}_K\epsilon_{KL} \approx 0 \,, &
    \Phi^{e_i}_{L}&=\Pi^{e_i}_L+\gamma h_{iK}\epsilon_{KL} \approx 0 \,, \nn \\
\Phi^{\bar{\tau}}_L&=\Pi^{\bar{\tau}}_L \approx 0 \,, &
    \Phi^{h_i}_{L}&=\Pi^{h_i}_L \approx 0 \,, \nn \\
    \Phi^{\omega}_{L}&=\Pi^{\omega}_L+\left(\sigma \tau_K-\frac{1}{2\mu}\omega_K\right)\epsilon_{KL}\approx 0 \,, &
\Phi^{\varpi_i}_{L}&=\Pi^{\varpi_i}_{L}-\sigma e_{iK}\epsilon_{KL} \approx 0\,.
\end{align}

%--- 7.9  the primary-constraint matrix and its kernel -----------------
The Poisson bracket matrix $C_{\alpha\beta}$ is constructed from the primary constraints,
\begin{equation} \label{c}
    C_{\alpha\beta}\delta^2(\Vec{x}-\Vec{y}):=\left\{ \Phi_{\alpha}(\Vec{x}),\Phi_{\beta}(\Vec{y}) \right\}\,.
\end{equation}
Since each one-form field possesses two spatial components, the index $\alpha$ is
understood as a composite index that labels both the field type and the spatial component.
Explicitly, the ordering of the constraints is taken as
\begin{equation} \label{ordering-Phi}
\left(\tau_1,\tau_2,\bar{\tau}_1,\bar{\tau}_2,\omega_1,\omega_2,
e_{11},e_{12},e_{21},e_{22},h_{11},h_{12},h_{21},h_{22},
\varpi_{11},\varpi_{12},\varpi_{21},\varpi_{22}\right)\,.
\end{equation}
Only the momenta enter $\Phi_{\alpha}$ with a field-dependent partner, so the brackets
\eqref{c} are field independent. Writing the composite index as $\alpha=(I,K)$ in the
scalar sector and $\alpha=(r,a,K)$ in the vector sector, they collapse to
\begin{equation} \label{C-closed}
C_{(IK)(JL)}=-\epsilon_{KL}\,g_{IJ}\,, \qquad
C_{(raK)(sbL)}=-\epsilon_{KL}\,\delta_{ab}\,g_{rs}\,, \qquad
C_{(IK)(sbL)}=0\,,
\end{equation}
with $g_{IJ}$ and $g_{rs}$ the flavor metrics of \eqref{flavor-metrics}. Note that
\eqref{flavor-metrics} is printed in the basis $r=(e,\varpi,h)$, whereas
\eqref{ordering-Phi} orders the vector components as $(e,h,\varpi)$; the last two rows and
columns of $g_{rs}$ must be exchanged before \eqref{C-closed} is read as a matrix in that
ordering. Thus $C_{\alpha\beta}$ factorizes, the two sectors do not mix, and
\begin{equation} \label{rankC}
\mathrm{rank}\,C=2\,\mathrm{rank}\,g_{IJ}+4\,\mathrm{rank}\,g_{rs}=2\cdot 3+4\cdot 2=14\,,
\end{equation}
so that $\dim\ker C=18-14=4$. The kernel is therefore not an accident of the Hamiltonian
analysis: it is the single null direction $\gamma\,\varpi+\sigma\,h$ of $g_{rs}$, counted
once for each of the two internal directions $a=1,2$ and each of the two spatial components
$K=r,\phi$. The null vectors satisfy
\begin{equation} \label{v}
\mathcal{W}^{\alpha}_{(A)}C_{\alpha\beta}=0\,, \qquad A=1,\ldots,4\,,
\end{equation}
and, in the ordering \eqref{ordering-Phi}, are
\begin{align} \label{4solns}
\mathcal{W}_{(1)}&=\left(0,0,0,0,0,0,0,0,0,0,0,0,0,\tfrac{\sigma}{\gamma},0,0,0,1\right)^T\,, \nn \\
\mathcal{W}_{(2)}&=\left(0,0,0,0,0,0,0,0,0,0,0,0,\tfrac{\sigma}{\gamma},0,0,0,1,0\right)^T\,, \nn \\
\mathcal{W}_{(3)}&=\left(0,0,0,0,0,0,0,0,0,0,0,\tfrac{\sigma}{\gamma},0,0,0,1,0,0\right)^T\,, \nn \\
\mathcal{W}_{(4)}&=\left(0,0,0,0,0,0,0,0,0,0,\tfrac{\sigma}{\gamma},0,0,0,1,0,0,0\right)^T\,,
\end{align}
which are the null directions labeled by $(i,K)=(2,2),(2,1),(1,2),(1,1)$
respectively. Each of them pairs $h_{iK}$ with $\varpi_{iK}$ along
$\gamma\,\varpi_{iK}+\sigma\,h_{iK}$, in agreement with \eqref{C-closed}.

%--- 7.15  total Hamiltonian -------------------------------------------
The total Hamiltonian is obtained from \eqref{separated-lagrangian} by a Legendre
transformation, with the primary constraints added with undetermined multipliers
$\mathcal{U}^{\alpha}$ \cite{Henneaux:1992ig},
\begin{equation} \label{H-total}
    H_T=\int_{\Sigma}d^2x\left(\mathcal{H}_C+\mathcal{U}^{\alpha}\Phi_{\alpha}\right)\,,
\end{equation}
where, up to the total spatial derivative already isolated in
\eqref{separated-lagrangian}, the canonical density is
\begin{equation} \label{H-canonical}
    \mathcal{H}_C=-\left(\tau_0 \mathcal{C}_{\tau_0}+\bar{\tau}_0 \mathcal{C}_{\bar{\tau}_0}+\omega_0 \mathcal{C}_{\omega_0}+e_{i0}\mathcal{C}_{e_{i0}}+h_{i0}\mathcal{C}_{h_{i0}}+\varpi_{i0}\mathcal{C}_{\varpi_{i0}}\right)\,.
\end{equation}
The multiplier term never contributes to the evolution of the null combinations
\eqref{4solns}: by \eqref{v} one has
$\mathcal{U}^{\beta}\mathcal{W}^{\alpha}_{(A)}C_{\alpha\beta}=0$ identically, not weakly,
since $C_{\alpha\beta}$ is field independent.

%--- 7.17  smeared constraints -----------------------------------------
Some of the brackets needed below produce derivatives of the two-dimensional Dirac delta
function, and these are conveniently handled by smearing the constraints. A local
constraint is smeared with an arbitrary test function $\lambda^{\alpha}(x)$ over the
spatial hypersurface $\Sigma$,
\begin{equation} \label{smearing}
 \Phi[\lambda]=\int_{\Sigma} d^2x\, \lambda^{\alpha}(x) \Phi_{\alpha}(x)\,,
\end{equation}
so that the derivatives of the delta function may be moved by integration by parts,
\begin{equation} \label{IBP}
    \int_{\Sigma} d^2y\, f(y)\, \partial^y_K \delta^2(x-y)=-\partial_K f(x)\,,
\end{equation}
up to a boundary contribution. In the bulk constraint analysis the boundary term is assumed
to vanish under appropriate boundary conditions; it must, however, be retained when
boundary charges are considered. As will be seen at once, the consistency conditions of
\eqref{Theta} require none of this, because the brackets entering them are ultralocal;
\eqref{smearing} and \eqref{IBP} are needed only for the mixed brackets \eqref{M-block}
below and for the boundary charges of Section~\ref{asympt}.

%--- 7.19  consistency of the null combinations ------------------------
The four null directions \eqref{4solns} define
$\Theta_A:=\mathcal{W}^{\alpha}_{(A)}\Phi_{\alpha}$. Labeling them by the pair $(m,M)$
that \eqref{4solns} singles out, and using \eqref{primary-constraints},
\begin{equation} \label{Theta}
\Theta^{mM}=\frac{\sigma}{\gamma}\Phi^{h_m}_M+\Phi^{\varpi_m}_M
=\frac{\sigma}{\gamma}\Pi^{h_m}_M+\Pi^{\varpi_m}_M-\sigma e_{mK}\epsilon_{KM}\,.
\end{equation}
Their consistency conditions turn out to be simpler than one might expect. The only
temporal constraint containing $\partial_Nh_{mM}$ or $\partial_N\varpi_{mM}$ is
$\mathcal{C}_{e_{j0}}$, through
$\epsilon_{NL}\left(-\sigma\partial_N\varpi_{jL}+\gamma\partial_Nh_{jL}\right)$, and the
null ratio $\sigma:\gamma$ annihilates precisely this pair,
\begin{equation} \label{no-boundary}
\frac{\sigma}{\gamma}\frac{\partial\mathcal{C}_{e_{j0}}}{\partial(\partial_Nh_{mM})}
+\frac{\partial\mathcal{C}_{e_{j0}}}{\partial(\partial_N\varpi_{mM})}
=\frac{\sigma}{\gamma}\,\gamma\,\delta_{jm}\epsilon_{NM}-\sigma\,\delta_{jm}\epsilon_{NM}=0\,,
\end{equation}
since the derivative part of $\mathcal{C}_{e_{j0}}$ is built from the same $g_{rs}$ that
appears in \eqref{C-closed}. No $\partial^y_N\delta^2(\Vec{x}-\Vec{y})$ therefore survives
in the bracket of $\Theta^{mM}$ with $\mathcal{C}_R$: it is ultralocal, and
$\dot{\Theta}^{mM}=\{\Theta^{mM},H_T\}$ can be evaluated with the delta function alone,
with no integration by parts and no boundary term. Using \eqref{H-total},
\eqref{H-canonical} and \eqref{fundamental-PB},
\begin{equation} \label{chiA}
\chi^{mM}:=\dot{\Theta}^{mM}
=\frac{q}{\mu}\,\epsilon_{KM}\,\epsilon_{im}\left(\bar{\tau}_0\,e_{iK}-e_{i0}\,\bar{\tau}_K\right)\,,
\end{equation}
with $q$ as in \eqref{genericity}. Every contribution of
$\mathcal{C}_{\omega_0}$, and likewise the $\omega_L$ terms of $\mathcal{C}_{e_{j0}}$,
cancels for the same reason as in \eqref{no-boundary}. Because $\epsilon_{KM}\epsilon_{im}$
is an invertible map on index pairs, the requirement $\chi^{mM}\approx0$ for all $(m,M)$ is
equivalent to $\bar{\tau}_0\,e_{iK}=e_{i0}\,\bar{\tau}_K$. For an invertible spatial
zweibein the left-hand side has rank two unless $\bar{\tau}_0=0$, whereas the
right-hand side is an outer product and has rank at most one. Hence $\bar{\tau}_0=0$, and then either
$e_{i0}=0$ or $\bar{\tau}_K=0$.

This alternative is settled by an exact identity among the temporal constraints. Along the
same null direction $\gamma\,\varpi+\sigma\,h$ all derivative terms cancel once more,
leaving
\begin{equation} \label{bartau-zero}
\sigma\,\mathcal{C}_{h_{i0}}+\gamma\,\mathcal{C}_{\varpi_{i0}}
=-\frac{\gamma q}{\mu}\,\epsilon_{KL}\,\epsilon_{ij}\,\bar{\tau}_K\,e_{jL}\,.
\end{equation}
Both constraints on the left vanish weakly, so for $q\neq0$ and invertible $e_{jL}$ one
obtains the ultralocal constraint
\begin{equation} \label{bartauK}
\bar{\tau}_K\approx 0\,,
\end{equation}
which is the canonical counterpart of the covariant statement $\bar{\tau}=0$ derived
from \eqref{Si=0} under \eqref{genericity}. Once \eqref{bartauK} holds,
$\chi^{mM}\approx0$ reduces to
$\frac{q}{\mu}\epsilon_{KM}\epsilon_{im}\bar{\tau}_0e_{iK}=0$, which fixes the multiplier
$\bar{\tau}_0=0$ and produces nothing further, so the $\Theta^{mM}$ generate no secondary
constraints. Note, however, that \eqref{bartauK} and the primary constraint
$\Phi^{\bar{\tau}}_L=\Pi^{\bar{\tau}}_L\approx0$ obey
$\{\bar{\tau}_K(\Vec{x}),\Pi^{\bar{\tau}}_L(\Vec{y})\}=\delta_{KL}\delta^2(\Vec{x}-\Vec{y})\neq0$
and therefore form a second-class pair. Every rank computed in what follows must accordingly
be evaluated on the surface \eqref{bartauK}, and not at a generic point of phase space.

%--- 7.24  the full set of constraints ---------------------------------
Now let us denote the complete set of constraints by
$\Xi_{\mathtt{I}}=\left(\Phi_{\alpha},\mathcal{C}_R\right)$, $\mathtt{I}=1,\ldots,27$.
Three kinds of bracket occur. The primary block is the field-independent, ultralocal matrix
\eqref{C-closed}. The temporal constraints \eqref{temporal-constraints} contain no momenta,
so that $\left\{\mathcal{C}_R(\Vec{x}),\mathcal{C}_S(\Vec{y})\right\}=0$ exactly, and
not merely weakly. The mixed brackets are the only ones carrying a derivative
of the delta function: since $\mathcal{C}_S$ depends on the fields and on their first
spatial derivatives,
\begin{equation} \label{M-block}
\left\{\Phi_{\alpha}(\Vec{x}),\mathcal{C}_S(\Vec{y})\right\}
=-\frac{\partial\mathcal{C}_S}{\partial q^{\alpha}}\,\delta^2(\Vec{x}-\Vec{y})
-\frac{\partial\mathcal{C}_S}{\partial(\partial_Kq^{\alpha})}\,
 \partial^y_K\delta^2(\Vec{x}-\Vec{y})\,,
\end{equation}
where $q^{\alpha}$ denotes the spatial field components ordered as in \eqref{ordering-Phi}.

%--- 7.25  the first-class combinations --------------------------------
Next, let us write $F$ for the number of first-class constraints among the twenty-seven
$\Xi_{\mathtt{I}}$ and denote the corresponding combinations by
\begin{equation} \label{GammaA}
\Gamma_A=u^{\alpha}_A\Phi_{\alpha}+v^R_A\,\mathcal{C}_R\,, \qquad
V^{\mathtt{I}}_A=\left(u^{\alpha}_A,v^R_A\right)\,, \qquad A=1,\ldots,F\,,
\end{equation}
so that the $V^{\mathtt{I}}_A$ are the null vectors of the matrix of all brackets among the
$\Xi_{\mathtt{I}}$.

%--- 7.26  counting the degrees of freedom -----------------------------
For a Chern--Simons-like model with $N$ one-form flavors in three dimensions there are $6N$
spatial field components, hence a $12N$-dimensional phase space, together with $6N$ primary
and $3N$ temporal constraints. If $F$ of these $9N$ constraints are first class and the
remaining $9N-F$ second class, the Dirac formula gives
\begin{equation} \label{dof-general}
\text{DOF}=\frac{1}{2}\Big(12N-2F-\left(9N-F\right)\Big)=\frac{1}{2}\left(3N-F\right)\,,
\end{equation}
so that the whole counting problem reduces to the single integer $F$, provided the Dirac
algorithm terminates at the $9N$ constraints just listed. For Einstein--Cartan gravity
($N=2$) it does. All six temporal constraints are first class, and \eqref{dof-general}
returns $F=6$ with $\text{DOF}=0$.

For relativistic TMG and MMG ($N=3$) it does not. There $g_{rs}$ is invertible, and the
algorithm produces one further constraint beyond the $9N$, so that the set numbers $28$
rather than $27$. The additional constraint is $\Delta^{ek}$ for TMG
\cite{Hohm:2012vh} and $\Delta^{eh}$ for MMG \cite{Bergshoeff:2014pca}, in the notation
of those references. With it one has $F=6$, hence $22$
second-class constraints, and
\begin{equation} \label{dof-TMG}
\text{DOF}=\frac{1}{2}\Big(36-2\cdot 6-22\Big)=1\,,
\end{equation}
the single massive graviton. Equation \eqref{dof-general} therefore does not apply to
those two models. The parity of $3N$ is the quickest way to see that it cannot: $3N=9$ is
odd, so \eqref{dof-general} would force $F$ to be odd, whereas the correct value is $6$.
What distinguishes the Carrollian model is established in Section~\ref{Fcount}. There the
algorithm does terminate at the $9N$ constraints \eqref{primary-constraints} and
\eqref{temporal-constraints}, so \eqref{dof-general} does apply.

The integer $F$ is determined in Section~\ref{Fcount} by an argument that evaluates no rank
and selects no background. It is confirmed in Section~\ref{rankcheck} by a direct
computation on the one background where the bracket matrix can be written down in full.

%=======================================================================
\subsection{The number of first-class constraints}\label{Fcount}
%=======================================================================

%--- 7.28  the strategy : bound F from both sides ----------------------
The integer $F$ of \eqref{dof-general} can be pinned without computing a rank at all,
and therefore without committing to a background. Since $\text{DOF}\geq0$, \eqref{dof-general} gives the upper
bound
\begin{equation} \label{Fbound}
F\leq 3N=9\,,
\end{equation}
so it is enough to exhibit nine independent first-class constraints. Six of them come from
the local symmetries of \eqref{carrollian-mmg-lagrangian} and three from the null directions
\eqref{4solns}. Neither count requires choosing a solution of \eqref{carroll-eom}, so the
argument applies on every background at once.

%--- the local symmetries of the Carrollian Lagrangian -----------------
Look for transformations of the Chern--Simons-like form
$\delta A^R_{\mu}=\delta^R_S\,\partial_{\mu}\lambda+c^R{}_{T}A^T_{\mu}\lambda$, with
constant $c^R{}_T$, that leave \eqref{carrollian-mmg-lagrangian} invariant up to a total
derivative. Exactly four of the nine flavor directions $S$ admit one, namely
$S=\tau,\omega,\varpi_1,\varpi_2$; the directions $\bar{\tau}$, $e_i$ and $h_i$ admit none.
The four are the rotation $\lambda$, the two boosts
$\kappa_i$ and the time translation $\varsigma$ of \eqref{rot-boost}. None of the three
involves $\Lambda_0$ except $\varsigma$, which exists for every value of it, and
$\varsigma$ is what makes $\Lambda^{\tau}$ a legitimate gauge parameter in
Section~\ref{asympt}. Diffeomorphisms supply three further
local symmetries, realized by $\Lambda^R_{\xi}=\xi^{\mu}A^R_{\mu}$ as in \eqref{gauge-parameters-xi}.

%--- 7.29  the span of the admissible gauge parameters -----------------
A local symmetry of the action yields one first-class combination \eqref{GammaA} whose
temporal part $v^R$ is the gauge parameter itself. This correspondence is all that is
needed here. It is the Castellani construction \eqref{G-eps}, carried out in
Section~\ref{asympt}. The number of first-class constraints with $v^R\neq0$ is therefore the dimension of
the span of the admissible $\Lambda^R$. Writing $\hat{X}$ for the unit vector in the flavor
direction $X$, the four transformations of \eqref{rot-boost} contribute
$\hat{\tau},\hat{\omega},\hat{\varpi}_1,\hat{\varpi}_2$ and the diffeomorphisms contribute
$\xi^{\mu}\mathring{A}^R_{\mu}$, so that
\begin{equation} \label{span}
\begin{aligned}
\text{flat vacuum:}&& \Lambda^R&\in\text{span}\left\{\hat{\tau},\,\hat{\omega},\,
 \hat{\varpi}_1,\,\hat{\varpi}_2,\,\hat{e}_1,\,\hat{e}_2\right\}\,, \\
\text{AdS vacuum:}&& \Lambda^R&\in\text{span}\left\{\hat{\tau},\,\hat{\omega},\,
 \hat{\varpi}_1,\,\hat{\varpi}_2,\,\hat{e}_1+c\,\hat{h}_1,\,
 \hat{e}_2+c\,\hat{h}_2\right\}\,,
\end{aligned}
\end{equation}
both six-dimensional. Seven parameters produce only six independent directions because the
time diffeomorphism repeats the direction $\hat{\tau}$ already supplied by $\varsigma$; the
two differ by a generator built purely from primary constraints. On the AdS
background $\mathring{\bar{\tau}}=0$ and $\mathring{h}_i=c\,\mathring{e}_i$, whence
$\Lambda^{\bar{\tau}}_{\xi}=0$ and $\Lambda^{h_i}_{\xi}=c\,\Lambda^{e_i}_{\xi}$: the
admissible set is rotated inside flavor space but its dimension is unchanged. The same
applies to the Carrollian BTZ solution, for which \eqref{BTZ-soln-of-h} again gives
$h_i=c\,e_i$ with $\bar{\tau}=0$. The first line of \eqref{span} is precisely
\eqref{Lambda-allowed} of Section~\ref{asympt}.

%--- the three surviving null directions -------------------------------
The remaining three first-class constraints are purely primary, and every step that
produced them is independent of the background and of $\Lambda_0$. The matrix
\eqref{C-closed} is field independent, so $\dim\ker C=4$ and the four combinations
$\Theta^{mM}$ of \eqref{Theta} exist on any background. Of the temporal constraints
\eqref{temporal-constraints}, only $\mathcal{C}_{\tau_0}$ and $\mathcal{C}_{e_{i0}}$ contain
$\Lambda_0$ at all. The identity \eqref{bartau-zero}, and with it $\bar{\tau}_K\approx0$,
therefore holds for every $\Lambda_0$. The quantity $\chi^{mM}$ of \eqref{chiA} is
$\Lambda_0$-independent as well. On the surface \eqref{bartauK} the requirement
$\chi^{mM}\approx0$ again reduces to fixing the single multiplier $\bar{\tau}_0=0$. One of the four $\Theta^{mM}$ is
thereby second class, paired with the constraint whose multiplier it fixes, and three
remain first class.

%--- the same three constraints, covariantly ---------------------------
These three are the shift \eqref{null-shift} of Section~\ref{mmgsec}, seen canonically.
There the parameter is a traceless $S_{ij}(x)$, it has three components, and it enters
undifferentiated, so the shift is a local symmetry that generates no secondary constraint.
Here the same three appear as the three purely primary first-class constraints. The two
routes agree on the number, on the direction $\gamma\varpi+\sigma h$ of \eqref{4solns},
and on the reduction from four to three, which is \eqref{traceless} there and the pairing
of one $\Theta^{mM}$ with $\bar{\tau}_0$ here. Neither route uses any input from the
other.

%--- 7.30  the count on any background ---------------------------------
Six first-class constraints with $v^R\neq0$ and three purely primary ones are manifestly
independent, so that $F\geq9$; together with \eqref{Fbound},
\begin{equation} \label{F-any}
\boxed{\;
F=9\,, \qquad \text{DOF}=\frac{1}{2}\left(3N-F\right)=0
\;}
\end{equation}
on every background of Carrollian MMG with $q\neq0$ and an invertible spatial zweibein: the
flat and AdS vacua and the Carrollian BTZ solution alike. The nine split as
$9=6+3$, into six generators carrying boundary charge and three purely primary ones, a
decomposition recovered in Section~\ref{rankcheck} by the one computation that reaches $F$
through a rank.

Nothing in the argument requires $\alpha\neq0$. At $\tilde{\alpha}=0$ the term
$\frac{\tilde{\alpha}}{2}\epsilon^{abc}\mathtt{e}_a\mathtt{h}_b\mathtt{h}_c$ is absent from
\eqref{mmg-lagrangian}, $\mathtt{h}^a$ becomes the multiplier that imposes vanishing
torsion, and the relativistic theory is TMG. Its Carroll contraction is
\eqref{carrollian-mmg-lagrangian} at $\alpha=0$, for which $q=\mu\gamma\neq0$, so that
\eqref{genericity} still holds and \eqref{F-any} applies unchanged. Relativistic TMG
propagates one massive graviton and Carrollian TMG none. The contraction removes the
propagating mode of TMG exactly as it removes that of MMG, and by the same mechanism in
both: it closes the Dirac algorithm at the $9N$ constraints and raises $F$ from $6$ to
$3N$. That the two counts agree is in fact forced, since by \eqref{cmmg-is-ctmg} the two
contracted theories differ by a change of variables and a total derivative, neither of
which the Dirac algorithm can see. Before the contraction there is no such identification.
Relativistic TMG and relativistic MMG both propagate one mode, but they do so as distinct
theories with the same constraint structure.

%--- 7.31  where the degeneracy of g_rs enters -------------------------
It is worth recording where the degeneracy of $g_{rs}$ enters the argument. The
four $\Theta^{mM}$ are an immediate consequence of that degeneracy, and so is
\eqref{Si=0}. The combination of \eqref{eom-varpi} and \eqref{eom-h} that cancels
$\mathcal{R}_i(e)$ is the one along the null direction $\gamma\,\varpi+\sigma\,h$ of
$g_{rs}$. In the normalization of \eqref{carroll-eom} it reads
\begin{equation} \label{null-contraction}
\gamma\left[\delta\varpi_i\right]-\sigma\left[\delta h_i\right]
 =\left(\gamma^{2}+\sigma\alpha\right)\bar{S}_i=\frac{\gamma q}{\mu}\,\bar{S}_i\,,
\end{equation}
where $\left[\delta X\right]$ denotes the right-hand side of the corresponding member of
\eqref{carroll-eom} and $\bar{S}_i$ is the two-form of \eqref{carroll-defs}. Every derivative of the fields has canceled, because in
\eqref{NLCSL-wedge} the derivative terms of the vector sector are carried by $g_{rs}$ alone.
This is the covariant origin of the canonical identity \eqref{bartau-zero}, and the reason
both are free of $\Lambda_0$: the couplings $\Lambda_0$ and $\alpha$ enter
\eqref{carrollian-mmg-lagrangian} only through $f_{rsI}$, never through $g_{rs}$.

%--- 7.32  cross-check from the linearized field equations -------------
The same conclusion can be read off the linearized field equations. Since
\eqref{carroll-eom} is at most quadratic in the fields, its linearization carries no
derivative of the background, and \eqref{null-contraction} gives
$\epsilon_{ij}\,\delta\bar{\tau}\,\mathring{e}_j=0$, hence $\delta\bar{\tau}=0$
algebraically by the argument that follows \eqref{Si=0}. The linearized torsion condition
$\delta\mathcal{R}_i(e)=0$ then determines $\delta\omega$ and constrains $\delta e_i$,
while the $\delta\tau$ and $\delta\omega$ equations together leave only
\begin{equation} \label{lin-leftover}
\epsilon_{ij}\,\mathring{e}_i\left(\delta h_j-c\,\delta e_j\right)=0\,.
\end{equation}
No equation anywhere in this chain is of second order in derivatives, in accordance with
\eqref{F-any}.

%--- what the count says -----------------------------------------------
The census behind \eqref{F-any} is short. A thirty-six-dimensional phase space carries the
eighteen primary constraints \eqref{primary-constraints}, whose bracket matrix
\eqref{C-closed} has rank fourteen and hence the four null directions \eqref{4solns},
together with the nine temporal constraints \eqref{temporal-constraints}. Of the
twenty-seven, nine are first class and eighteen second class.

The value $F=9=3N$ is the largest possible for three flavors. It is the pattern of
Einstein--Cartan gravity, where $F=3N$ with $N=2$. Relativistic MMG sits instead at $F=6$,
with one constraint more than the $9N$ and one propagating mode, so the contraction
restores the maximal gauge symmetry. The three directions by which $F=9$ exceeds that
$F=6$ are exactly the purely primary ones, which is to say the shift \eqref{null-shift}
that the degeneracy of $g_{rs}$ supplies. The assumption $q\neq0$ is essential throughout.
At $q=0$ the derivation of $\bar{\tau}_K\approx0$ in \eqref{bartau-zero} fails, the Dirac
algorithm no longer terminates at this level, and \eqref{dof-general} does not apply.
Section~\ref{rankcheck} records what the count returns if one ignores \eqref{genericity}.

%=======================================================================
\subsection{The rank at the flat vacuum}\label{rankcheck}
%=======================================================================

Section~\ref{Fcount} reached $F$ without computing a rank. On one background the rank can be
computed outright, by a route sharing no step with that argument.

%--- 7.33  the symbol matrix -------------------------------------------
Smear the constraints as in \eqref{smearing} with plane waves
$\lambda^{\mathtt{I}}(x)=\hat{\lambda}^{\mathtt{I}}e^{ik_Lx^L}$, which presupposes
Cartesian coordinates $x^K$ on the slice. With \eqref{IBP}, write
$\Xi[\lambda]=\int_{\Sigma}d^2x\,\lambda^{\mathtt{I}}\Xi_{\mathtt{I}}$. The requirement that it
have weakly vanishing brackets with every constraint becomes
$\hat{\lambda}^{\mathtt{I}}\mathbb{C}_{\mathtt{I}\mathtt{J}}(k)\approx0$, where
\begin{equation} \label{mathbbC}
\mathbb{C}_{\mathtt{I}\mathtt{J}}(k)=
\begin{pmatrix}
C_{\alpha\beta} & N_{\alpha S}(k) \\
P_{R\beta}(k) & 0_{9\times9}
\end{pmatrix}\,.
\end{equation}
The off-diagonal blocks are built from $\partial\mathcal{C}_S/\partial q^{\alpha}$ and
$\partial\mathcal{C}_S/\partial(\partial_Kq^{\alpha})$, and they carry opposite signs of
$ik_K$, because the derivative of the delta function in \eqref{M-block} acts on the
smearing function on one side of the bracket and on the constraint on the other.
Equivalently $\mathbb{C}(k)^{T}=-\mathbb{C}(-k)$. The $k_K$ stand for differential
operators and are not physical momenta. Then $F=\dim\ker\mathbb{C}(k)$ at generic $k$, and
the vectors $V^{\mathtt{I}}_A$ of \eqref{GammaA} span $\ker\mathbb{C}$.

%--- 7.34  the flat Carrollian vacuum ----------------------------------
The rank is safest to extract on a background where $\mathbb{C}$ is field independent. For
$\Lambda_0=0$ the configuration
\begin{equation} \label{flat-vacuum}
\mathring{\tau}=dt\,, \qquad \mathring{e}_{i}=\delta_{iK}\,dx^{K}\,, \qquad
\mathring{\bar{\tau}}=\mathring{\omega}=\mathring{\varpi}_i=\mathring{h}_i=0
\end{equation}
solves \eqref{carroll-eom} identically and is the flat Carrollian vacuum. It is constant,
so the second-order expansion of \eqref{carrollian-mmg-lagrangian} about it has constant
coefficients and every entry of \eqref{mathbbC} is known exactly, with no generic-point
ambiguity and no weak equality to interpret. One finds
\begin{equation} \label{rank-full}
\mathrm{rank}\,\mathbb{C}=18\,, \qquad F=\dim\ker\mathbb{C}=27-18=9\,,
\end{equation}
independently of $k_K$ and of $\sigma,\gamma,\alpha,\mu$, provided $q\neq0$. Then
\eqref{dof-general} gives $\tfrac{1}{2}(36-18-18)=0$ degrees of freedom, in agreement with
\eqref{F-any}, and the null vectors of \eqref{mathbbC} split the nine the same way as
Section~\ref{Fcount}: six carry $v^R_A\neq0$ and reproduce the first line of
\eqref{span}, while three are purely primary. The identical procedure applied to
Einstein--Cartan gravity returns $F=6$ and $\text{DOF}=0$, that being the one relativistic
model whose $9N$ constraints are the complete set.

The rank also shows what the genericity assumption \eqref{genericity} is doing. Drop it
and the count fails, in two different ways. At $\gamma^{2}+\sigma\alpha=0$ the rank falls
from $18$ to $14$, so \eqref{dof-general} returns $F=13$ and $\text{DOF}=-2$. A negative
number of degrees of freedom is not a result. It says that the surface being used is not
the constraint surface. At $q=0$ the failure comes earlier. The identity
\eqref{bartau-zero} no longer forces $\bar{\tau}_K\approx0$, so the constraints of
Section~\ref{Dirac} are not the full set and the algorithm has more to generate. The
gauge symmetry grows there as well, since the obstruction in \eqref{null-shift-boundary}
is proportional to $q$ and the shift \eqref{null-shift} gains a fourth parameter.

%--- why the rank cannot be re-evaluated elsewhere ---------------------
One limitation belongs with \eqref{rank-full}: the flat vacuum is the only background on
which it can be evaluated. Every entry of \eqref{mathbbC} is built from the spatial
components $A^R_K$, so a constant configuration is admissible only if it already satisfies
the temporal constraints \eqref{temporal-constraints}. For $\mathcal{C}_{\tau_0}$ the
derivative terms then drop out, and what is left is $\Lambda_0\det\mathring{e}_{iK}=0$.
Nothing can restore the balance, because the Carrollian rotation curvature
$\mathcal{R}(\omega)=d\omega$ is abelian and has no term quadratic in the connection,
unlike its relativistic ancestor in \eqref{decomposed}. With an invertible zweibein this
forces $\Lambda_0=0$, so the AdS branch of Section~\ref{adsbtz} has no constant
representative. The obstruction is to this method alone. The count on that branch is
settled by \eqref{F-any}.

The absence of local degrees of freedom makes Carrollian MMG a theory of boundary data, in
the same sense as three-dimensional Einstein gravity. Its entire physical content is
carried by the black hole charges \eqref{final-charges} and the asymptotic charges of the
next section. Third-way consistency is untouched by that conclusion, since the obstruction
identified in Section~\ref{mmgalg} is an off-shell failure of the Jacobi identity and is
unrelated to the number of propagating modes.

%=======================================================================
%  SECTION 8 :  ASYMPTOTIC CHARGES FROM THE CONSTRAINT GENERATOR
%=======================================================================
\section{Asymptotic charges from the constraint generator} \label{asympt}

%--- 8.1  the relativistic formula and why it fails --------------------
In \cite{Bergshoeff:2019rdb} the asymptotic charges of a generic Chern--Simons-like model
are calculated from
\begin{equation} \label{rel-asymptotic}
    (D\Lambda)^R=d\Lambda^R+f^R{}_{ST}A^S \times \Lambda^T=\delta_{\Lambda}A^R\,,
\end{equation}
where $\Lambda^R$ is the gauge parameter, $\delta_{\Lambda}A^R \rightarrow 0$ only at the
asymptotic boundary $r \rightarrow \infty$, and the indices of the totally symmetric tensor
$f_{RST}$ are raised with the inverse of the flavor metric. In the Carrollian limit the flavor metric
\eqref{flavor-metrics} has no inverse, so \eqref{rel-asymptotic} is unavailable.
The Hamiltonian analysis of Section~\ref{Dirac} supplies the replacement: the first-class
constraints \eqref{GammaA} generate the gauge symmetries, and their boundary terms are the
charges.

%--- 8.2  the Castellani generator -------------------------------------
Following \cite{Castellani:1981us}, the generator built from the first-class constraints is
\begin{equation} \label{G-eps}
G[\varepsilon]=\frac{1}{8\pi G}\int_{\Sigma}d^2x\;\varepsilon^A V^{\mathtt{I}}_A\Xi_{\mathtt{I}}
 =\frac{1}{8\pi G}\int_{\Sigma}d^2x\;\varepsilon^A\left(u^{\alpha}_A\Phi_{\alpha}
  +v^R_A\,\mathcal{C}_R\right)\,, \qquad A=1,\ldots,9\,,
\end{equation}
there being $F=9$ independent generators by \eqref{F-any}. It is convenient to collect
the parameters into one gauge parameter per temporal constraint,
\begin{equation} \label{Lambda-def}
\Lambda^R:=\varepsilon^A v^R_A\,,
\end{equation}
ordered as in \eqref{ordering-C}.

%--- 8.4  the boundary term --------------------------------------------
Varying \eqref{G-eps} and using $\Phi_{\alpha}\approx0$, $\mathcal{C}_R\approx0$ on the
constraint surface gives
\begin{equation} \label{delta-G}
\delta G[\varepsilon]=\frac{1}{8\pi G}\int_{\Sigma}d^2x\;\varepsilon^A
 \left(u^{\alpha}_A\,\delta\Phi_{\alpha}+v^R_A\,\delta\mathcal{C}_R\right)\,.
\end{equation}
Following the Regge--Teitelboim prescription for differentiable canonical generators
\cite{Regge:1974zd,Banados:1994tn} this splits into
\begin{equation} \label{delta-G-split}
    \delta G[\varepsilon]=\text{(bulk terms)}+\mathcal{B}[\varepsilon,A]\,,
\end{equation}
and a boundary term satisfying $\delta Q[\varepsilon]=-\mathcal{B}[\varepsilon,A]$ is added to
$G[\varepsilon]$ to make it differentiable; whenever $\delta Q$ is integrable there is an
associated conserved charge. The primary constraints \eqref{primary-constraints} contain no
spatial derivatives, so $\delta\Phi_{\alpha}$ contributes to the bulk only and the
$u^{\alpha}_A$ drop out of the charge altogether. The entire boundary contribution comes
from the derivative terms of \eqref{temporal-constraints}, each of which occurs in the
combination $\epsilon_{KL}\partial_KX_L$, whose integral over $\Sigma$ reduces to
$\oint d\phi\,X_{\phi}$. Hence
\begin{align} \label{master-charge}
\delta Q[\Lambda]=-\frac{1}{8\pi G}\oint d\phi\,\Big[&
 \Lambda^{\tau}\left(-\gamma\,\delta\bar{\tau}_{\phi}+\sigma\,\delta\omega_{\phi}\right)
 -\gamma\,\Lambda^{\bar{\tau}}\,\delta\tau_{\phi}
 +\Lambda^{\omega}\left(\sigma\,\delta\tau_{\phi}-\frac{1}{\mu}\,\delta\omega_{\phi}\right)
 \nn \\
 &+\Lambda^{e_i}\left(-\sigma\,\delta\varpi_{i\phi}+\gamma\,\delta h_{i\phi}\right)
 +\left(\gamma\,\Lambda^{h_i}-\sigma\,\Lambda^{\varpi_i}\right)\delta e_{i\phi}\Big]\,.
\end{align}
This is precisely the canonical boundary charge \eqref{NLCSL-charge} of
Section~\ref{CCSlike}, with the gauge parameters there identified with
\eqref{Lambda-def}. The coincidence is not accidental: \eqref{NLCSL-charge} follows from
varying the Hamiltonian and \eqref{master-charge} from varying the first-class generator,
so the Castellani construction is what justifies \eqref{NLCSL-charge} for a theory whose
flavor metric cannot be inverted and for which \eqref{rel-asymptotic} does not apply.

%--- 8.7  which generators carry charge --------------------------------
Not every $\Lambda^R$ is realized. The null vectors $V^{\mathtt{I}}_A$ must satisfy both
blocks of $\mathbb{C}_{\mathtt{I}\mathtt{J}}$. Evaluating them on the flat Carrollian
vacuum \eqref{flat-vacuum} shows that three of the nine generators are purely primary,
$v^R_A=0$. Those three carry no boundary charge at all. The remaining six realize
\begin{equation} \label{Lambda-allowed}
\Lambda^{\bar{\tau}}=0\,, \qquad \Lambda^{h_i}=0\,, \qquad
\Lambda^{\tau},\;\Lambda^{\omega},\;\Lambda^{e_i},\;\Lambda^{\varpi_i}\ \text{ free}\,.
\end{equation}
Equation \eqref{Lambda-allowed} is the form taken on the flat vacuum. By
Section~\ref{Fcount}, the count $F=9$ holds on every background of the theory. On a
background with $\mathring{\bar{\tau}}=0$ and $\mathring{h}_i=c\,\mathring{e}_i$, such as the
AdS vacuum or the Carrollian BTZ solution, the admissible parameters are instead
$\Lambda^{\bar{\tau}}=0$ and $\Lambda^{h_i}=c\,\Lambda^{e_i}$, by the second line of
\eqref{span}. The extra terms this contributes to \eqref{master-charge} are proportional to
$\delta e_{i\phi}$ and $\delta h_{i\phi}$, both of which vanish on the Carrollian BTZ
solution by \eqref{vanishing-BTZ-angular-variations}, so neither \eqref{BTZ-master} nor the
charges that follow from it are affected.

With \eqref{Lambda-allowed} the charge \eqref{master-charge} reduces to
\begin{align} \label{master-charge-reduced}
\delta Q[\Lambda]=-\frac{1}{8\pi G}\oint d\phi\Big[&
 \sigma\Lambda^{\tau}\,\delta\omega_{\phi}
 +\sigma\Lambda^{\omega}\,\delta\tau_{\phi}
 -\frac{1}{\mu}\Lambda^{\omega}\,\delta\omega_{\phi} \nn \\
 &-\sigma\Lambda^{e_i}\,\delta\varpi_{i\phi}
 +\gamma\Lambda^{e_i}\,\delta h_{i\phi}
 -\sigma\Lambda^{\varpi_i}\,\delta e_{i\phi}\Big]\,,
\end{align}
where $\delta\bar{\tau}_{\phi}=0$ has been used, as it holds on any solution by
\eqref{bartauK}. The $\delta\tau_{\phi}$ and $\delta h_{i\phi}$ terms survive this
reduction in general. They drop on the Carrollian BTZ solution, by
\eqref{vanishing-BTZ-angular-variations} of Section~\ref{bdrycharges}.

%--- 8.9  the charges of the Carrollian BTZ solution -------------------
On the Carrollian BTZ solution \eqref{BTZ-soln-of-vielbein},
\eqref{BTZ-solution-of-omega}, \eqref{BTZ-soln-of-h} the only non-vanishing angular
variations are those of \eqref{angular-variations}, all others vanishing by
\eqref{vanishing-BTZ-angular-variations}. In particular $\delta e_{i\phi}=0$ removes
$\Lambda^{\varpi_i}$ and $\delta\varpi_{1\phi}=0$ removes $\Lambda^{e_1}$, so that
\begin{equation} \label{BTZ-master}
\delta Q[\Lambda]=-\frac{1}{8\pi G}\oint d\phi\left[
 \left(\sigma\Lambda^{\tau}-\frac{1}{\mu}\Lambda^{\omega}\right)\delta N
 -\sigma\,\Lambda^{e_2}\,\delta\varpi_{2\phi}\right]\,.
\end{equation}
Exactly three of the nine asymptotic generators therefore act non-trivially on this
background. Since the Carrollian BTZ family is labeled by the two parameters $M$ and
$\mathcal{J}$,
and $\delta\omega_{\phi}=\delta N$ and $\delta\varpi_{2\phi}=-\delta\mathcal{J}/(2r)$ are the
only independent variations, \eqref{BTZ-master} yields two independent charges, and these
are the ones already obtained from \eqref{NLCSL-charge} and from
\eqref{master-charge}. For a time translation the
parameters are those of \eqref{energy-parameters}, that
is $\Lambda^{\tau}=N$ with $\Lambda^{\omega}=\Lambda^{e_2}=0$, and \eqref{BTZ-master} gives
\begin{equation} \label{E-from-master}
\delta E_{\rm C}=-\frac{1}{8\pi G}\int_0^{2\pi}\!\!d\phi\;\sigma N\delta N
 =\frac{\sigma}{8G}\,\delta M\,,
\end{equation}
which is \eqref{CMMG-Carroll-energy}. For a rotation $\Lambda^{\omega}=-N$ and
$\Lambda^{e_2}=-r$ with $\Lambda^{\tau}=0$, and
\begin{equation} \label{J-from-master}
\delta J_{\rm C}=-\frac{1}{8\pi G}\int_0^{2\pi}\!\!d\phi
 \left(-\frac{\delta M}{2\mu}-\frac{\sigma}{2}\,\delta\mathcal{J}\right)
 =\frac{1}{8G}\left(\sigma\,\delta\mathcal{J}+\frac{\delta M}{\mu}\right)\,,
\end{equation}
which is \eqref{CMMG-angular-momentum}. Both variations are integrable, and integrating
them with the reference configuration $M=\mathcal{J}=0$ reproduces \eqref{final-charges}.

%--- remarks -----------------------------------------------------------
This construction has two features worth drawing out. First, the charge depends on the gauge
parameters only through $\Lambda^R=\varepsilon^Av^R_A$, that is only through the temporal
constraints: the primary constraints, and with them the four null directions
\eqref{4solns} produced by the degeneracy of $g_{rs}$, are invisible at the boundary. The
degeneracy that dominates the bulk analysis of Section~\ref{Dirac} therefore leaves no
trace in the charges. Second, Carrollian MMG carries no local degrees of freedom by \eqref{F-any}. Hence
\eqref{master-charge} is not one part of the physical content of the theory but the whole
of it. The independence of \eqref{final-charges} from $\alpha$, $\gamma$ and $c$ is
accordingly a statement about the complete dynamics, not about a single sector of it.

%--- the two computations compared -------------------------------------
With this, \eqref{final-charges} has been reached twice over, and the two routes are
independent in a strong sense. Section~\ref{bdrycharges} varies the Hamiltonian of the
Chern--Simons-like form and contracts the result with gauge parameters
$\Lambda^R_{\xi}=\xi^{\mu}A^R_{\mu}$ read off from the Killing vectors. The present section
uses no Killing vector at all, only the first-class constraints and the
Regge--Teitelboim requirement that \eqref{G-eps} be differentiable. Beyond the flavor
metric \eqref{flavor-metrics} and the solution itself, the two share nothing.

The agreement therefore settles more than the arithmetic. The status of
\eqref{NLCSL-charge} for a theory whose flavor metric cannot be inverted is an assumption
in Section~\ref{bdrycharges} and a consequence of the Castellani construction here, as
noted after \eqref{master-charge}. And $\alpha$ leaves the answer by different doors:
through $g_{\varpi\varpi}=0$ and the state independence of the shift in
$\varpi_{\phi}^{2}$ in Section~\ref{bdrycharges}, and through the vanishing of the
boundary term that $\Lambda^{\varpi_i}$ would have to supply here. Two unrelated mechanisms give the same
answer. That is why we read the blindness of \eqref{final-charges} to the MMG deformation
as a property of the theory and not of a way of computing. Section~\ref{bms}
carries the same conclusion to the full asymptotic symmetry algebra, and
Section~\ref{thermo} to the entropy.

%=======================================================================
%  SECTION 9 :  THE ASYMPTOTIC SYMMETRY ALGEBRA
%=======================================================================
\section{The asymptotic symmetry algebra}\label{bms}

%--- 9.1  the on-shell connection --------------------------------------
On the AdS branch of Section~\ref{adsbtz} the whole field content collapses
on shell to a single flat connection. There $\bar{\tau}=0$ and $h_i=c\,e_i$, and
it is convenient to trade the boost connection for its torsion-free counterpart
\eqref{omega-from-Omega},
\begin{equation} \label{Omega-shift}
\Omega_i=\varpi_i+\frac{c\alpha}{\gamma}e_i\,.
\end{equation}
The six equations \eqref{carroll-eom} then reduce to
\begin{subequations} \label{flatness}
\begin{align}
de_i&=\epsilon_{ij}\,\omega\, e_j\,, &
d\omega&=\frac{1}{2\ell^{2}}\epsilon_{ij}e_ie_j\,, \label{flat-spatial} \\
d\tau&=\epsilon_{ij}e_i\Omega_j\,, &
d\Omega_i&=\epsilon_{ij}\,\omega\,\Omega_j+\frac{1}{\ell^{2}}\epsilon_{ij}\tau e_j\,,
 \label{flat-temporal}
\end{align}
\end{subequations}
the $\delta\omega$ equation \eqref{eom-omega} being implied by the other five once
$c=1/(2q\ell^{2})$ is used. Every trace of $\alpha$, $\gamma$ and $c$ has disappeared.
They entered \eqref{carroll-eom} only through the shift \eqref{Omega-shift}, and
\eqref{flatness} is blind to it, which is \eqref{cmmg-is-ctmg} seen on a solution. The first of \eqref{flat-spatial} says that $\omega$ is
the spin connection of the spatial zweibein, and the second that the spatial slices are
hyperbolic of radius $\ell$; the spatial curvature is fixed, not free.

%--- 9.3  the branch is standard plus exotic Chern--Simons -------------
The same substitution in \eqref{carrollian-mmg-lagrangian} makes the point at the level of
the action. Putting $\bar{\tau}=0$, $h_i=c\,e_i$ and \eqref{Omega-shift} into
\eqref{carrollian-mmg-lagrangian} gives, identically and with no total derivative
discarded,
\begin{equation} \label{std-ex}
L_{\text{CMMG}}\Big|_{\bar{\tau}=0,\,h_i=c\,e_i}=\sigma L_{\text{std}}
 +\frac{1}{2\mu}L_{\text{ex}}\,,
\end{equation}
with
\begin{equation} \label{std-ex-pieces}
L_{\text{std}}=\tau\,d\omega-e_i\left(d\Omega_i-\epsilon_{ij}\omega\Omega_j\right)
 +\frac{\Lambda_0}{2\sigma}\epsilon_{ij}\tau e_ie_j\,, \qquad
L_{\text{ex}}=-\omega\,d\omega-\frac{\Lambda_0}{\sigma}e_i\mathcal{R}_i(e)\,.
\end{equation}
These are the standard and the exotic AdS--Carroll Chern--Simons Lagrangians,
and the two invariant forms \eqref{two-forms} below are theirs. This is what the central
charges of \eqref{bms3} will see, and it is why they carry $\sigma$ and $1/\mu$ and
nothing else. The statement is local and confined to this branch. It does not make
\eqref{carrollian-mmg-lagrangian} a Chern--Simons theory off shell, where $g_{rs}$ is
degenerate and the Jacobi identity fails.

%--- 9.5  the AdS--Carroll algebra -------------------------------------
Equations \eqref{flatness} are the flatness conditions
$d\mathcal{A}+\frac{1}{2}\left[\mathcal{A},\mathcal{A}\right]=0$ for the one-form
\begin{equation} \label{connection}
\mathcal{A}=\omega J+e_iP_i+\tau H+\Omega_iK_i
\end{equation}
valued in the algebra
\begin{equation} \label{adscarroll}
\begin{aligned}
\left[J,P_i\right]&=\epsilon_{ij}P_j\,, &
\left[P_i,P_j\right]&=-\frac{1}{\ell^{2}}\epsilon_{ij}J\,, &
\left[J,K_i\right]&=\epsilon_{ij}K_j\,, \\
\left[P_i,K_j\right]&=-\epsilon_{ij}H\,, &
\left[H,P_i\right]&=\frac{1}{\ell^{2}}\epsilon_{ij}K_j\,, &
\left[H,J\right]&=\left[H,K_i\right]=\left[K_i,K_j\right]=0\,,
\end{aligned}
\end{equation}
which is the three-dimensional AdS--Carroll algebra. Its Chern--Simons gravity
formulation is given in \cite{Matulich:2019cdo}, and its $\mathcal{N}=1$ Chern--Simons
supergravity extension in \cite{Ravera:2019ize}. It satisfies the Jacobi
identity. This does not conflict with Section~\ref{mmgalg}: the Jacobiator that fails there
is a property of the off-shell flavor structure $f_{RST}$ of the Chern--Simons-like form,
whereas \eqref{adscarroll} is what remains once $\bar{\tau}$ and $h_i$ have been eliminated
by their own field equations. Third-way consistency is a statement about the former.

The generators $\left\{H,K_i\right\}$ span an abelian ideal on which
$\left\{J,P_i\right\}\cong\mathfrak{sl}(2,\mathbb{R})$ acts in the adjoint, so
\eqref{adscarroll} is $\mathfrak{isl}(2,\mathbb{R})\cong\mathfrak{iso}(2,1)$: the
isomorphism between the three-dimensional AdS--Carroll and Poincar\'e algebras
of \cite{Aviles:2024llx}. The on-shell content of Carrollian MMG is therefore that of a
Chern--Simons theory for the three-dimensional Poincar\'e algebra, which is consistent
with \eqref{F-any}: such a theory has no local degrees of freedom.

%--- 9.7  the charge as an invariant pairing ---------------------------
Under the shift \eqref{Omega-shift} the gauge parameters follow the fields,
$\Lambda^{\Omega_i}=\Lambda^{\varpi_i}+\frac{c\alpha}{\gamma}\Lambda^{e_i}$. Imposing
$\bar{\tau}=0$, $h_i=c\,e_i$ and the admissible parameters
$\Lambda^{\bar{\tau}}=0$, $\Lambda^{h_i}=c\,\Lambda^{e_i}$ of Section~\ref{asympt}, the
master charge \eqref{master-charge} becomes
\begin{equation} \label{pairing-charge}
\delta Q[\Lambda]=-\frac{1}{8\pi G}\oint d\phi\,
 \left\langle \Lambda,\delta\mathcal{A}_{\phi}\right\rangle\,, \qquad
\left\langle\;,\;\right\rangle=\sigma\left\langle\;,\;\right\rangle_{\rm I}
 -\frac{1}{\mu}\left\langle\;,\;\right\rangle_{\rm II}\,,
\end{equation}
where the two bilinear forms are
\begin{equation} \label{two-forms}
\left\langle H,J\right\rangle_{\rm I}=1\,, \quad
\left\langle P_i,K_j\right\rangle_{\rm I}=-\delta_{ij}\,; \qquad
\left\langle J,J\right\rangle_{\rm II}=1\,, \quad
\left\langle P_i,P_j\right\rangle_{\rm II}=-\frac{1}{\ell^{2}}\delta_{ij}\,,
\end{equation}
all unlisted pairings vanishing. Both are invariant,
$\left\langle\left[X,Y\right],Z\right\rangle+\left\langle Y,\left[X,Z\right]\right\rangle=0$,
and they are the two invariant forms \eqref{adscarroll} admits. The reduction of
\eqref{master-charge} to \eqref{pairing-charge} uses
$\sigma\frac{c\alpha}{\gamma}+\gamma c=\frac{cq}{\mu}=\frac{1}{2\mu\ell^{2}}$, with
$c=1/(2q\ell^{2})$ of \eqref{c-value}. This is where
the couplings $\alpha$, $\gamma$ and $c$ are traded for $1/\mu$, and it is the same
cancellation that made \eqref{final-charges} independent of them. The degeneracy of
$g_{rs}$ is what leaves exactly two invariant forms rather than three.

%--- 9.9  boundary conditions ------------------------------------------
Introduce the standard basis
\begin{equation} \label{LM-basis}
L_0=\ell P_1\,, \quad L_{\pm1}=J\pm\ell P_2\,, \qquad
M_0=K_1\,, \quad M_{\pm1}=\ell H\pm K_2\,,
\end{equation}
in which \eqref{adscarroll} reads
\[ \left[L_m,L_n\right]=(m-n)L_{m+n}\,, \qquad
\left[L_m,M_n\right]=(m-n)M_{m+n}\,, \qquad
\left[M_m,M_n\right]=0\,, \]
with $m,n\in\{-1,0,1\}$ throughout, these being the six generators of
$\mathfrak{isl}(2,\mathbb{R})$ rather than the modes of \eqref{bms3}. In this basis
\eqref{two-forms} becomes
\begin{equation} \label{forms-LM}
\left\langle L_0,M_0\right\rangle_{\rm I}=-\ell\,, \quad
\left\langle L_{\pm1},M_{\mp1}\right\rangle_{\rm I}=2\ell\,; \qquad
\left\langle L_0,L_0\right\rangle_{\rm II}=-1\,, \quad
\left\langle L_{\pm1},L_{\mp1}\right\rangle_{\rm II}=2\,.
\end{equation}
Writing $\mathcal{A}=b^{-1}\left(d+a\right)b$ with $b=\exp\left[\ln(r/\ell)\,L_0\right]$
strips the radial dependence, and because \eqref{two-forms} is invariant the charge
\eqref{pairing-charge} is insensitive to this conjugation. The Carrollian BTZ solution
\eqref{BTZ-soln-of-vielbein}--\eqref{BTZ-solution-of-omega} gives
\begin{equation} \label{btz-a}
a_{\phi}=L_1-\frac{M}{4}\,L_{-1}+\frac{\mathcal{J}}{4\ell}\,M_{-1}
 +\mathcal{O}\!\left(r^{-2}\right)\,,
\end{equation}
which we adopt, in the Ba\~{n}ados sense \cite{Banados:1998gg}, as the boundary condition
\begin{equation} \label{bc}
a_{\phi}=L_1+\mathcal{J}_L(\phi)\,L_{-1}+\mathcal{J}_M(\phi)\,M_{-1}\,,
\end{equation}
the Carrollian BTZ solution being the constant representative
$\mathcal{J}_L=-M/4$, $\mathcal{J}_M=\mathcal{J}/(4\ell)$.

%--- 9.13  residual transformations ------------------------------------
A gauge parameter $\lambda$ preserves \eqref{bc} if the $L_1$, $L_0$, $M_1$ and $M_0$
components of $\delta a_{\phi}=\partial_{\phi}\lambda+\left[a_{\phi},\lambda\right]$
vanish. This fixes all but two functions, a prime
denoting $d/d\phi$ in this section,
\begin{equation} \label{residual}
\lambda=\varepsilon L_1-\varepsilon'L_0
 +\left(\mathcal{J}_L\varepsilon+\tfrac{1}{2}\varepsilon''\right)L_{-1}
 +\chi M_1-\chi'M_0
 +\left(\mathcal{J}_L\chi+\mathcal{J}_M\varepsilon+\tfrac{1}{2}\chi''\right)M_{-1}\,,
\end{equation}
with $\varepsilon(\phi)$ and $\chi(\phi)$ free, and the remaining components of
$\delta a_{\phi}$ give
\begin{subequations} \label{transf}
\begin{align}
\delta\mathcal{J}_L&=\varepsilon\mathcal{J}_L'+2\mathcal{J}_L\varepsilon'
 +\tfrac{1}{2}\varepsilon'''\,, \\
\delta\mathcal{J}_M&=\varepsilon\mathcal{J}_M'+2\mathcal{J}_M\varepsilon'
 +\chi\mathcal{J}_L'+2\mathcal{J}_L\chi'+\tfrac{1}{2}\chi'''\,.
\end{align}
\end{subequations}
Both transform with weight two. Here $\varepsilon$ generates superrotations and $\chi$
supertranslations. Neither carries a radial component, so the generator that would detect
the deformation \eqref{beta-shift}, by \eqref{beta-charge}, is not among them. Whether a
different choice of boundary conditions admits one is taken up in
Section~\ref{conclusions}.

%--- 9.15  the charges -------------------------------------------------
With \eqref{residual} and \eqref{bc}, the pairing \eqref{pairing-charge} reduces to
\begin{equation} \label{bms-charge}
\delta Q\left[\varepsilon,\chi\right]=-\frac{1}{4\pi G}\oint d\phi
 \left[\sigma\ell\left(\varepsilon\,\delta\mathcal{J}_M
 +\chi\,\delta\mathcal{J}_L\right)-\frac{1}{\mu}\varepsilon\,\delta\mathcal{J}_L\right]\,,
\end{equation}
which is manifestly integrable. Setting
$\mathcal{L}_m=Q\!\left[e^{im\phi},0\right]$ and
$\mathcal{M}_m=Q\!\left[0,e^{im\phi}\right]$ and evaluating on \eqref{btz-a} gives
\begin{equation} \label{zero-modes}
\mathcal{L}_0=-J_{\rm C}\,, \qquad \mathcal{M}_0=\ell\,E_{\rm C}\,,
\end{equation}
with $E_{\rm C}$ and $J_{\rm C}$ of \eqref{final-charges}. This is not a fourth
independent computation of the charges since \eqref{bms-charge} descends from
\eqref{master-charge} but it fixes the normalization of the generators and confirms
that \eqref{bc} admits the black hole.

%--- 9.17  the algebra -------------------------------------------------
Using $\left\{Q[\lambda_1],Q[\lambda_2]\right\}=\delta_{\lambda_2}Q[\lambda_1]$ with
\eqref{transf}, and writing $\left[\;,\;\right]=i\left\{\;,\;\right\}$,
\begin{equation} \label{bms3}
\boxed{\;
\begin{aligned}
\left[\mathcal{L}_m,\mathcal{L}_n\right]
 &=(m-n)\mathcal{L}_{m+n}+\frac{c_L}{12}m^{3}\delta_{m+n,0}\,, &
 c_L&=-\frac{3}{\mu G}\,, \\
\left[\mathcal{L}_m,\mathcal{M}_n\right]
 &=(m-n)\mathcal{M}_{m+n}+\frac{c_M}{12}m^{3}\delta_{m+n,0}\,, &
 c_M&=\frac{3\sigma\ell}{G}\,, \\
\left[\mathcal{M}_m,\mathcal{M}_n\right]&=0\,. &&
\end{aligned}
\;}
\end{equation}
This is $\mathfrak{bms}_3$ \cite{Barnich:2006av}. The $m^{3}$ rather than $m(m^{2}-1)$ is the usual shift of
$\mathcal{L}_0$ and $\mathcal{M}_0$ by constants. The factor $\ell$ in $c_M$ is the
normalization of $M_{\pm1}$ in \eqref{LM-basis}; rescaling $M_n\to M_n/\ell$, which leaves
the algebra unchanged, gives $c_M=3\sigma/G$. For $\sigma=+1$ this is the $3/G$ obtained
for three-dimensional AdS--Carroll gravity in \cite{Aviles:2025ygw}, whose Chern--Simons
level $k=1/(4G_{\rm C})$ gives $c_M=12k$. The vacuum comparison made after
\eqref{final-charges} is not independent of this. Both configurations sit at
$\mathcal{M}_0=-c_M/24$, ours because $\mathcal{M}_0=\ell E_{\rm C}$ by \eqref{zero-modes}
and theirs because the vacuum carries $-k/2$.

%--- what is new in the central charges --------------------------------
The appearance of $\mathfrak{bms}_3$ here is not itself new. It was obtained from
three-dimensional AdS--Carroll gravity in \cite{Aviles:2024llx}, and a larger set of
boundary conditions there gives the generalized $\mathfrak{bms}_3$ of
\cite{Aviles:2025ygw}, which contains \eqref{bms3} as a subalgebra. What is new is the
second central charge, and the reason is visible in \eqref{two-forms}. The AdS--Carroll
algebra admits two invariant forms, written in \cite{Aviles:2024llx} with coefficients
$\beta_1$ and $\beta_2$, and the analysis there is carried out at $\beta_2=0$. In
\eqref{pairing-charge} the bulk couplings fix both, $\beta_1$ by $\sigma$ and $\beta_2$ by
$-1/\mu$. The Lorentz--Chern--Simons term of \eqref{carrollian-mmg-lagrangian} turns
the second form on, and $c_L$ is what it produces.

The resulting pair is the one flat-space TMG carries. There the $\mathfrak{bms}_3$ central
charges are $c_1=3/(\mu G)$ and $c_2=3/G$ \cite{Bagchi:2012yk}, with the Einstein term
supporting one and the gravitational Chern--Simons term the other, exactly as in
\eqref{bms3}. By \eqref{cmmg-is-ctmg} this compares two limits of the same
relativistic theory, the flat-space limit of TMG and its Carroll contraction, and the two
agree up to the overall sign and normalization of the generators.

%--- what the central charges depend on --------------------------------
Three features of \eqref{bms3} are worth recording. The two charges are carried by
different terms of \eqref{carrollian-mmg-lagrangian}. Here $c_M$ comes from the Einstein
term through $\sigma$ and $c_L$ from the Lorentz--Chern--Simons term through $1/\mu$, each
supported by exactly one of the invariant forms \eqref{two-forms}. Neither depends on
$\alpha$, $\gamma$ or $c$, which extends to the whole asymptotic algebra the statement
made for the black hole charges in Section~\ref{bdrycharges}. The relativistic combination
$\tilde{\sigma}+\frac{\tilde{\alpha} \tilde{c}}{\tilde{\gamma}\tilde{\mu}}$ of \eqref{rel-charges} is replaced by $\sigma$, as
\eqref{cmmg-is-ctmg} leads one to expect. And since the theory carries no local degrees of
freedom by \eqref{F-any}, \eqref{bms3} is not the symmetry algebra of a boundary sector
but the complete physical content, shared with Carrollian TMG.

%=======================================================================
%  SECTION 10 :  THERMODYNAMICS OF THE CARROLLIAN BLACK HOLE
%=======================================================================
\section{Thermodynamics of the Carrollian black hole}\label{thermo}

%--- where the Carrollian structure degenerates ------------------------
The first law established below needs only Sections~\ref{adsbtz} and \ref{bdrycharges}.
Section~\ref{bms} enters in the remarks that follow it.

\subsection{The Carroll extremal surface and its surface gravity}
The Carrollian frame \eqref{BTZ-soln-of-vielbein} degenerates where $N=0$. Since $\mathcal{J}$ is absent from
$N^{2}$ of \eqref{lapse}, there is for $M>0$ exactly one such locus,
\begin{equation} \label{rplus}
r_+=\ell\sqrt{M}\,,
\end{equation}
and no second one however large the rotation. The inner horizon of the relativistic BTZ
solution has no counterpart in the Carrollian frame.

%--- 10.2  what degenerates there, and what does not -------------------
What degenerates at \eqref{rplus} needs care, because the chart hides it. In the
coordinates of \eqref{BTZ-soln-of-vielbein} the component $e_{1r}=N^{-1}$ blows up, but
that is a property of $r$ rather than of the geometry. Setting
\begin{equation} \label{xcoord}
r=r_+\cosh\!\left(x/\ell\right)\,, \qquad N=\sqrt{M}\,\sinh\!\left(x/\ell\right)\,,
\end{equation}
turns \eqref{BTZ-soln-of-vielbein} into
\begin{equation} \label{xframe}
\tau=\sqrt{M}\,\sinh\!\left(x/\ell\right)dt\,, \qquad
e_1=dx\,, \qquad e_2=r_+\cosh\!\left(x/\ell\right)d\phi\,,
\end{equation}
so that $\det e_{iK}=r_+\cosh(x/\ell)$ is finite and non-zero at $x=0$. The spatial
zweibein is therefore invertible at \eqref{rplus}, and the Hamiltonian analysis of
Section~\ref{Dirac}, which needs exactly that, holds there as well. What does vanish is
the clock form $\tau$, and the Carroll vector $v=N^{-1}\partial_t$ diverges with it. The
degeneracy is in the Carroll structure as a whole and not in the spatial metric
\cite{Ecker:2023uwm}. Of the two statements collected in Section~\ref{mmgsec}, the
invertibility of $e_{iK}$ survives at \eqref{rplus} and the basis property of
$(\tau,e_1,e_2)$ does not.

The divergence of $v$ is also what makes \eqref{rplus} a Carroll extremal surface, the
locus where the Carroll vector field diverges being the definition used in
\cite{Ecker:2023uwm,Ecker:2026noz}. The same criterion is applied to three-dimensional
AdS--Carroll gravity in \cite{Aviles:2025ygw}, where the surface exists provided the
energy is positive. Here that condition reads $M>0$ at $\sigma=+1$ by
\eqref{final-charges}, which is what \eqref{rplus} already requires. The surface does not
move with the rotation, since $\mathcal{J}$ is absent from $N^{2}$, and the same is true
of the rotating Carroll--Schwarzschild solution of \cite{Ecker:2026noz}.

There is no Killing horizon here and no light cone, a Carrollian geometry having no
spacetime metric to supply either. There is an intrinsic Carroll surface gravity, and
\eqref{rplus} is where it is evaluated,
\begin{equation} \label{kappa}
\kappa=NN'\big|_{r_+}=\frac{r_+}{\ell^{2}}=\frac{\sqrt{M}}{\ell}\,,
\end{equation}
the prime denoting $d/dr$.

\subsection{The horizon branch}
%--- 10.5  the black hole does not rotate ------------------------------
The shift function of the relativistic solution has dropped out of the Carrollian frame.
Equation \eqref{BTZ-soln-of-vielbein} gives $e_2=r\,d\phi$ with no $dt$ component, and the
rotation parameter survives only in the connection \eqref{BTZ-solution-of-omega}. The
Carroll structure is static, and the angular velocity attached to \eqref{rplus} vanishes,
\begin{equation} \label{Omega-zero}
\Omega_{\rm H}=0\,,
\end{equation}
consistent with the energy \eqref{final-charges} being independent of $\mathcal{J}$. We
call the thermodynamics built on \eqref{rplus} the horizon branch. A second branch,
attached to a holonomy condition rather than to a locus and carrying $\Omega\neq0$,
appears below. A rotating Carrollian black hole that is static as far as its frame is
concerned is what \cite{Ecker:2026noz} constructs in four dimensions, and the mechanism
is the one at work here. The staticity theorem of \cite{Kolar:2025ebv} constrains the
metric data and says nothing about the connection, which is where the angular momentum of
\eqref{BTZ-solution-of-omega} resides. The absence of frame dragging is then expected
rather than surprising, and \eqref{Omega-zero} is its Carrollian expression.
Three dimensions also admit the topological rotation of \cite{Kolar:2025ebv}, obtained by
a Carroll boost and a re-identification of $\phi$. Our $\mathcal{J}$ is that rotation
written in another gauge. Sending $t\to t+\nu\phi$ in the non-rotating solution produces a
second flat \eqref{connection}, with $\tau=N(dt+\nu d\phi)$ and
$\Omega_2=\left(r/\ell^{2}\right)(dt+\nu d\phi)$. The two are compared by the invariants
of $\mathcal{A}_{\phi}$ under \eqref{adscarroll}. Both give $M/2$ for the invariant built
from the $L_n$ part alone, while the mixed invariant is $-\mathcal{J}/(4\ell)$ for
\eqref{BTZ-solution-of-omega} and $\nu M/(2\ell)$ for the topological rotation. These agree
exactly when
\begin{equation} \label{top-rotation}
\mathcal{J}=-2\nu M\,,
\end{equation}
and for $M>0$ the pair is a complete invariant, the $L_n$ part of $\mathcal{A}_{\phi}$
being regular. The two holonomies are then conjugate. The conjugation is not a
transformation of \eqref{bc}. In the radial gauge the topological rotation carries an
$M_1$ component $\nu/\ell$ in $a_{\phi}$, which is a supertranslation source and not a
state. The two configurations are therefore the same geometry presented with different
boundary data. That is how \eqref{final-charges} can assign an angular momentum to a
solution for which \eqref{Omega-zero} holds.

%--- 10.7  temperature from the surface gravity ------------------------
The temperature follows from \eqref{kappa} rather than being postulated,
\begin{equation} \label{temperature}
T=\frac{\kappa}{2\pi}=\frac{\sqrt{M}}{2\pi\ell}\,.
\end{equation}
With \eqref{Omega-zero} and the charges \eqref{final-charges}, the first law
\begin{equation} \label{first-law}
\boxed{\;
\delta E_{\rm C}=T\,\delta S+\Omega_{\rm H}\,\delta J_{\rm C}
\;}
\end{equation}
then fixes the entropy up to an additive constant, which $S\to0$ as $M\to0$ removes,
\begin{equation} \label{entropy}
S=\frac{\sigma\pi\ell\sqrt{M}}{2G}
 =\frac{\sigma}{4G}\oint_{r=r_+}\!\!e_2\,.
\end{equation}
The second form is the length of \eqref{rplus} in the spatial metric
$\delta_{ij}e_ie_j$, weighted by the Einstein coupling. The Bekenstein--Hawking area law
is thus an output of \eqref{kappa} and \eqref{first-law} and not an assumption. Since
$\Omega_{\rm H}=0$ and $S$ does not depend on $\mathcal{J}$, \eqref{first-law} reduces to
the single relation $\delta E_{\rm C}=T\,\delta S$. On this branch the rotation drops out
of the thermodynamics as completely as it drops out of the frame.

%--- 10.10  the relativistic check -------------------------------------
Nothing relativistic has entered. As a check, each of \eqref{temperature},
\eqref{entropy} and \eqref{Omega-zero} is also the contraction of its relativistic
counterpart. The outer-horizon data of the parent solution are the charges
\eqref{rel-charges}, the Hawking temperature
$\tilde{T}_+=(\tilde{r}_+^{2}-\tilde{r}_-^{2})/(2\pi\ell^{2}\tilde{r}_+)$, the angular velocity
$\tilde{\Omega}_+=\tilde{r}_-/(\ell \tilde{r}_+)$ and the entropy
\begin{equation} \label{rel-entropy}
\tilde{S}_+=\frac{\pi}{2G}\left(\bar{\sigma}\,\tilde{r}_++\frac{\tilde{r}_-}{\tilde{\mu}\ell}\right)\,, \qquad
\bar{\sigma}=\tilde{\sigma}+\frac{\tilde{\alpha}\tilde{c}}{\tilde{\gamma}\tilde{\mu}}\,,
\end{equation}
with $\tilde{r}_+^{2}+\tilde{r}_-^{2}=M\ell^{2}$ and $\tilde{r}_+\tilde{r}_-=\tilde{\mathcal{J}}\ell/2$
\cite{Solodukhin:2005ah,Tachikawa:2006sz,Setare:2015pva}. The combination $\bar{\sigma}$ is
the one carried by the charges \eqref{rel-charges}. Take $\tilde{t}=t/\rho$ and
$\tilde{\mathcal{J}}=\mathcal{J}/\rho$ as in Section~\ref{btzsub}, together with the scalings of
Section~\ref{mmgsec}, so that $\bar{\sigma}=\rho\sigma+\mathcal{O}(\rho^{-1})$ and
$\tilde{r}_-=\mathcal{J}/(2\rho\sqrt{M})+\mathcal{O}(\rho^{-3})$. Then
\begin{equation} \label{outer-limits}
E_{\rm C}=\lim\frac{\tilde{E}}{\rho}\,, \quad
J_{\rm C}=\lim \tilde{J}\,, \quad
S=\lim\frac{\tilde{S}_+}{\rho}\,, \quad
T=\lim \tilde{T}_+\,, \quad
\Omega_{\rm H}=\lim\frac{\tilde{\Omega}_+}{\rho}=0\,,
\end{equation}
every limit being $\rho\to\infty$, and dividing the relativistic first law by $\rho$
returns \eqref{first-law}. One caveat belongs with \eqref{outer-limits}. The $T$ appearing
there is the finite temperature conjugate to $S$, and not the one read off the rescaled
Euclidean time, which is $\tilde{T}_+/\rho\to0$.

%--- 10.12  the Chern--Simons term on this branch ----------------------
The second term of \eqref{rel-entropy} is easy to misplace and is worth a sentence. It
belongs to the entropy of the outer horizon and is proportional to the inner-horizon
radius. It is not an entropy carried by the inner horizon
\cite{Solodukhin:2005ah,Tachikawa:2006sz,Setare:2015pva}. Since $\tilde{r}_-$ vanishes in the
contraction,
\begin{equation} \label{no-CS-entropy}
\lim_{\rho\to\infty}\frac{1}{\rho}\,\frac{\pi \tilde{r}_-}{2G\tilde{\mu}\ell}=0\,,
\end{equation}
and the Lorentz--Chern--Simons coupling contributes nothing to \eqref{entropy}. On the
horizon branch the Carrollian black hole obeys the area law exactly. This is worth
contrasting with the two places where $1/\mu$ does survive, the angular momentum
\eqref{final-charges} and the central charge $c_L$ of \eqref{bms3}. It is the same
statement as the absence of a $\tilde{\mathcal{J}}^{2}/(4r^{2})$ term from \eqref{lapse},
and it holds verbatim for Carrollian TMG. It fails on the second branch, treated next.

\subsection{The holonomy branch}
%--- 10.13  the holonomy branch ----------------------------------------
The inner horizon has no locus in the Carrollian frame, but its thermodynamic data have a
finite limit of their own, and the entropy needs no rescaling to reach it. The
relativistic inner-horizon quantities are
$\tilde{T}_-=(\tilde{r}_+^{2}-\tilde{r}_-^{2})/(2\pi\ell^{2}\tilde{r}_-)$, $\tilde{\Omega}_-=\tilde{r}_+/(\ell \tilde{r}_-)$ and
$\tilde{S}_-=\frac{\pi}{2G}\left(\bar{\sigma}\tilde{r}_-+\frac{\tilde{r}_+}{\tilde{\mu}\ell}\right)$
\cite{Castro:2012av,Detournay:2012ug}. The limits that gave \eqref{outer-limits} give, for
$\mathcal{J}>0$,
\begin{align} \label{inner-branch}
T_{\rm hol}&=\lim\frac{\tilde{T}_-}{\rho}=\frac{M^{3/2}}{\pi\mathcal{J}}\,, &
\Omega_{\rm hol}&=\lim\frac{\tilde{\Omega}_-}{\rho}=\frac{2M}{\mathcal{J}}\,, \nn \\
S_{\rm hol}&=\lim \tilde{S}_-
 =\frac{\pi}{2G}\left(\frac{\sigma\mathcal{J}}{2\sqrt{M}}+\frac{\sqrt{M}}{\mu}\right)\,, &&
\end{align}
and these satisfy
\begin{equation} \label{first-law-hol}
\delta E_{\rm C}=-T_{\rm hol}\,\delta S_{\rm hol}+\Omega_{\rm hol}\,\delta J_{\rm C}
\end{equation}
identically in $M$ and $\mathcal{J}$. Two things separate \eqref{inner-branch} from the
horizon branch. The Lorentz--Chern--Simons coupling survives in $S_{\rm hol}$, and the
sign in \eqref{first-law-hol} is reversed, so that the inverse temperature of this branch
is negative.

Both features are known from the Chern--Simons side, where demanding trivial holonomy
around a thermal cycle fixes the chemical potentials in terms of the charges
\cite{Bunster:2014mua,Aviles:2025ygw}. Section~\ref{holsub} carries that computation out
inside the Carrollian theory and recovers \eqref{inner-branch} from it. At $\sigma=+1$ and
$\mu\to\infty$, where \eqref{carrollian-mmg-lagrangian} loses its Lorentz--Chern--Simons
term, \eqref{inner-branch} becomes
\begin{equation} \label{hol-aviles}
S_{\rm hol}=\pi J_{\rm C}\sqrt{\frac{2k}{E_{\rm C}}}\,, \qquad
T_{\rm hol}=\frac{1}{\pi J_{\rm C}}\sqrt{\frac{2E_{\rm C}^{3}}{k}}\,, \qquad
\Omega_{\rm hol}=\frac{2E_{\rm C}}{J_{\rm C}}\,, \qquad k=\frac{1}{4G}\,,
\end{equation}
which are the Carroll thermal entropy, temperature and angular velocity of
\cite{Aviles:2025ygw} term by term, negative inverse temperature included. That sign is
read there as the signature of a configuration of cosmological rather than black hole
type.

\subsection{The holonomy branch from the Chern--Simons connection}\label{holsub}
%--- 10.16  the connection, the thermal cycle and the entropy ----------
The data \eqref{inner-branch} were read off a limit. They can also be derived inside the
Carrollian theory, from the connection \eqref{connection} and the holonomy prescription of
\cite{Bunster:2014mua,Aviles:2025ygw}, with no relativistic input at any step. The two
routes share nothing, so their agreement is a check on both.

Evaluating \eqref{connection} on \eqref{BTZ-soln-of-vielbein} and \eqref{BTZ-Omega} and
collecting the coefficients of $dt$, $dr$ and $d\phi$ gives
\begin{equation} \label{A-components}
\mathcal{A}_t=NH+\frac{r}{\ell^{2}}K_2\,, \quad
\mathcal{A}_r=\frac{1}{N}P_1+\frac{\mathcal{J}}{2Nr^{2}}K_1\,, \quad
\mathcal{A}_{\phi}=NJ+rP_2-\frac{\mathcal{J}}{2r}K_2\,,
\end{equation}
This one-form is flat, which is \eqref{flatness} again. Strip the radial dependence
as in Section~\ref{bms}, with $\mathcal{A}=b^{-1}(d+a)b$ and
$b=\exp\left[\ln(r/\ell)L_0\right]$. Since $b$ depends on $r$ alone, the $t$ and $\phi$
components are simply conjugated. The angular one is \eqref{btz-a} and the
temporal one is
\begin{equation} \label{a-t}
a_t=\frac{1}{\ell}\left(M_1-\frac{M}{4}M_{-1}\right)+\mathcal{O}\!\left(r^{-2}\right)\,.
\end{equation}

The thermal cycle is generated by $\xi_{\rm th}=\partial_t+\Omega\partial_{\phi}$ with
$\Omega$ not yet fixed, so the connection along it is $a_{\rm th}=a_t+\Omega a_{\phi}$.
The prescription asks for the holonomy around that cycle to lie in the center, and two
conditions follow. The first removes the $M_n$ part by a conjugation. For $g=e^{zM_0}$ the
series $g\,a_{\rm th}\,g^{-1}=a_{\rm th}+z\left[M_0,a_{\rm th}\right]$ terminates, since
$\left[M_m,M_n\right]=0$ in \eqref{LM-basis}. The $M_1$ component is then removed by
$z=1/(\ell\Omega)$, which leaves $(\Omega\mathcal{J}-2M)/(4\ell)$ in the $M_{-1}$ direction.
Requiring that to vanish fixes
\begin{equation} \label{Omega-hol-cs}
\Omega_{\rm hol}=\frac{2M}{\mathcal{J}}
\end{equation}
and reduces the thermal connection to
$\Omega_{\rm hol}\left(L_1-\frac{M}{4}L_{-1}\right)$. The second condition is the usual one
on the Euclidean period. That generator has eigenvalues $\pm\sqrt{M}/2$ in the
two-dimensional representation, so a central holonomy along the contractible cycle needs
\begin{equation} \label{T-hol-cs}
\beta_{\rm phys}\,\Omega_{\rm hol}\sqrt{M}=2\pi\,, \qquad \text{hence} \qquad
T_{\rm hol}=\beta_{\rm phys}^{-1}=\frac{M^{3/2}}{\pi\mathcal{J}}\,.
\end{equation}

The entropy comes from the same connection and needs no further input. With the pairings
\eqref{forms-LM},
\begin{equation} \label{pairings-th}
\left\langle a_{\rm th},a_{\phi}\right\rangle_{\rm I}=\Omega\mathcal{J}-M\,, \qquad
\left\langle a_{\rm th},a_{\phi}\right\rangle_{\rm II}=-\Omega M\,,
\end{equation}
so the combination entering \eqref{pairing-charge}, evaluated at \eqref{Omega-hol-cs}, is
$\sigma M+2M^{2}/(\mu\mathcal{J})$. The oriented Euclidean period of this branch is
$\beta_{\rm hol}=-\beta_{\rm phys}$, which is the reversed sign of \eqref{first-law-hol}
seen at its source. In the normalization already fixed by \eqref{pairing-charge}, the
Chern--Simons entropy is then
\begin{equation} \label{S-hol-cs}
S_{\rm hol}=-\frac{\beta_{\rm hol}}{8\pi G}\oint d\phi\,
 \left\langle a_{\rm th},a_{\phi}\right\rangle
 =\frac{\pi}{2G}\left(\frac{\sigma\mathcal{J}}{2\sqrt{M}}+\frac{\sqrt{M}}{\mu}\right)\,.
\end{equation}

Equations \eqref{Omega-hol-cs}, \eqref{T-hol-cs} and \eqref{S-hol-cs} reproduce
\eqref{inner-branch} term by term, the Lorentz--Chern--Simons contribution included.
Nothing relativistic was used. The holonomy branch is therefore a configuration of the
Carrollian theory in its own right. The inner horizon of the relativistic solution is
where it comes from, not what it is.

\subsection{Cardy forms and caveats}
%--- 10.22  Cardy forms on the two branches ----------------------------
Each branch has a Cardy-like form, and the two are not the same one. Eliminating $M$
between \eqref{entropy} and \eqref{final-charges} gives, for $\sigma=+1$,
\begin{equation} \label{cardy}
S=\pi\ell\sqrt{\frac{2E_{\rm C}}{G}}=2\pi\sqrt{\frac{c_M\mathcal{M}_0}{6}}\,,
\end{equation}
with $c_M$ of \eqref{bms3} and $\mathcal{M}_0=\ell E_{\rm C}$ of \eqref{zero-modes}. Both
signs of $\sigma$ are covered by \eqref{entropy}, the energy and the entropy changing sign
together at $\sigma=-1$. Only the supertranslation central charge appears in
\eqref{cardy}, consistent with the absence of $c_L$ and of $\mathcal{L}_0$ from
\eqref{entropy}. It should be read as an algebraic rewriting of \eqref{entropy} and not as
a counting of states, and a relation of the same shape appears for Carroll
Jackiw--Teitelboim gravity in \cite{Ecker:2023uwm}.

The holonomy branch is the one that uses both central charges. From \eqref{inner-branch},
\eqref{bms3} and \eqref{zero-modes},
\begin{equation} \label{cardy-hol}
S_{\rm hol}=\pi\left(J_{\rm C}\sqrt{\frac{c_M}{6\mathcal{M}_0}}
 -c_L\sqrt{\frac{\mathcal{M}_0}{6c_M}}\right)\,,
\end{equation}
identically in $M$ and $\mathcal{J}$ and for both signs of $\sigma$. This has the shape of
the $\mathfrak{bms}_3$ Cardy formula of
\cite{Barnich:2012xq,Bagchi:2012xr,Riegler:2014bia}, in which a term carried by $c_M$ and
a term carried by $c_L$ add. The second term is where the Lorentz--Chern--Simons coupling
enters. What the two branches share is the algebra \eqref{bms3}. What separates
them is the configuration the counting is applied to.

%--- what T is and is not ----------------------------------------------
Two reservations should be recorded. First, \eqref{temperature} is not merely an
integrating factor for \eqref{first-law}. It is $\kappa/2\pi$ with $\kappa$ the intrinsic
Carroll surface gravity \eqref{kappa}, so the split of the product $T\,\delta S$ between
its two factors is fixed by the geometry. The contraction \eqref{outer-limits} then agrees
with that split rather than supplying it.
The bookkeeping that makes $E_{\rm C}$ and $S$ finite is the one that places the factor
$\rho$ on the couplings. Held instead at fixed Newton constant, the contraction sends the
temperature conjugate to the Carrollian time to zero and the entropy to infinity with
their product finite, which is the behavior of the magnetic branch found in
\cite{Xu:2026vpj}. The two descriptions differ only in where the factor of $\rho$ is
carried.

Second, \eqref{entropy} and \eqref{inner-branch} are statements about the charges and the
field equations. Nothing here counts states, and the absence of local degrees of freedom
\eqref{F-any} means that whatever they count is not carried by bulk excitations.

A first law for a rotating Carroll black hole whose angular momentum is carried by the
connection is listed as an open problem in \cite{Ecker:2026noz}. The obstacle named there
is an independent determination of the temperature. In three dimensions \eqref{kappa} supplies
that temperature and \eqref{first-law} is the resulting first law. Whether the same route is available in four
dimensions we do not know. There the surface gravity of a Carroll extremal surface would
have to be built from the connection rather than read off a lapse.

%=======================================================================
%  SECTION 11 :  CONCLUSIONS AND OUTLOOK
%=======================================================================
\section{Conclusions and outlook}\label{conclusions}

%--- what the limit turns out to be ------------------------------------
We have constructed the Carrollian limit of MMG and asked what becomes of third-way
consistency. The answer splits in two, and the halves go opposite ways. The closure
structure survives the contraction. The coupling that carries it does not.

The contraction is singular for the algebra and regular for the Chern--Simons-like data.
The Lagrangian \eqref{carrollian-mmg-lagrangian} is finite while the contracted brackets
\eqref{algdec} are not, and the reason is not accidental: the single divergent bracket
points along the null direction of the contracted bilinear form. What survives of
third-way consistency is then both sharper than the relativistic statement and weaker. The
relativistic closure condition $\Psi=0$ of \eqref{ehzero} is not inherited. The Carrollian
field equations force only $e_i\wedge\Psi=0$, which is precisely what the closure of
\eqref{DFleft} needs. Written intrinsically this is \eqref{thirdway-cov}, whose obstruction
carries the same $q^{2}$ as the Jacobiators \eqref{carroll-jacobiators}, with a second
condition \eqref{eDh} at order $q$. The closure requirement is a genuine condition on the
solutions rather than an identity, which is the content of the name.

%--- the one fact that organizes the rest ------------------------------
Downstream of this, one fact organizes everything, and it is the degeneracy of the vector
flavor metric \eqref{flavor-metrics}. It gives the matrix of primary constraint brackets a
four-dimensional kernel \eqref{4solns}, and the Lagrangian the local shift symmetry
\eqref{null-shift}. It forces $\bar{\tau}_K\approx0$ through \eqref{bartau-zero}. It
obstructs the relativistic formula \eqref{rel-asymptotic} for asymptotic charges, which
needs the inverse flavor metric and has to be replaced by the canonical construction of
Section~\ref{asympt}. And it is one of the two reasons the couplings $\alpha$, $\gamma$ and
$c$ are missing from every boundary charge the theory possesses, the other being the
redefinition \eqref{alpha-removal}.

That brings us back to the split announced at the start. The third-way coupling is the
half that does not survive. By \eqref{cmmg-is-ctmg} it can be absorbed into $h_i$ and
$\varpi_i$ whenever $q\neq0$, so that Carrollian MMG and Carrollian TMG are the same
theory, carrying $\sigma$, $\Lambda_0$ and $\mu$ alone. That one statement accounts for
the absence of $\alpha$ from the energy, from the angular momentum, from both central
charges of \eqref{bms3} and from the entropy, and for the two counts of
Section~\ref{Dirac} agreeing. The contraction removes the massive graviton of TMG and of
MMG alike, and \eqref{dof-general} shows the mechanism: the Dirac algorithm closes at the
$9N$ constraints while $F$ reaches its maximal value $3N$. What none of this does is make
the contracted theory a Chern--Simons theory. The obstruction \eqref{thirdway-cov} carries
$q^{2}$ and not $\alpha^{2}$, and $q=\mu\gamma\neq0$ already at $\alpha=0$, so the system
still closes only on shell. The limit thus separates two things that the relativistic
discussion tends to run together. Being Chern--Simons-like rather than Chern--Simons
survives the contraction. Carrying a deformation that is not a field redefinition does
not. The one place where the two part company is the critical locus $q=0$, where
\eqref{alpha-removal} degenerates.

The gravitational Chern--Simons coupling supplies the contrast. Unlike $\alpha$, the
parameter $1/\mu$ cannot be redefined away. It survives in the angular momentum and in
$c_L$, and even so it leaves no trace in the entropy of the horizon branch
\eqref{no-CS-entropy}, its relativistic contribution being proportional to an
inner-horizon radius that the contraction sends to zero. It reappears on the holonomy
branch \eqref{inner-branch}, where no such factor is present. The two branches
therefore differ on exactly the coupling that cannot be redefined away.

%--- the critical locus ------------------------------------------------
The critical locus $q=0$, excluded throughout by \eqref{genericity}, is a different theory
rather than a degenerate corner of this one. By \eqref{null-shift-general} the auxiliary
field is pure gauge there, so the model loses a vector flavor. This is the one member of
the family that is not Carrollian TMG in disguise. Both \eqref{rel-jacobiators} and
\eqref{carroll-jacobiators} are proportional to $q^{2}$, so the parent and its contraction
are ordinary Chern--Simons theories on that locus, the parent splitting into AdS
Einstein gravity and an independent $\mathfrak{sl}(2,\mathbb{R})$ sector whose level is
carried by the Lorentz--Chern--Simons coupling $1/\mu$. The constraint count follows suit,
on the factors of $q$ already carried by \eqref{chiA} and \eqref{bartau-zero}. Two of the
nine temporal constraints become dependent, the four null combinations \eqref{4solns} turn
first class, and $\mathbb{C}$ has rank $14$ rather than the $18$ of \eqref{rank-full}. The
count then runs over $25$ independent constraints with $F=11$, and again returns no local
degrees of freedom. The solutions change character as well. Equations \eqref{eom-tau} and
\eqref{eom-omega} force $\mathcal{R}(\omega)=0$ and
$\gamma\,d\bar{\tau}=\frac{\Lambda_0}{2}\epsilon_{ij}e_ie_j$, so $\bar{\tau}$ cannot
vanish for $\Lambda_0\neq0$, reversing \eqref{Si=0}. The black hole
\eqref{BTZ-soln-of-vielbein} survives this, with $\bar{\tau}=-(\sigma/\gamma)Nd\phi$,
$\omega=0$ and $\varpi_i$ the torsion-free connection \eqref{BTZ-Omega}. Its charges and
thermodynamics are not treated here.

%--- open problems -----------------------------------------------------
Several questions are left open.

The most immediate is whether the mechanism is particular to MMG. MMG
is the simplest member of an infinite sequence of third-way consistent theories of
three-dimensional gravity, whose Chern--Simons-like formulations use more one-form flavors
\cite{Ozkan:2018cxj,Dedeoglu:2024ies,Dedeoglu:2026ceb}. It would be worth carrying the
contraction through the further members of the sequence, those with $N\geq4$ flavors. One
would then want to know whether the divergent brackets again align with the null directions
of the contracted bilinear form, whether $F$ again saturates at $3N$, and whether the
boundary charges again lose the third-way couplings. The sharpest form of the question is
whether \eqref{alpha-removal} has a counterpart there. That is, one would ask whether the
contraction again turns the third-way couplings into field redefinitions, and if so how
many of the $N$ flavors it can remove. One constraint is immediate from the
counting formula \eqref{dof-general}, whenever the Dirac algorithm terminates at the $9N$
constraints. Since $\text{DOF}=\frac{1}{2}(3N-F)$ must then be a non-negative integer,
$F$ and $N$ have the same parity, and $F$ cannot exceed $3N$. A Carrollian member of the sequence that saturates the bound, $F=3N$, has no
local degrees of freedom, whatever $N$ is.

Section~\ref{third-way} establishes third-way consistency in its first-order form and
deliberately stops short of the metric-level statement. Producing that statement would mean
eliminating $h_i$ and $\bar{\tau}$ in favor of a single equation for the Carrollian
geometry and exhibiting the identity whose failure off shell and validity on shell is what
the name refers to. This is the natural completion of Section~\ref{third-way} and we expect
it to be more subtle than its relativistic counterpart, since a Carrollian geometry has no
metric to write that equation for. One possible route is suggested by
\cite{Broccoli:2021pvv}, where relativistic MMG is recovered by shifting the Christoffel
connection,
$\Gamma^{\nu}_{\mu\rho}\to\Gamma^{\nu}_{\mu\rho}+\kappa\,\varepsilon_{\mu}{}^{\nu\alpha}S_{\alpha\rho}$,
with $S_{\alpha\rho}$ the Schouten tensor. Not every such shift works, and the same
construction with the Ricci tensor in place of the Schouten tensor does not. Contracting
that statement, rather than the first-order system, may be the more tractable route.

The contraction performed here is the magnetic one. The electric Carroll limit
\cite{Henneaux:2021yzg} rescales canonical rather than gauge variables, and the argument
that keeps \eqref{carrollian-mmg-lagrangian} finite, i.e.\ that the divergence lies along the null
direction of the contracted bilinear form, has no evident analogue there. Whether an
electric Carrollian MMG exists, and whether it is third-way consistent, is open. The same question can be asked of the Galilean limit.

On the boundary side, the algebra \eqref{bms3} was obtained with one choice of boundary
conditions, and with that choice the third-way coupling is invisible. Whether that is a
property of the theory or of the choice is open. Some other consistent set, a near-horizon
one or one that relaxes the highest-weight form of \eqref{bc}, might make $\alpha$ appear
in the asymptotic algebra. The question is sharper for the deformation
\eqref{beta-shift}, whose charge \eqref{beta-charge} is non-zero only for a generator with
$\xi^{r}\neq0$; its status therefore turns on which radial components the boundary
conditions admit. More generally, $\mathfrak{bms}_3$ admits deformations that interpolate
toward two commuting Virasoro algebras. It is natural to ask whether the third-way
deformation of the bulk theory is related to any of them, since a positive answer would
give the boundary algebra a handle on exactly the structure the charges computed here
cannot see.

Finally, the two thermodynamic branches of Section~\ref{thermo} are not on the same
footing. The holonomy branch comes with a Cardy-like counting \eqref{cardy-hol} in both
central charges, and Section~\ref{holsub} derives it a second time from the Carrollian
Chern--Simons connection alone. The horizon branch has only the algebraic rewriting
\eqref{cardy}, in which $c_L$ does not appear. Which configuration the
$\mathfrak{bms}_3$ counting is meant to apply to, and whether a counting exists for the
horizon branch at all, is open.

%=======================================================================
%  ACKNOWLEDGMENTS
%=======================================================================
\acknowledgments
%%% PLACEHOLDER: thanks, grant numbers and institutional support go here.
\"O.~S. is grateful to Ozan Tu\u{g}luk for his intangible support. U.Z.
is supported by the Scientific and Technological Research Council of
T\"urkiye (T\"UB\.ITAK) under Grant Nos. 125F467 (ARDEB 1001) and
125F024 (ARDEB 3501).

\paragraph*{Use of artificial intelligence.}
An LLM (Claude, Anthropic) was used in the preparation of this work. It verified the field
equations, the Carroll limits, the third-way identities and the boundary charges
symbolically. It also checked the reference list against the INSPIRE-HEP records. It
assisted in the derivation of the asymptotic symmetry algebra of Section~\ref{bms} and of
the thermodynamic relations of Section~\ref{thermo}, and it drafted and edited much of the
text. Every result reported here was subsequently rederived and checked by the authors, who take
full responsibility for the content of this paper.

%=======================================================================
%  DECLARATIONS
%=======================================================================
\section*{Declarations}

\paragraph*{Data availability.}
This manuscript has no associated data. [Authors' comment: Data sharing not applicable
to this article as no datasets were generated or analyzed during the current study.]

\paragraph*{Competing interests.}
The authors declare that they have no competing interests.

\paragraph*{Author contributions.}
All authors contributed to the conception of the work and to the calculations reported in
it, all authors participated in writing the manuscript, and all authors read and approved
the final version.

%=======================================================================
%  REFERENCES
%=======================================================================

\end{document}